\documentclass[lettersize,journal]{IEEEtran}
\usepackage{amsmath,amsfonts}
\usepackage{algorithmic}
\usepackage{algorithm}
\usepackage{array}
\usepackage{balance}
\usepackage[caption=false,font=footnotesize]{subfig}
\usepackage{textcomp}
\usepackage{stfloats}
\usepackage{url}
\usepackage{verbatim}
\usepackage{graphicx}
\usepackage{cite}
\usepackage{xcolor}
\usepackage{pdfsync}
\usepackage{makecell}
\usepackage{multirow}
\usepackage{booktabs}
\usepackage{etoolbox}
\usepackage{eurosym}
\usepackage{threeparttable}
\usepackage{tabularx}
\usepackage[table]{colortbl}
\usepackage[colorlinks,citecolor=blue,urlcolor=black,linkcolor=black]{hyperref} 
\begin{document}

\title{Communication-Efficient Large-Scale Distributed Deep Learning: A Comprehensive Survey}
\author{Feng~Liang,~\IEEEmembership{Member,~IEEE},
        Zhen~Zhang,
        Haifeng~Lu, 
        Victor~C.~M.~Leung,~\IEEEmembership{Life Fellow,~IEEE},
        Yanyi~Guo$^{\ast}$,
        Xiping~Hu$^{\ast}$,~\IEEEmembership{Member,~IEEE}
\thanks{$^{\ast}$ Corresponding authors.}%
\thanks{Feng Liang and Xiping Hu are with a) Artificial Intelligence Research Institute, Shenzhen MSU-BIT University, Shenzhen 518000, China, and b) Guangdong-Hong Kong-Macao Joint Laboratory for Emotional Intelligence and Pervasive Computing, Shenzhen MSU-BIT University, Shenzhen 518107, China. (e-mail: \{fliang, huxp\}@smbu.edu.cn)}%
\thanks{Zhen Zhang and Haifeng Lu are with a) Gansu Provincial Key Laboratory of Wearable Computing, School of information Science and Engineering, Lanzhou University, Gansu 730000, China, and b) Guangdong-Hong Kong-Macao Joint Laboratory for Emotional Intelligence and Pervasive Computing, Shenzhen MSU-BIT University, Shenzhen 518107, China. (e-mail: \{zhangzhen19, luhf18\}@lzu.edu.cn)}%
\thanks{Victor C. M. Leung is with a) Artificial Intelligence Research Institute, Shenzhen MSU-BIT University, Shenzhen 518000, China, and b) the Department of Electrical and Computer Engineering, The University of British Columbia, Vancouver, Canada. (e-mail: vleung@ieee.org)}%
\thanks{Yanyi Guo is with a) Frontier Cross Disciplinary Research Institute, Shenzhen MSU-BIT University, Shenzhen 518000, China, and b) the School of Mechanical and Electrical Engineering, Beijing Institute of Technology, Beijing 10081, China. (e-mail: guoyy@smbu.edu.cn)}%
}

\markboth{IEEE Communications Surveys \& Tutorials,~Vol.~14, No.~8, August~2021}%
{Shell \MakeLowercase{\textit{et al.}}: A Sample Article Using IEEEtran.cls for IEEE Journals}


\maketitle

\begin{abstract}
With the rapid growth in the volume of data sets, models, and devices in the domain of deep learning, there is increasing attention on large-scale distributed deep learning.
In contrast to traditional distributed deep learning, the large-scale scenario poses new challenges that include fault tolerance, scalability of algorithms and infrastructures, and heterogeneity in data sets, models, and resources. 
Due to intensive synchronization of models and sharing of data across GPUs and computing nodes during distributed training and inference processes, communication efficiency becomes the bottleneck for achieving high performance at a large scale.
This article surveys the literature over the period of 2018-2023 on algorithms and technologies aimed at achieving efficient communication in large-scale distributed DL at various levels, including algorithms, frameworks, and infrastructures. 
Specifically, we first introduce efficient algorithms for model synchronization and communication data compression in the context of large-scale distributed training.
Next, we introduce efficient strategies related to resource allocation and task scheduling for use in distributed training and inference. 
After that, we present the latest technologies pertaining to modern communication infrastructures used in distributed deep learning, including GPU interconnects, programmable network devices, collective communication protocols, and communication topologies.
We focus our discussion of these topics on examining the impact of the communication overhead in a large-scale and heterogeneous setting.
Finally, we conduct a case study on the distributed training of large language models at a large scale to illustrate how to apply these technologies in real cases.
This article aims to offer researchers a comprehensive understanding of the current landscape of large-scale distributed deep learning and to reveal promising future research directions toward communication-efficient solutions in this scope.
\end{abstract}

\begin{IEEEkeywords}
Communication-efficient, distributed deep learning, scalability, federated learning, heterogeneity, LLM, large models, large scale, pipeline parallelism.
\end{IEEEkeywords}

\section{Introduction}
Due to the rapid advancements in computational power of GPUs~\cite{gpu13} and the development of foundation artificial neural network models such as ResNet~\cite{resnet16} and Transformer~\cite{transformer17}, deep learning (DL) has become the state-of-the-art approach across diverse fields over the past decade. The fields include natural language processing (NLP)~\cite{transformerTranslate15,bert19}, multimedia processing~\cite{imageRecog20,wav2vec20}, biomedical engineering~\cite{aiMolecular23,deepSkinLesion23}, and autonomous driving solutions~\cite{mapLessILAuto22,DRLAuto22}.
Traditional DL models and the associated training data sets can run on a single GPU or server node without inter-GPU or inter-node communication. 
However, with the increase in the sizes of data sets and DL models, a standalone GPU or node cannot handle DL tasks efficiently, and distributed DL emerges to help.

Distributed DL has become the state-of-the-art solution for addressing challenges present in complex scenarios of artificial intelligence.
Distributed DL entails the training or inference of deep neural network (DNN) models on multiple CPUs or GPUs in one or multiple computing nodes to handle large training data sets and extensive learning models. 
The benefits of distributed DL are threefold.
Firstly, distributed DL enhances the training parallelism on large data sets and optimizes hyperparameters tuning. 
For instance, numerous DL models used in the fields of remote sensing~\cite{ddlRemoteSensing20,ddlRemoteSensing21} require processing vast volumes of high-revolution multimedia data across multiple modalities to improve classification accuracy. 
To enhance training parallelism, data sets can be distributed across multiple nodes, with each node training models independently and sharing its efforts through a specific synchronization mechanism. 
Secondly, distributed DL facilitates the training and inference of large artificial neural network models. 
In particular, to train a large language model (LLM)~\cite{lamda22} with tens of billions of parameters, a large number of GPUs and nodes are required to collaborate in accommodating the entire model and to perform parallel training with distributed data for rapid convergence. 
Thirdly, the evolution of Internet of Things (IoT) and Internet of Vehicles (IoV) has led to intricate scenarios that require use of distributed DL. 
Distributed DL empowers IoT and IoV solutions~\cite{semisupervisedAIoT20} by enabling them to make intelligent decisions by leveraging the computational and communication capabilities of the servers, network infrastructures, and end devices.

Communications and distributed DL are intricately intertwine themes, forming integral components of each other's core functionalities.
On the one hand, distributed DL has emerged as a prominent technique for addressing and optimizing diverse communications problems.
With the surge in communication devices and network traffic volume, distributed DL offers a real-time and agile approach for tasks encompassing traffic analytics, routing, and network resource management across diverse communications domains, such as wireless communications, IoT, and network security.
The inherent distributed nature of this approach also contributes to robustness and fault tolerance, thereby ensuring reliable communication in dynamic environments.
On the other hand, efficient communications technologies play a pivotal role in achieving high-performance distributed DL.
Given the collaborative and coordination-driven nature of distributed DL, communications permeate nearly all aspects of this domain and act as its driving force.
Technologies aimed at optimizing high-performance distributed DL fall predominantly into the category of efficient communications technologies.
Because of this mutual dependence, advancements in one domain impact significantly the capability of the other. 
While the former case, exploring distributed DL for communications, has been investigated extensively in numerous studies~\cite{dlWirelessSurvey19,drlCommSurvey19,dlSecuritySurvey21,dlIoTSecuSurvey21,aiIoTSurvey22}, 
this article emphasizes strategically the latter case, communications for distributed DL, for focused discussion.

Efficient communication is crucial for achieving high performance at different levels in distributed DL. 
1) At the algorithm level, synchronizing models during distributed training involves intensive inter-GPU and inter-node communication to ensure model consistency and convergence performance~\cite{layerSgdFL20}. 
Optimizing communication frequency and traffic volume to reduce communication overhead can help models converge swiftly with significantly less training time. 
2) At the framework level, underutilizing communication resources and having inferior task scheduling cause communication traffic congestion and straggler problems in distributed DL~\cite{energyEfficientFL20}. 
The performance of distributed DL can be enhanced greatly by fully utilizing communication resources, balancing the allocation of computational and communication resources, and overlapping computational and communication tasks to prevent blocking communication.
3) At the infrastructure level, low-performance communication links, devices, protocols, and topologies can offset the power of high-performance computing units and are prone to becoming the bottleneck of the overall system of distributed DL~\cite{topoImpactDML19}. 
A high-performance and cost-effective solution for all these communication infrastructure layers~\cite{rdmaDDL19} can maximize the computational and communication capacity of the entire distributed DL system.
Addressing these communication challenges at various levels in diverse environments is crucial for high-performance distributed DL.

Recently, the growing number of computational and communication devices in systems has led to the emergence of large-scale distributed DL and posed additional challenges for this context.
The first challenge pertains to algorithmic complexity. 
In large-scale distributed DL, the increased number of devices and volume of workloads introduce additional computational and communication overhead. 
Efficient algorithms used for model synchronization and communication data compression that can scale linearly are crucial.
Otherwise, large-scale distributed DL can introduce more overhead than benefits.
Simultaneously, given that the optimization solution search space for distributed DL algorithm can grow exponentially with the scale, designing optimal algorithms for larger-scale distributed DL can be significantly more challenging.
The second challenge concerns heterogeneity. 
In large-scale distributed DL, heterogeneity is prevalent in various aspects, including the data distribution, model specification, and geographical location and resource capacity of computational and communication devices.
The heterogeneity not only degrades the convergence performance of distributed DL, but also impacts communication efficiency, and is exacerbated by problems of resource underutilization and stragglers.
The heterogeneous factors further contribute to the algorithmic complexity in designing communication-efficient distributed DL algorithms. 
The third challenge concerns large models. 
The size of large models has grown exponentially compared to previous sizes.
Distributed training of large models at scale has raised both theoretical and practical concerns about various factors, including convergence performance, training efficiency, and the costs of computational and communication resources.
Given the increased number of devices and extended training time required, fault tolerance has also become more critical for large-model training compared to traditional cases with medium-sized models.

With the rapid development of large-scale distributed DL, 
there is an urgent need for conducting a comprehensive survey on communication-efficient large-scale distributed DL technologies that aim to empower researchers in the fields of communications, computer science, and artificial intelligence to understand critical research problems in this domain and to make valuable contributions.
Existing surveys on distributed DL do express concerns about communication acting as a bottleneck at various levels. However, they lack a systematic and comprehensive investigation into the communication problems and solutions in the large-scale scenario.


This article surveys the literature over the period of 2018-2023 on communication-efficient technologies that operate at various levels of large-scale distributed DL, including the algorithm, framework, and infrastructure levels. 
For a comprehensive understanding of the development history of a specific topic, some milestone works before 2018 may be included. 
The topics covered in this article include model-synchronization and communication-data-compression algorithms, resource-allocation and task-scheduling strategies, and communication infrastructures. 
The discussion of these topics focuses on two dimensions: 1) \textit{how resolving the bottleneck of communication can enhance the performance of distributed DL;} 
and 2) \textit{how they can be high-performance in large-scale settings.} 
We also summarize some lessons learned from each topic to emphasize promising future research directions toward high-performance large-scale distributed DL. 
At the end this article, we conduct a case study on the large-scale distributed training of LLM to explore how these communication-efficient solutions can be applied practically in real cases. 



\newcolumntype{Y}{>{\centering\arraybackslash}X}
\newcommand\tabTitle[1]{\cellcolor{blue!15}\textbf{#1}}
\begin{table*}[!t]
    \caption{A Comparison of Related Surveys}\label{tab:survey}
    \centering
    \begin{tabularx}{\textwidth}{ cc *{8}{Y} }
    \Xhline{2\arrayrulewidth}
        \tabTitle{Ref.} & \tabTitle{Year} & \tabTitle{Comm-Efficient Synchro-nization} &\tabTitle{Comm Data Compression} & \tabTitle{Comm-Efficient Resource Allocation} & \tabTitle{Comm-Efficient Task Scheduling} & \tabTitle{Comm \mbox{Infrastructure}} & \tabTitle{Edge-Cloud Heterogeneity} & \tabTitle{Large-Scale Distributed DL} & \tabTitle{Distributed Training of Large Models}\\
        \Xhline{2\arrayrulewidth}
        \cite{ddl20} & 2020 & \checkmark & \checkmark & & ~ & ~ & ~ & ~ & \\ 
        \rowcolor{gray!15}
        \cite{paralleDDL19} & 2019 & \checkmark & \checkmark & ~ & \checkmark & ~ & ~ & ~ & \\ 
        \cite{scalableDL20} & 2020 & \checkmark & \checkmark & ~ & \checkmark & \checkmark &  & \checkmark & \\ 
        \rowcolor{gray!15}
        \cite{principlesML16} & 2016 & \checkmark & ~ & ~ & \checkmark & ~ &  & \checkmark & \\ 
        \cite{commOptDDL21} & 2021 & \checkmark & \checkmark & ~ & \checkmark & \checkmark & ~ & ~ & \\ 
        \rowcolor{gray!15}
        \cite{ddlCommAlgoSurvey23} & 2023 & \checkmark & \checkmark & ~ & \checkmark & ~ & ~ & ~ & \\ 
        \cite{commDL23} & 2023 & \checkmark & \checkmark & \checkmark & ~ & ~ & ~ & ~ & \\ 
        \rowcolor{gray!15}
        \cite{commDDL23} & 2023 & \checkmark & \checkmark & ~ & \checkmark & ~ & ~ & ~ & \\ 
        \cite{wirelessEdge19} & 2019 & \checkmark & ~ & ~ & ~ & \checkmark & \checkmark & \checkmark & \\ 
        \rowcolor{gray!15}
        \cite{EdegAI19} & 2019 & \checkmark & \checkmark & \checkmark & ~ & ~ & \checkmark & \checkmark & \\ 
        \cite{commEdgeAI20} & 2020 & \checkmark & \checkmark & ~ & \checkmark & ~ & \checkmark & \checkmark & \\ 
        \rowcolor{gray!15}
        \cite{edgeAI20} & 2020 & \checkmark & \checkmark & ~ & ~ & ~ & \checkmark & ~ & \\ 
        \cite{communicationFL21} & 2021 & \checkmark & \checkmark & \checkmark & ~ & ~ & \checkmark & ~ & \\
        \rowcolor{gray!15}
        \cite{MLEdge21} & 2021 & \checkmark & \checkmark & ~ & ~ & \checkmark & \checkmark & ~ & \\ 
        \cite{dmlToFL22} & 2022 & \checkmark & \checkmark & ~ & ~ & ~ & \checkmark & ~ & \\ 
        \rowcolor{gray!15}
        \cite{EndEdgeCloud22} & 2022 & \checkmark & \checkmark & ~ & ~ & ~ & \checkmark & \checkmark & \\ 
        \cite{edgeCloudAI22} & 2022 & ~ & \checkmark & ~ & ~ & \checkmark & \checkmark & \checkmark & \\ 
    \Xhline{2\arrayrulewidth}
        Ours & - & \checkmark & \checkmark & \checkmark & \checkmark & \checkmark & \checkmark & \checkmark & \checkmark  \\ 
    \Xhline{2\arrayrulewidth}
    \end{tabularx}
\end{table*}

\subsection{Related Surveys}
Table~\ref{tab:survey} compares our survey with other related surveys on various topics. 
An empty cell denotes insufficient coverage of the topic in the corresponding survey. 
Existing surveys explore distributed DL from various perspectives, but do not address thoroughly key aspects regarding the large-scale setting, especially the critical role of communication in handling large volumes of data, models, devices, and infrastructures. 
They typically do not explore sufficiently the scalability of algorithms and infrastructures at different levels of the distributed DL ecosystems in the large-scale setting. 

Generally, existing surveys on distributed DL have the following drawbacks regarding the large-scale setting: 
\begin{itemize}
        \item \textbf{About heterogeneity.} Few surveys, when discussing technologies for communication optimization in distributed DL, highlight the impacts of heterogeneity, including model synchronization and compression of models for communication.
        The issue of heterogeneity in data, models, devices, and infrastructures is pervasive and salient in modern distributed DL scenarios, especially in large-scale distributed DL.
        \item \textbf{About resource allocation.} Existing surveys lack a thorough study on communication-efficient resource-allocation strategies. In a multi-tenant environment with a large number of servers and devices for computing and network infrastructures for communication, optimizing the allocation of computational and communication resources for better resource utilization is critical to the performance of large-scale distributed DL.
        \item \textbf{About communication infrastructures.} Few surveys study the infrastructure for efficient communication in large-scale distributed DL. 
        Communication infrastructures, encompassing interconnects, network devices, collective communication libraries, and network topologies, are of paramount importance for high-performance large-scale distributed DL. 
        The rapid development of these communication infrastructure technologies deserves timely attention and follow-up.
        \item \textbf{About large models.} Existing surveys lack a focus on distributed DL with large models. Given the success of extremely large foundation models in fields such as NLP, there is an urgent need for the study of various technologies for distribute DL in the scenario of large models. 
\end{itemize}

Specifically, several surveys~\cite{ddl20,paralleDDL19,scalableDL20} have focused on distributed DL from various perspectives. 
In particular, Verbraeken \textit{et al.}~\cite{ddl20} provide a review of distributed DL algorithms and frameworks.
Ben-Nun and Hoefler~\cite{paralleDDL19} provide an analysis on the concurrency of parallel and distributed DL architectures and models.
Mayer and Jocobsen~\cite{scalableDL20} focus on parallelization methods to enable scalable distributed training and cover various topics such as distributed resource and multi-tenant management.
However, these surveys do not emphasize adequately the communication as the critical bottleneck when discussing the ecosystem of distributed DL.

Several surveys~\cite{principlesML16,commOptDDL21,ddlCommAlgoSurvey23,commDL23,commDDL23} have addressed communication topics in distributed DL at various levels. 
Particularly, Xing \textit{et al.}~\cite{principlesML16} provide a survey on strategies and principles for distributing DL algorithms in a cluster, aiming to enhance inter-node communication efficiency. 
Ouyang \textit{et al.}~\cite{commOptDDL21} focus on communication strategies aimed at reducing network communication traffic through algorithm-level optimization and accelerating network communication speed through network-level optimization.
Both Yu \textit{et al.}~\cite{ddlCommAlgoSurvey23} and Cao \textit{et al.}~\cite{commDL23} further summarize and categorize communication-optimization strategies for distributed DL such as communication frequency reduction and communication data compression, at the algorithm level.
In addition, Cao \textit{et al.}~\cite{commDL23} also explore radio-resource-management strategies and game theory approaches for improving communication efficiency in distributed DL. 
Tang \textit{et al.}~\cite{commDDL23} conduct a survey of efficient communication technologies used in distributed DL at the architecture and application levels. 
Organized in a presentation structure that is partially similar to this paper, 
the survey focuses on topics like communication synchronization, system architectures, compression techniques, and parallelism of computational and communication tasks. 
However, it do not tackle the influence of heterogeneity issues and the scalability of the related technologies when discussing these topics. It also overlooks certain important topics for large-scale distributed DL, such as resource allocation and communication infrastructures. 

Several surveys address the communication challenges in specific distributed DL paradigms, including cloud-edge-based~\cite{wirelessEdge19,EdegAI19,commEdgeAI20,edgeAI20,communicationFL21,MLEdge21}, P2P-based~\cite{dmlToFL22}, and hybrid~\cite{EndEdgeCloud22,edgeCloudAI22} paradigms.
Notably, surveys that focus on the edge-based paradigm cover different aspects of communication-efficient technologies, including edge-specific architectures~\cite{wirelessEdge19,EdegAI19}, algorithms~\cite{commEdgeAI20,edgeAI20,communicationFL21}, and infrastructures~\cite{MLEdge21}.

To fill the gap in existing surveys on distributed DL, this article particularly surveys communication-efficient technologies in large-scale scenarios. 
Specifically, this article focuses on:
\begin{itemize}
        \item Communication-efficient distributed DL algorithms in heterogeneous settings. 
        We highlight the impact of heterogeneity in data, model, and resources when presenting various optimization and compression algorithms, discussing how to improve communication performance in this context.
        \item Strategies for resource allocation and task scheduling aimed at fully utilizing computational and communication resources and increasing distributed DL throughput in large-scale settings.
        We explore various resource-allocation strategies for large-scale distributed training and inference, considering framework and container levels. We also study various task-scheduling strategies for overlapping computational and communication tasks, aiming to increase distributed training and inference throughput with diverse heterogeneous resources and workloads.
        \item Contemporary communication infrastructures for high-performance communication. 
        We study various state-of-the-art communication infrastructure technologies on different layers, including GPU interconnects, programmable network devices, collective communication interfaces, and communication topologies.
        \item A case study of various technologies for large foundation DL models in large-scale distributed DL.
        We use a question-and-answer approach to discuss applying these communication-efficient technologies to the distributed training of LLMs in real cases. This case study helps researchers identify practical, high-performance, and cost-effective solutions for large-model training in a large-scale and heterogeneous environment. 
\end{itemize}

\begin{figure*}[!t]
\centering
\includegraphics[width=1\textwidth]{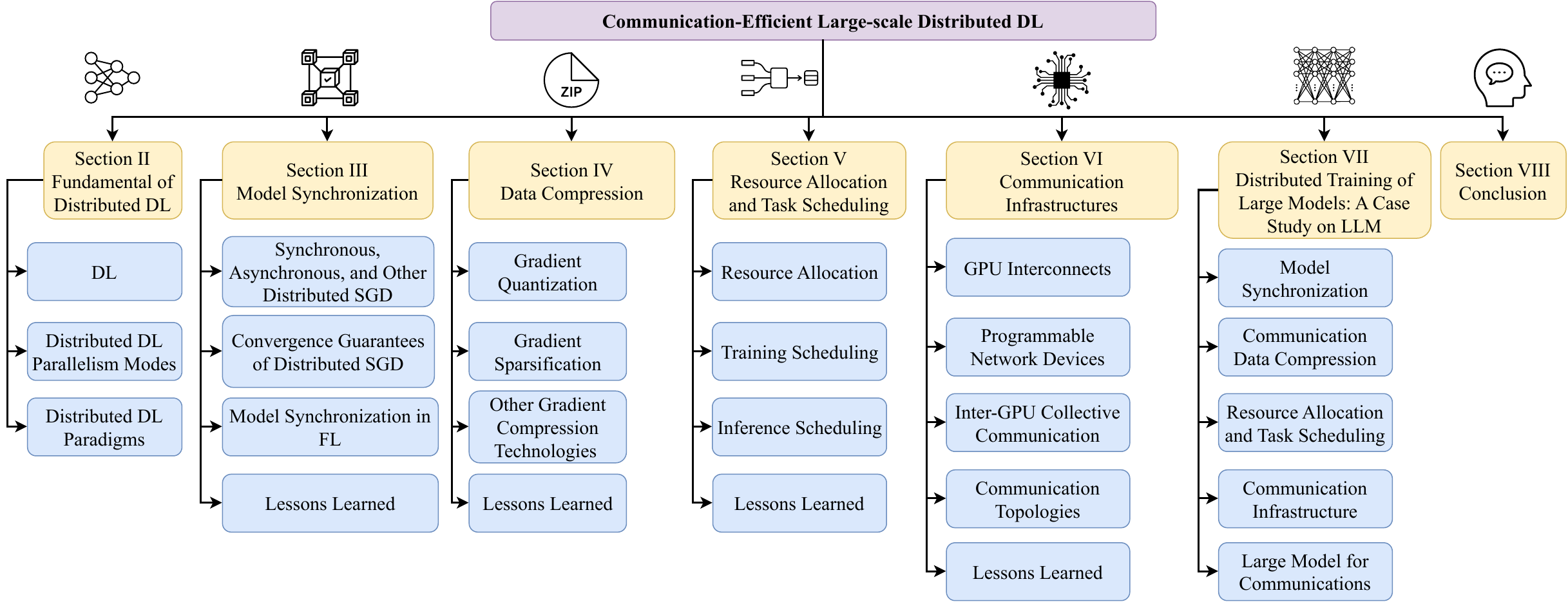}
\caption{The organization of the survey}
\label{fig:organization}
\end{figure*}

\subsection{Survey Organization}
Fig.~\ref{fig:organization} outlines the detailed organization of the remaining sections in this survey. 
Section~\ref{sec:background} provides fundamental knowledge about distributed DL. 
Sections~\ref{sec:synchronization} and~\ref{sec:compression} present works on communication-efficient algorithms for model synchronization and communication data compression in large-scale distributed DL, respectively.
Section~\ref{sec:resourceTask} examines various communication-efficient strategies for resource allocation and task scheduling in large-scale distributed training and inference.
Section~\ref{sec:infrastructure} introduces works on communication infrastructure technologies at different system layers for high-performance communication in large-scale DL clusters.
We present large-scale distributed training of large foundation DL models as a case study in Section~\ref{sec:caseStudy}, 
and conclude this survey in Section~\ref{sec:conclusion}.

\begin{table}[!t]
\caption{A List of Common Abbreviations}\label{tab:abbr}
\centering
\begin{tabular}{cl}
\Xhline{2\arrayrulewidth}
\cellcolor{blue!15}\textbf{Abbreviation} & \cellcolor{blue!15}\textbf{Description}\\
\Xhline{2\arrayrulewidth}
AIMD & Additive-Increase Multiplicative-Decrease\\
CNN & Convolutional Neural Network\\
DL & Deep Learning\\
DLRM & Deep Learning Recommendation Model\\
DNN & Deep Neural Network\\
FL & Federated Learning \\
FPGA & Field Programmable Gate Array\\
GPU & Graphics Processing Unit\\
GRU & Gated Recurrent Unit\\
IID & Independent and Identically Distributed\\
INA & In-Network Aggregation\\
IoT & Internet of Things\\
IoV & Internet of Vehicles\\
LARS & Layer-wise Adaptive learning Rate Scaling\\
LLM & Large Language Model\\
LSTM & Long Short-Term Memory\\
MPI & Message Passing Interface\\
MPS & Multiple Process Sharing\\
NCCL & NVIDIA Collective Communications Library\\
NIC & Network Interface Card\\
NLP & Natural Language Processing\\
PaaS & Platform as a Service\\
PCIe & Peripheral Component Interconnect Express\\
PS & Parameter Server\\
P2P & Peer-to-Peer\\
RNN & Recurrent Neural Network\\
SGD & Stochastic Gradient Descent\\
SiP & Silicon Photonic\\
SLO & Service-Level Objective\\
\Xhline{2\arrayrulewidth}
\end{tabular}
\end{table}

\section{Fundamentals of DL and Distributed DL}\label{sec:background}
In this section, we present the fundamentals of DL and distributed DL.
Table~\ref{tab:abbr} includes common abbreviations used in this survey.

\subsection{Deep Learning}
DL is a subfield of machine learning that utilizes deep artificial neural networks, also known as deep neural networks (DNN), to extract complex patterns from training data in a hierarchical manner. 
The trained DNN is capable to recognize/predict patterns in unseen data. 
DL has been used in various fields, including NLP~\cite{transformerTranslate15}, computer vision~\cite{imageRecog20}, and biomedical engineering~\cite{deepSkinLesion23}.

\begin{figure*}[!t]
\centering
\subfloat[Fully Connected DNN]{\includegraphics[width=0.27\textwidth]{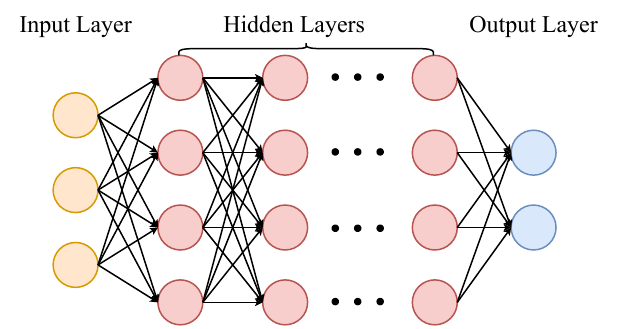}%
\label{fig:model_dnn}}
\hfil
\subfloat[CNN]{\includegraphics[width=0.3\textwidth]{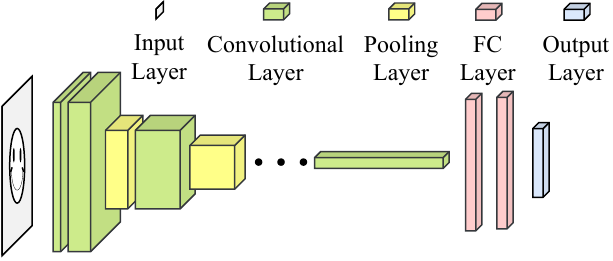}%
\label{fig:model_cnn}}
\hfil
\subfloat[RNN]{\includegraphics[width=0.4\textwidth]{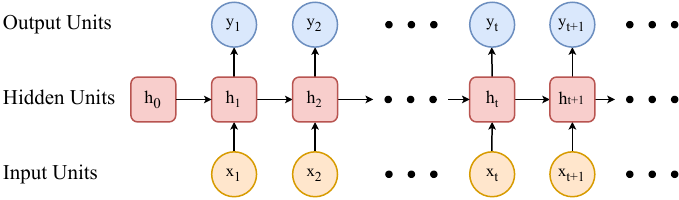}%
\label{fig:model_rnn}}
\caption{Common artificial neural network models for DL}
\label{fig:model}
\end{figure*}

\subsubsection{DL models}
A DNN consists of multiple hidden layers. Each layer is comprised of neurons, which are typically activated by non-linear functions. 
Based on the connections between neurons within and between layers, there can be various types of DNN models.
In this survey, when referring to models or DL models, we mean DNNs unless the context otherwise specifies.
Fig.~\ref{fig:model} illustrates three basic DNN models: the fully connected DNN, convolutional neural network (CNN), and recurrent neural network (RNN). 

$\bullet$ \textit{Fully connected DNN:} The fully connected DNN, also known as the feedforward neural network, constitutes a dense network with an input layer, a number of hidden layers, and an output layer, as depicted in Fig.~\ref{fig:model_dnn}. 
Neurons in a preceding layer connect to all neurons in the subsequent layer, and each connection has a learnable weight parameter indicating the strength of the connection. 
This architecture enables the fully connected DNN to capture complex relationships within data, finding extensive application in tasks such as classification~\cite{commDnnClassification23}, regression~\cite{commDnnRegression22}, and feature representation embedding~\cite{commDnnFeature20}. 

$\bullet$ \textit{CNN:} CNN stands as a prevalent model designed for feature extraction and classification, primarily tailored for image and video data. 
As depicted in Fig.~\ref{fig:model_cnn}, in addition to the input and output layers, CNN comprises a stack of convolutional layers and pooling layers for feature extraction, succeeded by fully connected (FC) layers for classification.
Unlike the fully connected layer, which assigns a weight parameter to each neuron connection, the convolutional layer substantially reduces the number of weight parameters by utilizing a number of kernels, or filters, each containing shared weights for feature extraction. 
The feature extraction process of the convolutional layer's feature extraction process is empowered by the convolution operation, wherein kernels traverse the receptive fields of an image, extracting new features through weighted summations followed by a non-linear activation function. 
The pooling layer, typically using max-pooling or average-pooling functions, downsamples the data in the convolutional layer to reduce feature dimensions and alleviate overfitting issues. 
CNN has found widespread applications in various computer vision tasks, including image classification~\cite{snsCnnImage19}, semantic segmentation~\cite{transformerCnnRemoteSensing22}, and object detection~\cite{vehicleCnnObject22}. 

$\bullet$ \textit{RNN:} RNN, a DL model that deals with sequential data like time-series data, natural language, and speech audio, is illustrated in Fig.~\ref{fig:model_rnn}.
The general architecture of RNN 
includes hidden units that capture and propagate temporal context from the input sequence to subsequent hidden unites. 
It updates continuously and utilizes the temporal context based on the current input and previous temporal context to make predictions.
To address the challenge of capturing long-range temporal dependencies, 
two common variants of RNNs, known as Long Short-Term Memory (LSTM) and Gated Recurrent Unit (GRU), have been developed,
providing a trade-off between modeling such dependencies and reducing computation complexity effectively.
Common applications of RNN include tasks such as time series forecasting~\cite{timeSeriesRnn19}, NLP~\cite{secureNlp20}, and automated planning~\cite{automatedPlanRnn21}. 

\subsubsection{Training and inference}
The training of a DL model is the process of optimizing its parameters to minimize the prediction error on a training data set, as determined by a specified loss function, or objective function. 
Loss functions can be either convex or non-convex, leading to convex or non-convex optimization problems.
Training can be decomposed into two key processes: feedforward and backpropagation. 
In the feedforward process, training data are passed into the model's input layer, and the output prediction is computed by forwarding data through the network using the current model parameters. 
In the backpropagation process, the prediction error and gradients are calculated with respect to the loss function, and trainable parameters are updated iteratively in a backward manner, optimizing the model for the minimum loss.
Common optimizers for backpropagation updating include minibatch Stochastic Gradient Descent (SGD)~\cite{minibatchSGD14}, SGD with momentum~\cite{momentumSgd13}, Adagrad~\cite{adagrad11}, and Adam~\cite{adam15}.
The training process usually operates on batches of training data iteratively over multiple epochs until the model converges.
A model is said to have converged when the training error settles to within a predefined error range, and additional training will not further decrease the error. 
After completing the training process, the weight parameters in the DL model are learned and fixed. 
Following the training process, there is typically a validation process for validating the performance of the trained model, providing information for fine-tuning hyperparameters and retraining the model for better performance.

The inference process passes forward unseen data through the trained DL model to make predictions. 
Depending on the specific requirements of an application, the resulting prediction can be extracted either either from the output layer or from the predicted latent representation in an intermediate hidden layer. 
For example, in the context of network traffic analysis, an end-to-end DNN model may be trained to classify traffic types directly~\cite{trafficClassificaiton20}, 
Alternatively, an encoder-decoder model trained on traffic data can utilize the latent representation generated by the encoder for subsequent tasks such as attack detection~\cite{latentRepresentationIot20}. 

Computing tasks related to a specific portion of the DL model for specific epochs during the training or inference process are generally referred to as DL tasks in this survey, when it is not necessary to distinguish training tasks and inference tasks in the context.  

\begin{figure}[!t]
\captionsetup[subfloat]{width=0.4\columnwidth}
\centering
\subfloat[Residual Block]{\includegraphics[width=0.40\columnwidth]{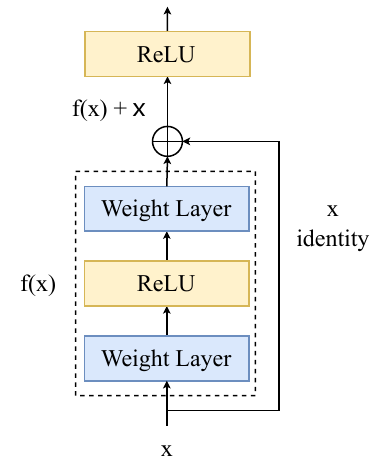}%
\label{fig:block_res}}
\vfil
\subfloat[Transformer Block]{\includegraphics[width=0.8\columnwidth]{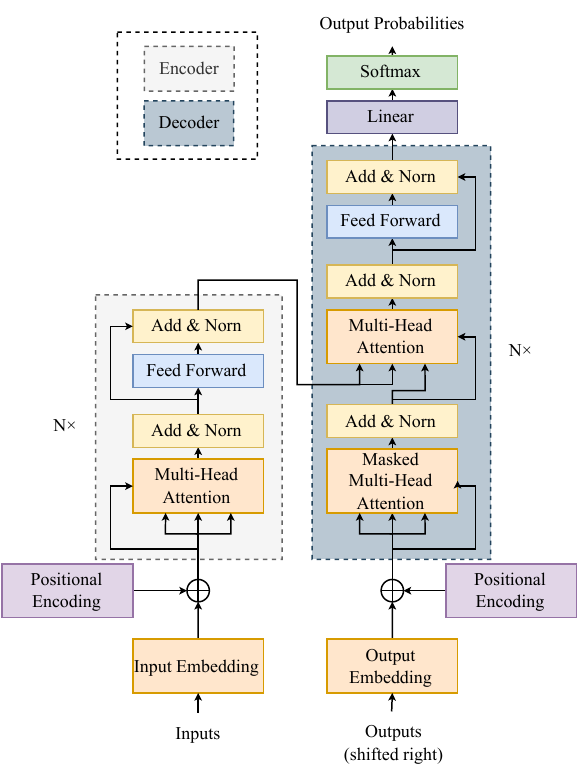}%
\label{fig:block_transformer}}
\caption{Fundamental blocks for deep neural networks}
\label{fig:block}
\end{figure}

\subsubsection{Fundamental neural network blocks}
DL models have experienced an exponential increase in terms of both depth and scale, with some models comprising thousands of neural layers~\cite{thousandLayer21} or tens of billions of parameters~\cite{lamda22}. 
However, the growth in the complexity of models has introduced various challenges, including the vanishing gradient problem~\cite{vanishingGradient18} and issues related to training and inference efficiency.
To overcome these challenges, certain neural network structures have been developed as fundamental blocks for building various DNN models that can capture complex patterns efficiently. 
In practice, a DNN model can be built easily by stacking these fundamental blocks on top of each other. 
Fig.~\ref{fig:block} illustrates two widely recognized fundamental neural network blocks: the Residual~\cite{resnet16} and Transformer~\cite{transformer17} blocks.  

$\bullet$ \textit{Residual:} The Residual block, depicted in Fig.~\ref{fig:block_res}, features a shortcut connection that adds the identity input to its mapping after passing through several layers. 
The shortcut connection facilitates the smooth flow of data through multiple layers, ensuring that important gradient updates can be propagated efficiently back to shallower layers. 
As a result, the Residual block addresses the issue of vanishing gradients efficiently, enabling efficient training in very deep models. 

$\bullet$ \textit{Transformer:} The Transformer block is a ubiquitous choice for DL models in learning tasks with sequential data, such as NLP problems. 
It adopts an autoencoder architecture characterized by a combined encoder and decoder, as shown in Fig.~\ref{fig:block_transformer}.
The Transformer block benefits from the self-attention mechanism, aligning the data with the context and enabling the data to attend to the important parts within that context. 
This mechanism can be calculated in parallel, making the Transformer block less time-consuming to train compared to previous RNN architectures such as LSTM and GRU.
This facilitates the building and training of large DL models, particularly giving rise to LLMs~\cite{gpt20,lamda22}, which serve as foundation models for various NLP tasks. 

\subsection{Distributed DL Parallelism Modes}
In distributed DL, data, models, and training or inference tasks are partitioned across multiple processing units (commonly GPUs in the DL context) within a single computing node or across multiple nodes in a cluster. 
Training DNN models in distributed environments involves intensive computation and communication within a computing cluster, making distributed training a focus of significant research.  
During the distributed training process, leveraging computational capabilities of available GPUs and nodes enhances training parallelism, thereby improving efficiency. 
Furthermore, it enables the successful completion of training tasks that involve extensive data or models exceeding the capacity of a single GPU or node.
Distributed Training requires intensive exchanges of model parameters or gradients across GPUs and nodes, where the communication process typically becomes the performance bottleneck. 
Crucially, designing communication-efficient algorithms for model synchronization, resource management, task scheduling, and infrastructures is essential for making distributed training a versatile solution for large-scale distributed DL tasks.

\begin{figure*}[!t]
\centering
\subfloat[Data Parallelism]{\includegraphics[width=0.3\textwidth]{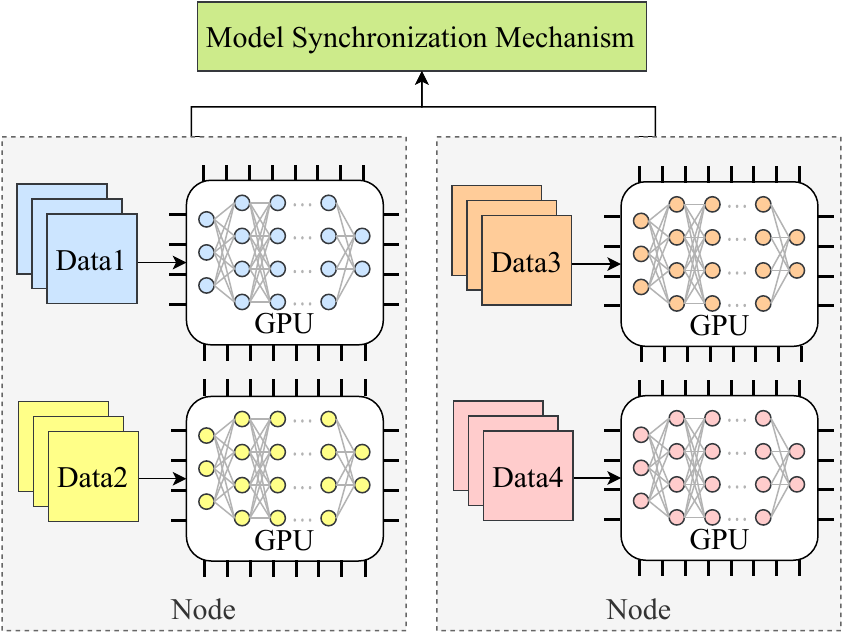}%
\label{fig:parallelism_data}}
\hfil
\subfloat[Model Parallelism]{\includegraphics[width=0.26\textwidth]{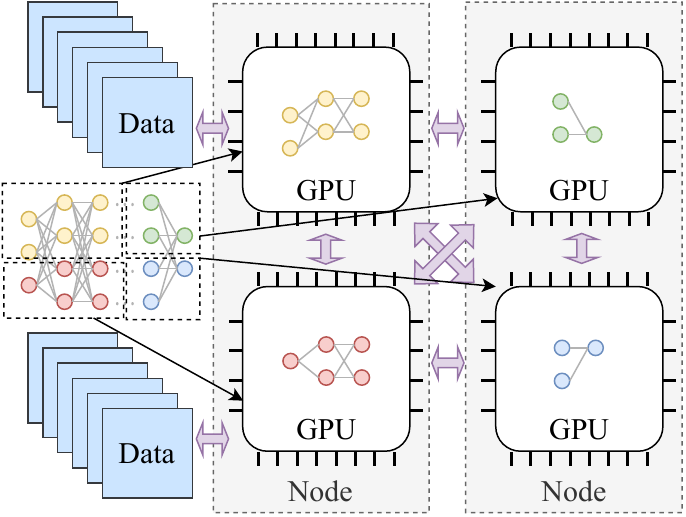}%
\label{fig:parallelism_model}}
\hfil
\subfloat[Pipeline Parallelism]{\includegraphics[width=0.33\textwidth]{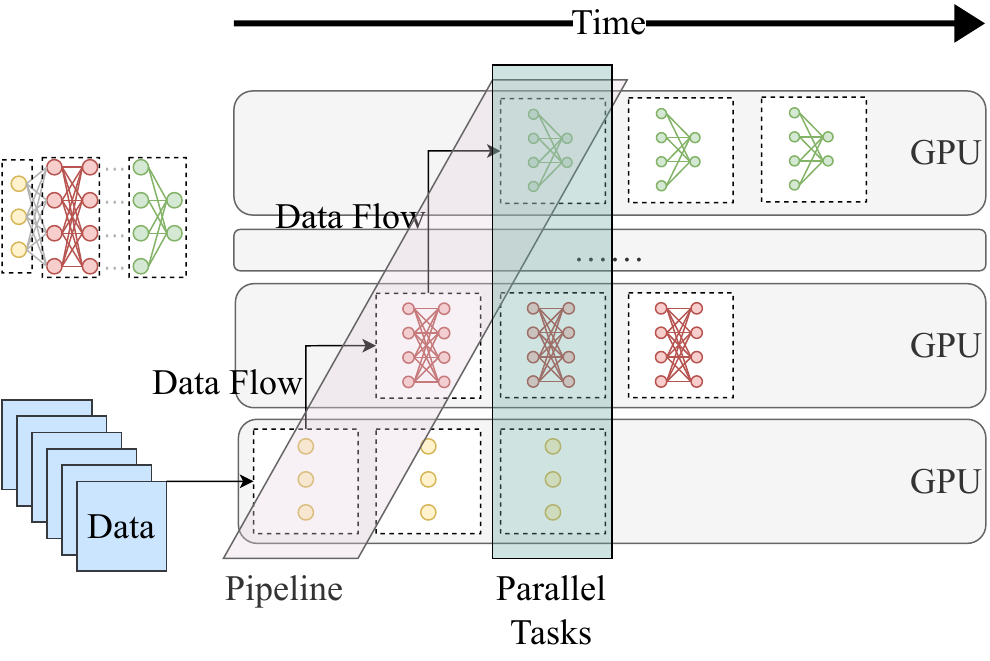}%
\label{fig:parallelism_pipeline}}
\caption{Parallelism modes of distributed DL}
\label{fig:parellelism}
\end{figure*}

Various partitioning strategies in distributed DL result in three prevalent parallelism modes, as illustrated in Fig.~\ref{fig:parellelism}:
data parallelism (Fig.~\ref{fig:parallelism_data}), model parallelism (Fig.~\ref{fig:parallelism_model}), and pipeline parallelism (Fig.~\ref{fig:parallelism_pipeline}).

\subsubsection{Data parallelism} 
In this mode, the entire training data set is divided into several splits, which are then distributed across multiple GPUs within a cluster~\cite{dataParallelPytorch20}, aiming to increase the level of training parallelism and reduce training time.
The DL model is replicated on each GPU, and each GPU trains a local model with an identical structure, concentrating on a specific partition of the data set. 
Throughout the distributed training process, these local models share their knowledge to update a global model using a specific model-synchronization mechanism. 
The convergence rate of distributed training is tied to the communication efficiency of the synchronization mechanism, considering factors such as the communication pattern and the underlying network infrastructure.
Moreover, it is influenced significantly by the trade-off between the model computation and synchronization communication performance, including factors such as communication frequency and the volume of data to be transferred across GPUs and computing nodes. 

\begin{figure}[!t]
\centering
\subfloat[Centralized]{\includegraphics[width=0.4\columnwidth]{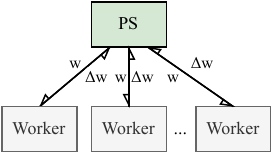}%
\label{fig:sgd_ps}}
\hfil
\subfloat[Decentralized]{\includegraphics[width=0.3\columnwidth]{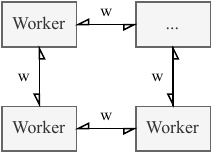}%
\label{fig:sgd_p2p}}
\caption{Distributed SGD}
\label{fig:distributedSgd}
\end{figure}

Distributed SGD is a ubiquitous model-synchronization mechanism employed for distributed training in the context of data parallelism. 
Fig.~\ref{fig:distributedSgd} illustrates two common architectures of distributed SGD: parameter-server-based centralized SGD and gossip-based decentralized SGD.
In centralized SGD with a parameter server (PS)~\cite{ps14} and multiple workers, SGD updates from local workers are transmitted to the parameter servers, either synchronously or asynchronously, to determine up-to-date parameters. 
Subsequently, these updated parameters are sent back to the workers for local model updating. 
In decentralized SGD without a central PS, workers synchronize their models with other workers either in a gossip manner or via a hierarchical structure of multiple parameter servers.

\subsubsection{Model parallelism} 
When an entire DL model exceeds the capacity of a single GPU or node, the model parallelism mode~\cite{modelParallelLargeModel22} employs a specific model-division strategy to separate the model into multiple neural network chunks, or called submodels. 
These submodels are then distributed across multiple GPUs within a cluster, 
with connections maintaining data communication for feedforward and backpropagation.
This enables a seamless flow of data and gradients throughout the entire model between GPUs and nodes, framing the training and inference process as if it were operating within a unified, black-box GPU. 
This parallelism mode introduces several crucial communication considerations. Initially, there is a need for optimizing the division of the DL model to minimize communication overhead. 
Additionally, establishing an efficient strategy to locate submodels is essential to ensure an optimized communication pattern with respect to submodel dependency.

\subsubsection{Pipeline parallelism} 
The pipeline parallelism mode further enhances the parallelism of DL tasks by reducing the complexity of submodel dependency in the model parallelism mode and preventing computational resources from idling~\cite{li2021chimera,liu2022autopipe,oh2022out}.
In this mode, distributed training tasks are decomposed layer-by-layer into multiple subtasks, with their completions depending on previous layers. 
The associated submodels for handling the subtasks are distributed across the GPU cluster.
Thus, when multiple batches of data flow through all submodels in a pipeline manner, 
each submodel is trained by different batches simultaneously. 
This mode is particularly applicable in the domains of edge-computing, IoT, and IoV, where devices have heterogeneous computational and communication capabilities to handle various distributed DL subtasks.  

In practice, these parallelism modes can be combined~\cite{tarnawski2021piper} to tackle complex DL tasks with large model structures, e.g., LLMs, which we will explore in a case study in Section~\ref{sec:caseStudy}.

\begin{figure*}[!t]
\centering
\subfloat[Cluster/Cloud]{\includegraphics[width=0.4\textwidth]{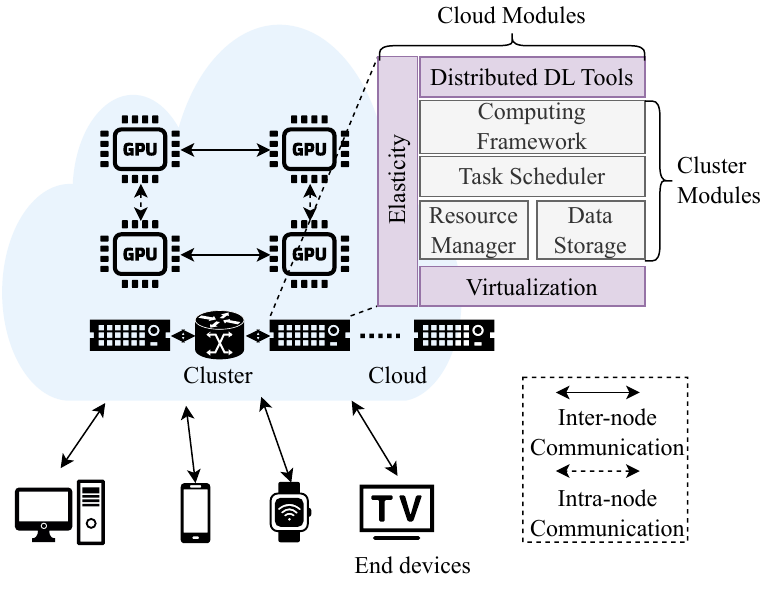}%
\label{fig:paradigm_cloud}}
\hfil
\subfloat[Edge/Federated]{\includegraphics[width=0.35\textwidth]{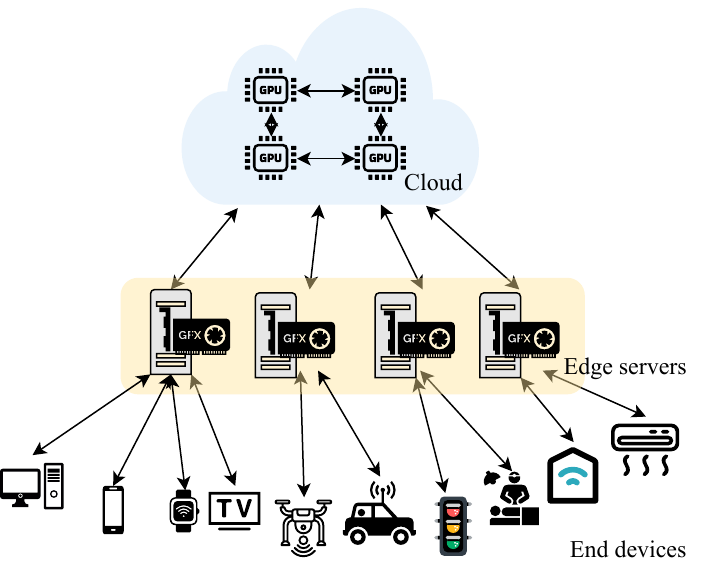}%
\label{fig:paradigm_edge}}
\hfil
\subfloat[P2P]{\includegraphics[width=0.25\textwidth]{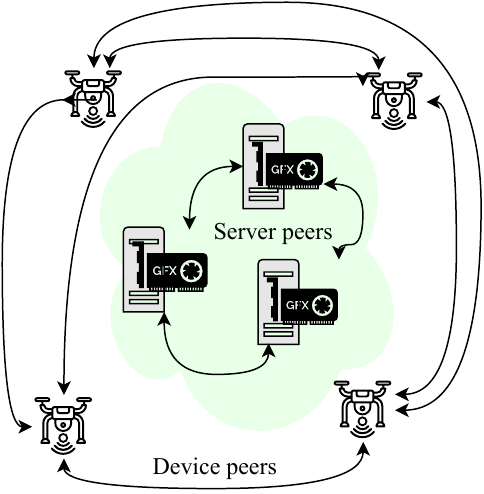}%
\label{fig:paradigm_p2p}}
\caption{Typical distributed DL paradigms}
\label{fig:paradigm}
\end{figure*}

\subsection{Distributed DL Paradigms}
We introduce mainstream paradigms of distributed DL, discussing their characteristics, research challenges, and applications.
Based on the communication pattern, training and inference locality, and hardware platform,
we classify these paradigms into three types: 
cluster/cloud-based, edge-based, and peer-to-peer-based (P2P-based).
Fig.~\ref{fig:paradigm} depicts each paradigm, using solid and dashed arrows 
to represent inter-node communications and intra-node communications, respectively.

\subsubsection{Cluster/Cloud-based Centralized Distributed DL}
The cluster-based paradigm, depicted in Fig.~\ref{fig:paradigm_cloud}, is a traditional scale-out solution for centralized distributed DL. 
A cluster of computing servers, interconnected via local networks, 
provide extended computing capability for training and inference 
beyond what a single computing server can achieve.
End devices, such as desktop computers, perceive the cluster as a unified entity, accessing its computational resources through unified network interfaces without being concerned with the communication mechanisms within the cluster.
The cluster comprises four key modules: the resource manager, data storage manager, task scheduler, and computing framework.
The resource manager oversees the computational resources (e.g., GPU, CPU, and memory) 
and network communication resources (e.g., network topology and bandwidth) in the cluster, allocating them to DL tasks. 
The data storage manager partitions and manages large training and testing data in a distributed manner, considering factors such as storage resource and network topology to ensure 
data availability, efficiency, and reliability.
The task scheduler breaks down DL tasks into various levels of granularity and schedules them within the cluster, taking into account resource consumption, data locality, workloads, and task characteristics to improve parallelism and failure tolerance.  
The computing framework implements the kernel and API to run the distributed DL algorithms efficiently with available resources allocated by the resource manager. 
This paradigm is commonly employed for accelerating model training~\cite{unityAccelerateTraining}, accommodating large models~\cite{modelParallelLargeModel22}, and facilitating hyperparameter tuning~\cite{antTuneHyper23}. 
In this context, efficient strategies for resource management and task scheduling are essential, aiming to maximize computational and communication-resource utilization, as well as the parallelism of task execution. This optimization is crucial for achieving high performance in this domain.

The cloud-based distributed DL paradigm can be viewed as an extension of the cluster-based paradigm, as shown in Fig.~\ref{fig:paradigm_cloud}.
Cloud service providers encapsulate the cluster-based distributed DL platform 
into Platform-as-a-Service (PaaS) products and deliver the distributed DL services in a pay-as-you-go manner. 
End devices are charged by the resources they actually use, 
including storage for the training data, computational resources for training and inference,
and traffic for network communication.
Compared to the cluster-based paradigm, the advantages of the cloud-based paradigm are threefold:
virtualization, elasticity, and utility. 
First, the cloud employs virtualization technologies that include resource isolation and network virtualization
to facilitate the sharing of computational and communication resources
among multiple tenants in the cloud, thus reducing the resource costs for tenants. 
Second, the cloud can adjust elastically resource capacities based on runtime learning workloads.
This allows the distributed DL platform to expand to a larger scale to handle larger models and peak workloads
or contract when not necessary to conserve resources. 
Third, the cloud typically provides more utility tools for managing and monitoring 
data, models, resources, and tasks. 
For instance, AWS SageMaker~\cite{sageMaker20} offers visualization and model versioning tools 
for rapidly fine-tuning distributed DL models. 
The cloud-based paradigm is commonly applied in enterprise solutions across various industries, 
such as image and video intelligence~\cite{imageCloud22}, medical diagnosis~\cite{healthCloud22}, and cloud manufacturing~\cite{cloudManufactureAttend22,cloudManufacture22}. 

\subsubsection{Edge-based Distributed DL and Federated Learning}
With the rapid development of IoT and IoV technologies, the edge-based distributed DL paradigm emerges. 
Compared to the cluster/cloud-based paradigm, the edge-based paradigm follows a hierarchical architecture, which includes an additional edge layer connecting the cluster/cloud
and heterogeneous end devices (as shown in Fig.~\ref{fig:paradigm_edge}). 
The key idea of this paradigm is to place training and inference capabilities close to data to enhance the efficiency and latency of distributed DL. 
End devices, such as mobile devices and vehicles, typically have limited storage and computational capacities
and unstable wireless network connections but require real-time data collection and DL inference results.
The cloud paradigm may not be the ideal solution for such real-time tasks 
due to the high communication latency. 
The edge layer places edge servers close to 
the end devices to expand the storage, training, and inference capability of the end devices 
in an efficient manner for communication.
The edge-based paradigm is applied widely in scenarios that demand high resource efficiency and low latency, such as AR/VR~\cite{edgeMetaverse22}, intelligent transportation systems~\cite{edgeTransport22}, and smart industry and manufacturing~\cite{edgeIndustry23}. 

Federated Learning (FL) has recently emerged as a rapidly growing research topic, primarily exploiting the capacity of edge-based paradigms in collaborative DL.
To uphold privacy and data security while multiple parties train a global model collaboratively, each party trains a model with local data on its edge servers and shares only gradients, avoiding sharing raw data with other parties and the central cluster.  
FL finds applications in various domains, with a particular focus on privacy and data security issues across multiple parties, including healthcare~\cite{fedHome22,healthFL23}, finance~\cite{flBlockchain21}, and autonomous driving~\cite{autoDrivingFL22}. 
Nevertheless, FL is characterized by the heterogeneity of 1) geographical locations and computational and communication capacities of devices,  2) model structures, and 3) non-independent-and-identically-distributed (non-IID) data within each party.
This heterogeneity poses a major challenge for achieving high-performance FL.

\subsubsection{P2P-based Decentralized Distributed DL}
The P2P-based distributed DL paradigm (as shown in Fig.~\ref{fig:paradigm_p2p}) utilizes decentralized networks for collaborative DL model training.
Unlike the above paradigms that follow a between-layer communication pattern,
the P2P-based distributed DL paradigm is distinguished by the communication between peers at each layer. 
Peers collaborate to complete training tasks by exchanging learning information, such as the data, models, parameters, 
or gradients, in a gossip manner. 
Due to its decentralized nature, the P2P-based paradigm is expected to demonstrate scalability and fault tolerance.
Addressing concerns related to privacy protection, data integrity, 
and model security is crucial in both design and implementation of this paradigm.
It finds applications in decentralized trading systems~\cite{p2pTradeFL22}, swarm intelligence of autonomous vehicles~\cite{decentralizedSwarmDRL22}, and mobile robotic systems~\cite{p2pFLMobileRobotic23}. 

There are also hybrid distributed DL paradigms, which are mixed architectures combining the above paradigms for complex DL scenarios~\cite{mobileEdgeCloudTraining20}. 
In a hybrid paradigm, communication can happen between adjacent or skipping layers, or among peers within each layer. 
It capitalizes on the computational power of the cluster and cloud, while maintaining low-latency through the edge and ensuring
privacy and network robustness through the P2P-based component. 


\newcommand\sCate[1]{\multirow{2}{*}{\parbox[t]{1.5cm}{\centering #1}}}
\newcommand\sRef[1]{\parbox[t]{2.3cm}{#1}}
\newcommand\sYear[1]{\parbox[t]{0.6cm}{\centering #1}}
\newcommand\sContent[1]{\parbox[t]{11.2cm}{#1 \vspace{0.6pt}}}
\begin{table*}[!t]
\caption{Studies on communication-efficient model-synchronization algorithms for large-scale distributed DL}\label{tab:algorithms}
\centering
\begin{tabular}{|c|c||l|c|l|}
\Xhline{2\arrayrulewidth}
\multicolumn{2}{|c||}{\cellcolor{blue!15}\textbf{Category}} &\cellcolor{blue!15}\textbf{Strategy\&Ref.}  & \cellcolor{blue!15}\textbf{Year}& \cellcolor{blue!15}\textbf{Highlight}\\
\Xhline{2\arrayrulewidth}
\parbox[t]{2mm}{\multirow{29}{*}{\rotatebox[origin=c]{90}{Distributed SGD variants (\ref{sec:distributedSgd})}}} & \parbox[t]{1.5cm}{\centering Synchronous (\ref{sec:synchronous})}  & \sRef{OSP~\cite{overlapSync23}}& 2023& \sContent{Overlapping unimportant gradient synchronization with computation of the next iteration.} \\
\cline{2-5}
& \sCate{Asynchronous (\ref{sec:asynchronous})} & \sRef{Downpour SGD~\cite{downpourSGDAsync12}} & 2012& \sContent{Asynchronous with independent workers and PS shards.} \\
\cline{3-5}
  & & \sRef{PS+~\cite{psPlusAsyncSgd22}}& 2022& \sContent{Pulling the global model eagerly before pushing the latest local updates.}  \\
\cline{2-5}
& \sCate{Stale synchronous (\ref{sec:stale})} & \sRef{SSP~\cite{staleSynchronous13}} & 2013& \sContent{Allowing asynchronous model synchronization within a limited staleness bound.} \\
\cline{3-5}
  & & \sRef{AdaptiveRevision~\cite{adaptiveLRStaleSgd14}} & 2014& \sContent{Adaptive learning rate for minimum regret bound in stale SGD.} \\
\cline{3-5}
  &  & \sRef{R$^2$P~\cite{roundRobinSync19}}& 2019& \sContent{Workers synchronize in a fixed round-robin order with adaptive minibatch size to avoid network contention.}\\
\cline{3-5}
  & & \sRef{HSP~\cite{roundRobinPsPull22}} & 2022& \sContent{Workers synchronize in a fixed round-robin order and switch between synchronous and stale modes.} \\
\cline{2-5}
&\sCate{Adaptive local (\ref{sec:adaptiveLocal})} & \sRef{EASGD~\cite{elasticAvergeSGD15}}  & 2015& \sContent{ Incrementing synchronization frequencies iteratively.}\\
\cline{3-5}  
  &  & \sRef{AdaComm~\cite{adaptUpdateLocalSgd19}}& 2019& \sContent{Low synchronization frequency first for fast convergence and high frequency later for lower errors.} \\
\cline{3-5}
  & & \sRef{Post-local~\cite{postLocalSgd20}} & 2020& \sContent{Adding large-minibatch local SGD as the second stage; hierarchical local SGD at different infrastructure levels.} \\
\cline{3-5}
  & & \sRef{SlowMo~\cite{slowMo20}}  & 2020& \sContent{Local SGD with momentum updates.}\\
\cline{2-5}
&\sCate{Event-triggered local (\ref{sec:eventLocal})}& \sRef{LAG~\cite{lazyGradient18}}  & 2018& \sContent{Triggering synchronization when accumulated parameter changes are smaller than current parameter changes.} \\
\cline{3-5}
  & & \sRef{DETSGRAD~\cite{eventSGD20}} & 2020& \sContent{Triggering synchronization when the current parameter change in the PS is larger than a threshold.}\\
\cline{3-5}
  && \sRef{FSP~\cite{flexibleSync23}}  & 2023& \sContent{Fitting loss into an empirical two-stage pattern and triggering synchronization at the end of the first stage.}\\
\cline{2-5}
&\sCate{Decentralized (\ref{sec:decentralized})} & \sRef{D-PSGD~\cite{decentralizedSgd17}} & 2017& \sContent{Theoretically and experimentally showing that decentralized SGD has linear speedup performance for convergence.}\\
\cline{3-5}
  & & \sRef{D$^2$~\cite{decentralDataSGD18}}  & 2018& \sContent{Aggregating gradients based on differences between parameters and between gradients for heterogeneous data.}\\
\cline{2-5}
&\sCate{Hybrid (\ref{sec:hybrid})}& \sRef{GSSP~\cite{groupGossipSgd22}}  & 2022& \sContent{Clustering workers with similar performance into multiple groups; stale SGD within a group and decentralized SGD across groups.}\\
\cline{3-5}
  && \sRef{A2S~\cite{adaptiveSgdJIS22}}& 2022& \sContent{Stale SGD for fast workers and asynchronous SGD for slow workers.}\\
\cline{3-5}
  & & \sRef{ASHL~\cite{adaptiveHybrid23}}  & 2023& \sContent{Asynchronous local SGD for faster convergence at an early stage and stale SGD for model consistency at a late stage.}\\
\Xhline{2\arrayrulewidth}
 \parbox[t]{2mm}{\multirow{4}{*}{\rotatebox[origin=c]{90}{Guarantee}}} & \sCate{Local (\ref{sec:localConvergence})}& \sRef{\cite{localSgdConvergence18,localSgdConverge19,paraRestartedSgd19,localSgdAdaptive19,dynamicBatchLocalSgd19,localVsMinibatch20,oneShotSgd21,cooperativeSgd21}} & \sYear{2018-2021} & \sContent{Convergence guarantees of local SGD w.r.t. the number of workers, local updating iterations, entire iterations, etc.}  \\
\cline{2-5}
  & \sCate{Asynchronous (\ref{sec:asyncConvergence})} & \sRef{\cite{delayedSgdConverge20,asgdDelayed21,asyncDelaySgdCohen21,asgdConverge22,asyncSgdDelay22}}  & \sYear{2020-2022} & \sContent{Convergence guarantees of asynchronous SGD w.r.t. the gradient delay, gradient noise, stationary point, etc.}\\
\Xhline{2\arrayrulewidth}
\parbox[t]{1mm}{\multirow{35}{*}{\rotatebox[origin=c]{90}{Heterogeneous FL (\ref{sec:flSgd})}}} & \sCate{Random workers (\ref{sec:flRandom})}& \sRef{FedAvg~\cite{fedAvg17}}  & 2017& \sContent{Randomly selecting a fraction of workers in every synchronization round.}  \\
\cline{3-5}
  &  & NetMax~\cite{selectedWorkerFL21} & 2021& \sContent{Optimizing probabilities of selecting a peer w.r.t network bandwidth status in decentralized SGD.}\\
\cline{2-5}
  & \sCate{Model breaking-down (\ref{sec:flBreakdown})}& \sRef{ASTW\_FedAvg~\cite{layerSgdFL20}} & 2020& \sContent{Synchronizing shallow layers more frequently than deep layers.}  \\
\cline{3-5}  
  & & \sRef{APF~\cite{paramFreezeFL21}} & 2021& \sContent{Freezing stable parameters with AIMD frozen periods.} \\
\cline{3-5}
  & & \sRef{APF\#~\cite{paramFreezeFL23}}  & 2023& \sContent{Aggressive APF by freezing unstable parameters with a probability.}  \\
\cline{3-5}
  && \sRef{YOGA~\cite{layerAggrFL23}}  & 2023& \sContent{Selecting certain layers and peers based on data discrepancy and bandwidth for synchronization in decentralized SGD.}\\
\cline{2-5}
  & \sCate{FL-tailored aggregation (\ref{sec:flAggr})} & \sRef{FedProx~\cite{fedProxSgd20}}& 2020& \sContent{Finding local parameters that minimize the local loss and magnitude of local updates inexactly.}\\
\cline{3-5}  
  && \sRef{FedNova~\cite{fedNovaFLObj20}} & 2020& \sContent{Computing global gradients as a scaled weighted sum of normalized local gradients of randomly selected workers.}\\
\cline{3-5}  
  & & \sRef{CGA~\cite{crossDecentralizedNonIID21}}  & 2021& \sContent{Addressing non-IID data in decentralized SGD by training local and neighbor models at each worker.}\\ 
\cline{3-5}  
  & & \sRef{AsyNG~\cite{decentralFLNonIID23}} & 2023& \sContent{Addresses non-IID data in decentralized SGD by selecting appropriate neighbors based on loss difference and model staleness.}\\  
\cline{2-5}
  & \sCate{Hierarchical aggregation (\ref{sec:flHierarchical})}& \sRef{Two-level~\cite{scaleFL19}} & 2019& \sContent{A global aggregator spawning sub-aggregators for membership maintenance and encrypted aggregation.}\\
\cline{3-5}  
  &  & \sRef{FedCH~\cite{hierarchiSgdFL23}} & 2023& \sContent{Optimizing allocation of workers to groups in two-level aggregation to minimize loss with resource and time constraints.}\\
\cline{3-5}  
  & & \sRef{TT-HF~\cite{semiDecentralFL21}}& 2021& \sContent{Adaptive learning rate and synchronization frequency for decentralized SGD within an edge and local SGD across edges.}\\
\cline{3-5}  
  & & \sRef{Moshoit SGD~\cite{decentralAllReduceFL21}} & 2021& \sContent{Clustering pairs of peers into different groups iteratively for decentralized SGD in environments with unstable networks.}\\
\cline{2-5}
  & \sCate{Adaptive training (\ref{sec:flAdaptive})} & \sRef{Adaptive FL~\cite{adatpiveLocalFLRes19}}& 2019& \sContent{Optimizing synchronization frequency to minimize errors with heterogeneous resource constraints.}\\
\cline{3-5}  
  &  & \sRef{FedLamp~\cite{adaptiveSyncCompFL23}} & 2023& \sContent{Jointly optimizing synchronization frequency and compression ratio with convergence and resource constraints.}\\
\cline{3-5}  
  & & \sRef{AdaSFL~\cite{freqBatchFL23}}& 2023& \sContent{Jointly optimizing synchronization frequency and minibatch size with convergence and resource constraints.}\\
\cline{3-5}  
  & & \sRef{AAFL~\cite{adaptiveResFL23}}& 2023& \sContent{Optimizing synchronization frequency with bandwidth and convergence constraints via deep reinforcement learning.}\\
\cline{3-5}  
  & & \sRef{FAST~\cite{adaptiveSamplingFL23}} & 2023& \sContent{Optimizing synchronization frequency and data sampling jointly with resource and time constraints.}\\
\cline{3-5}  
  & & \sRef{AMBLE~\cite{adjustBatchEpochFL22}}& 2022& \sContent{Different learning rate, minibatch size, and synchronization frequency for fast and slow workers empirically.}\\
\cline{2-5}
  & \parbox[t]{1.5cm}{\centering Unlearning (\ref{sec:flUnlearning})} & \sRef{FL unlearning~\cite{forgetFL22}}  & 2022& \sContent{Efficient unlearning for non-stationary data distribution over time.} \\
\Xhline{2\arrayrulewidth}
\end{tabular}
\end{table*}

\section{Communication-Efficient Model Synchronization}\label{sec:synchronization}
Distributed SGD stands as the state-of-art algorithm for model synchronization in the distributed training of DNN models. 
The performance of different algorithms of distributed SGD is primarily associated to the trade-off between communication efficiency and model consistency.
In this section, we first introduce various distributed SGD algorithms. Subsequently, we present theoretical analyses on the convergence guarantee of distributed SGD. Moreover, we discuss distributed SGD algorithms in large-scale FL environments, where heterogeneity across various aspects dominates the research focus. 
Finally, we summarize key lessons learned from existing literature, guiding future endeavors in achieving high-performance model synchronization for large-scale distributed DL.
Table~\ref{tab:algorithms} provides a summary of the related studies. 

\begin{figure}[!t]
\centering
\includegraphics[width=0.66\columnwidth]{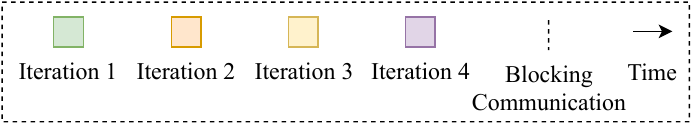}%
\vfil
\subfloat[Synchronous SGD]{\includegraphics[width=0.5\columnwidth]{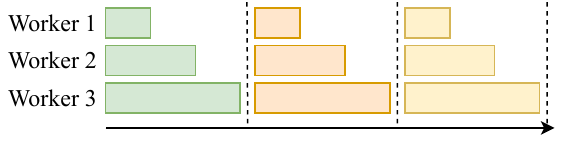}%
\label{fig:sgd_sync}}
\hfil
\subfloat[Asynchronous SGD]{\includegraphics[width=0.5\columnwidth]{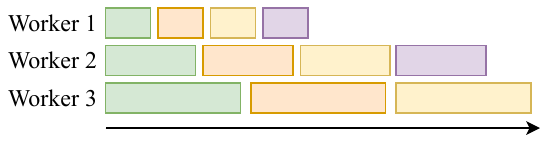}%
\label{fig:sgd_async}}
\vfil
\subfloat[Stale SGD]{\includegraphics[width=0.5\columnwidth]{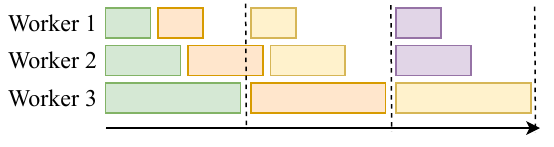}%
\label{fig:sgd_stale}}
\hfil
\subfloat[Local SGD]{\includegraphics[width=0.5\columnwidth]{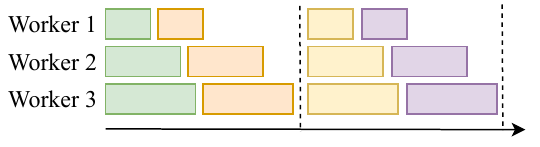}%
\label{fig:sgd_local}}
\caption{Typical variants of distributed SGD}
\label{fig:sgd}
\end{figure}

\subsection{Synchronous, Asynchronous, and Other Distributed SGD}\label{sec:distributedSgd}
This subsection introduces optimizations for various distributed SGD algorithms, including synchronous, asynchronous, stale, local, decentralized, and hybrid SGD algorithms.
Fig.~\ref{fig:sgd} illustrates some typical distributed SGD algorithms.

\subsubsection{Synchronous SGD}\label{sec:synchronous}
In fully synchronous SGD, also known as minibatch distributed SGD, every gradient iteration serves as a synchronization round (see Fig.~\ref{fig:sgd_sync}). Workers retrieve the most up-to-date model parameters from the PS during every gradient iteration, where the PS updates the global model based on the gradients from all workers, resulting in more consistent models across all workers and a more deterministic and coordinated training progress. 
However, since the training process on any worker is blocked until all other workers have synchronized to the same progress, this strictly consistent-pace strategy introduces significant communication overhead, such as idle time caused by straggling workers. 
Some work, such as Overlapped Synchronization Parallel (OSP)~\cite{overlapSync23}, introduces a technique of overlapping computation and communication aimed at alleviating this blocking communication overhead of fully synchronous SGD. 
OSP divides the synchronization communication of synchronous SGD into two stages, a predecessor synchronization stage for important gradients and a successor stage for unimportant gradients.
It enables the successor communication stage to overlap with the feedforward and backpropagation computation of the next training iteration. 
However, the blocking communication overhead cannot be eliminated completely in synchronous SGD because of the existence of a strict consistency constraint. This motivates the exploration of numerous variants of distributed SGD algorithms that relax the consistency constraint. 

\subsubsection{Asynchronous SGD}\label{sec:asynchronous}
The impact of blocking communication overhead of synchronous SGD can be challenging in large-scale distributed DL when many heterogeneous workers participate in the synchronization. 
Asynchronous SGD addresses the blocking communication overhead by allowing workers to synchronize their models with the PS independently, without waiting for gradient updates from other workers (see Fig.~\ref{fig:sgd_async}). 
Compared to synchronous SGD, asynchronous SGD eliminates the synchronization blocking time and is more suitable for large-scale distributed training in heterogeneous environments. 
To enhance asynchronous SGD for large data, Downpour SGD~\cite{downpourSGDAsync12}, a fully asynchronous SGD algorithm for data-parallel training with a centralized PS, segments the PS into multiple shards, each of which is responsible for storing and applying SGD updates to a portion of the global model parameters. 
Downpour SGD is asynchronous in two aspects: model replicas in workers operate independently of each other, which is the common form of asynchronous SGD, and PS shards also operate independently of each other. 
Downpour SGD has been verified experimentally to utilize network bandwidth efficiently and stabilize volatile parameters using the Adagrad optimizer for non-convex objectives.

Efficient asynchronous SGD is facilitated by a PS that optimizes computation and communication parallelism.
PS+~\cite{psPlusAsyncSgd22} accelerates asynchronous SGD by decoupling gradient pushing and the successive parameter pulling operations, enabling workers to pull the global parameters eagerly from the PS for the next gradient iteration even before pushing the latest local gradients. 
This early-pulling strategy overlaps computational and communication workloads in every synchronization round, though it introduces additional model staleness. 
Nevertheless, the impact of staleness can be mitigated by a delay-adaptive strategy for the global learning rate, adjusting it smaller dynamically when the staleness grows large throughout the training process. 

\subsubsection{Stale synchronous SGD} \label{sec:stale}
However, asynchronous SGD introduces staleness among models in different workers, leading to less consistent models across workers, causing oscillations during the training process potentially. 
Stale synchronous SGD, a family of distributed SGD variants, relaxes model consistency among workers within a bounded model staleness (Fig.~\ref{fig:sgd_stale}).
This trade-off between model consistency and communication overhead aims for improved synchronization performance.
In Stale Synchronous Parallel (SSP)~\cite{staleSynchronous13}, the PS maintains historical model parameters and gradients, enabling workers to retrieve a historical global model that is a bounded number of synchronization intervals ago. 
Consequently, faster workers can proceed without waiting for slower workers in every synchronization round. They use the latest updated model, as long as slower workers are not falling behind beyond the bounded interval. 
Ensuring a theoretical convergence guarantee, the bounded staleness reduces synchronization waiting time among workers and results in a faster convergence rate than fully synchronous SGD.
To limit the impact of model staleness, AdaptiveRevision~\cite{adaptiveLRStaleSgd14} adjusts the learning rate appropriately and efficiently for gradient updates between periodic synchronization rounds.
This adjustment minimizes the regret bound, defined based on the difference between the actual and optimal model parameters.

Workers synchronizing models with the PS at the consistent pace can lead to the thundering herd problem in bulk synchronization of synchronous SGD, causing network contention in the PS bandwidth. 
To address this network contention issue, R$^2$P~\cite{roundRobinSync19} and HSP~\cite{roundRobinPsPull22} introduce similar schemes for stale synchronous SGD, where workers update models to the PS in a fixed round-robin order. 
Consequently, this minimizes contention in the PS network bandwidth, and models in workers are evenly staggered, indicating bounded model staleness. 
Additionally, to utilize computational resources efficiently in heterogeneous computing clusters, R$^2$P assigns larger minibatch sizes for faster workers dynamically, keeping them engaged throughout the round-robin iteration when it is other worker's turn to synchronize with the PS.
In contrast, HSP allows workers to switch between synchronous and stale modes based on the resource utilization status.

\subsubsection{Local SGD}\label{sec:local}
The above synchronous and asynchronous SGD algorithms demand frequent communication between workers and the PS, leading to significant communication overhead, particularly in large-scale clusters. 
To decrease communication frequency in large-scale distributed training, an alternative model-synchronization algorithm, known as local distributed SGD or local SGD, has been proposed. 
Local SGD allows workers to update local models for multiple iterations and only synchronize models with the PS periodically to maintain the most current global model (Fig.~\ref{fig:sgd_local}).
In addition to concerns about synchronization overhead, allowing local updates in the worker implies more exploration in the existence of local optima with distributed SGD, potentially resulting in enhanced optimization results.
A general form of local SGD can be characterized by four parameters: synchronous or asynchronous synchronization patterns, the faction of workers performing updates in each synchronization round, the number of local gradient update iterations in each synchronization round (corresponding to the synchronization frequency), and the local minibatch size and learning rate used for each local iteration. 

$\bullet$ \textit{Adaptive local SGD:} \label{sec:adaptiveLocal}
Various local SGD algorithms use all workers in each synchronization round and adaptive settings for the synchronization frequency and minibatch size, whether synchronously or asynchronously.
In asynchronous Elastic Averaging SGD (EASGD)~\cite{elasticAvergeSGD15}, the synchronization frequency of each worker is controlled by a locally maintained clock, which is incremented by one after every local update iteration.
The PS updates the global model in a moving average style for every synchronization from a worker.
The stability of asynchronous EASGD is guaranteed theoretically, and experimental results demonstrate its significantly faster convergence compared to Downpour SGD.
By exploring the trade-off between convergence and wall-clock runtime with different numbers of local update iterations, AdaComm~\cite{adaptUpdateLocalSgd19} adjusts the number of local update iterations over time to minimize the optimization error at a given wall-clock time.
AdaComm suggests an optimal strategy for local SGD convergence concerning synchronization frequency: starting with infrequent synchronization can reduce communication delay and improve convergence rate initially; increasing synchronization frequency gradually over time contributes to achieving lower optimization error. 
Lin \textit{\textit{et al.}}~\cite{postLocalSgd20} propose two local SGD variants: Post-local SGD, aiming to reach high generalization, and Hierarchical Local SGD, to optimize the trade-off between computation and communication.
Large-minibatch distributed SGD has advantages in training speed, but it often does not generalize as well as local SGD~\cite{measureEffectParallel19}. 
Post-local SGD closes this generalization gap by adding large-minibatch local SGD as the second training phase after the initial phase of standard minibatch distributed SGD.
The initial phase can be viewed as the warm-up phase for tuning the learning rate. 
Hierarchical Local SGD employs local SGD as an inner loop on each layer of a hierarchical system comprising GPUs, computing nodes, racks, or even data centers.
This hierarchical approach limits intensive model synchronization within each layer and mitigates communication overhead between different layers.
SlowMo~\cite{slowMo20} is a momentum-based variant of local SGD, where the global model in the PS is updated through SGD with local update momentum. 
Supported by experimental results on various image and NLP benchmarks,
the momentum-based local SGD is believed to converge faster and generalize better than vanilla local SGD. 

$\bullet$ \textit{Event-triggered local SGD:}\label{sec:eventLocal}
One category of asynchronous local SGD algorithms is event-triggered SGD.
Representative works in this category include Lazily Aggregated Gradient (LAG)~\cite{lazyGradient18} and Distributed Event-Triggered Stochastic GRAdient Descent (DETSGRAD)~\cite{eventSGD20}, with model synchronization triggered by a specific condition on gradients. 
Workers and the PS synchronize models only when the gradients of the local or global model change significantly. 
In the case of LAG, the synchronization is triggered in the PS or a worker when the accumulated parameter change within a specific period is smaller than the most recent parameter change in the PS or gradient change in a worker.
LAG is proven theoretically to converge for objectives that are smooth and strongly convex, convex, or non-convex.
In the case of DETSGRAD, the triggering condition is met in the PS when the current parameter change exceeds a threshold, which decreases as the training iteration increases.
The convergence of DETSGRAD is proved under a series of assumptions for the non-convex objectives.  
Rather than using a synchronization trigger conditioned on gradient changes, Flexible Synchronous Parallel (FSP)~\cite{flexibleSync23} uses a trigger conditioned on optimization gains.
This is based on an empirical observation that the loss change w.r.t the objective function can be divided into two stages, where the loss decreases linearly faster in the former stage than in the latter stage. 
FSP fits the loss pattern of the local SGD process into this two-stage pattern and identifies the barrier between these two stages as the optimal moment for model synchronization, as further local model synchronization in the latter stage delivers inferior optimization gain. 
FSP provides convergence guarantee for smooth and strongly convex objectives with Lipschitz-continuous gradients.

\subsubsection{Decentralized SGD}\label{sec:decentralized}
The distributed SGD algorithms mentioned above are centralized approaches, featuring a central PS.
This centralized setup is susceptible to traffic jam and single-point failure in a large-scale distributed DL setting.
Decentralized SGD addresses this bottleneck by eliminating the central PS during model synchronization and typically relies on gossip-based communication among workers. 
To explore the advantage of decentralized SGD over centralized SGD, Lian \textit{\textit{et al.}}~\cite{decentralizedSgd17} conduct a theoretical and experimental comparison of the convergence performance between these two approaches.
The results demonstrate that decentralized SGD converges as effectively as centralized SGD but outperforms it in terms of the maximum communication cost at any single point.
To the best of our knowledge, this is the first work to offer a theoretical analysis of the linear speedup performance concerning the number of workers for decentralized SGD.
Nevertheless, this result assumes similar data distributions across workers and its robustness to data in heterogeneous distributions remains unknown. 
By investigating the convergence performance of decentralized SGD with data variance among workers, D$^2$~\cite{decentralDataSGD18} is proposed as a decentralized SGD variant where the parameters shared among peers for aggregation are calculated based on the difference between local parameters and the difference between gradients of consecutive iterations. 
This parameter updating strategy yields a theoretically improved convergence rate for decentralized SGD, particularly when dealing with heterogeneous data across workers, assuming Lipschitz-continuous and variance-bounded gradients.

\subsubsection{Hybrid SGD for the straggler problem} \label{sec:hybrid}
In heterogeneous large-scale distributed DL, the issue of straggling worker is exacerbated, impacting both synchronous and asynchronous SGD.
Workers lagging significantly behind others can impede the synchronization in the case of synchronous SGD and contribute to increased model staleness in asynchronous SGD.
To address the straggler problem, a few studies have employed hybrid SGD approaches. 

The straggler problem can be mitigated by identifying stragglers during training and organizing workers into groups with varied model-synchronization strategies. 
Grouping Stale Synchronous Parallel (GSSP)~\cite{groupGossipSgd22} categorizes workers with comparable performance into specific groups and implements distinct strategies for intra- and inter-group model synchronization.
Within each group, workers apply stale synchronous SGD with the group PS, whereas each group synchronizes its model with other groups using an asynchronous gossip approach.
Consequently, GSSP mitigates stragglers during intra-group synchronization and improves convergence performance through inter-group synchronization.
In contrast, A2S~\cite{adaptiveSgdJIS22} applies distinct model-synchronization strategies to fast and slow workers.
These groups are identified based on maintaining synchronization speeds of workers in the PS, fast and slow workers are separated dynamically into synchronous and asynchronous groups, respectively. 
Fast workers apply synchronous SGD with a relaxed adaptive synchronization barrier, akin to stale synchronous SGD.
In contrast, slow workers apply asynchronous SGD.
This approach limits the model staleness through stale synchronous SGD and mitigates the communication delay caused by stragglers through asynchronous SGD. 
The convergence performance of these grouping SGD approaches is guaranteed theoretically and experimentally. 

Some studies address the straggler problem by implementing various model-synchronization strategies at different training stages.
To accelerate distributed training in heterogeneous environments with stragglers, 
ASHL~\cite{adaptiveHybrid23} divides the training process into coarse- and fine-grained stages depending on whether the global loss exceeds a predefined threshold.
In the coarse-grained stage, where the model needs to converge quickly to a certain extent, ASHL employs asynchronous local SGD to reduce communication frequency and waiting time for faster convergence; in the fine-grained stage, where the model should be refined to enhance the convergence bound, ASHL employs stale synchronous SGD with bounded staleness to ensure model consistency.
The synchronization frequency of each worker in both stages are determined by profiling the updating speeds of workers, aiming for a roughly consistent synchronization pace for both fast and slow workers.

\begin{table*}[!t]
\caption{A Comparison of convergence analyses of local SGD}\label{tab:sgdConvergence}
\centering
\begin{tabular}{cccl}
\Xhline{2\arrayrulewidth}
\cellcolor{blue!15}\textbf{Ref.} & \cellcolor{blue!15}\textbf{Convergence Rate} & \cellcolor{blue!15}\textbf{Synchronization Round (R)} & \cellcolor{blue!15}\textbf{Setting}\\
\Xhline{2\arrayrulewidth}
Zhou and Cong (2018)~\cite{localSgdConvergence18} & $\mathcal{O}(\frac{1}{\sqrt{NT}})$ & $\Omega(T)$ & Non-convex; Lipschitz-continuous gradients\\
Stich (2019)~\cite{localSgdConverge19} & $\mathcal{O}(\frac{G^2}{NT})$ & $\Omega(N^\frac{1}{2}T^\frac{1}{2})$ & Bounded gradient; strongly convex\\
Yu \textit{et al.} (2019)~\cite{paraRestartedSgd19} & $\mathcal{O}(\frac{G^2}{\sqrt{NT}})$ & $\Omega(N^\frac{3}{4}T^\frac{3}{4})$ &  Bounded gradient; non-convex\\
Haddadpour \textit{et al.} (2019)~\cite{localSgdAdaptive19} & $\mathcal{O}(\frac{1}{NT})$ & $\Omega(N^\frac{1}{3}T^\frac{1}{3})$ &  Non-convex under PL Condition\\
Yu and Jin (2019)~\cite{dynamicBatchLocalSgd19} & $\mathcal{O}(\frac{1}{NT})$ & $\Omega(logT)$ &  Increasing minibatch size; non-convex under PL Condition\\
Woodworth \textit{et al.} (2020)~\cite{localVsMinibatch20} & $\mathcal{O}(\frac{1}{\sqrt{NT}}+\frac{1}{\sqrt[3]{TR}})$ & $\Omega(N\times poly(log T))$ &  Convex\\
Spiridonoff \textit{et al.} (2021)~\cite{oneShotSgd21} & $\mathcal{O}(\frac{1}{NT})$ & $\Omega(N)$ &  Smooth and strongly convex under PL Condition\\
Wang and Joshi (2021)~\cite{cooperativeSgd21} & $\mathcal{O}(\frac{1}{\sqrt{NT}})$ & $\Omega(N^\frac{3}{2}T^\frac{1}{2})$ &  Non-convex\\
\Xhline{2\arrayrulewidth}
\end{tabular}
\begin{tablenotes}
\item Convergence rate, the expected different between the optimization objectives with the averaged model and minibatch SGD: $\mathbb{E}[f(\bar{x}_T) - f(x^{\ast})]$, where $f$ is the objective function, $x_t$ is the local model after $t$ local SGD update iterations, and $x^{\ast}$ is the optimal model; \\
\item $G$, the uniform upper bound for the L2-norm of gradients; \\
\item $N$, the number of worker; \\
\item $T$, the total number of SGD update iterations at each worker.
\end{tablenotes}
\end{table*}

\subsection{Convergence Guarantees of Distributed SGD} \label{sec:convergence}
Many studies analyze theoretical convergence guarantees of different distributed SGD algorithms across various optimization objectives.
Two major branches of these analyses focus on local SGD and asynchronous SGD, 

\subsubsection{Convergence of local SGD} \label{sec:localConvergence}
A significant portion of these studies focus on local SGD. 
These works formalize local SGD across various optimization objective settings, studying the convergence performance concerning the number of workers ($N$), number of local updating iterations within each synchronization round, entire updating iterations ($T$) at each worker, and minibatch size. 
Therefore, they prove the scalability advantage of local SGD over other distributed SGD algorithms when applied in large-scale distributed DL.
Table~\ref{tab:sgdConvergence} presents a comparison of analysis results of local SGD in various objective settings.

These analyses provides bounds on the convergence rate and synchronization round. 
Zhou and Cong~\cite{localSgdConvergence18} analyze the convergence of local SGD for non-convex objectives with Lipschitz-continuous gradients, under various settings of the learning rate and minibatch size. 
The analysis shows that local SGD allows larger learning rate, which is a common practice in large-batch training for large-scale distributed DL, and scales more efficiently than asynchronous SGD.
Stich~\cite{localSgdConverge19} demonstrates that both synchronous and asynchronous local SGD algorithms share the same convergence rate but reduce synchronization rounds by a factor of the root square of the number of local gradient updates compared to synchronous SGD for strongly convex objectives.
Assuming a uniform upper bound on the L2-norm of gradients, local SGD can scale with a linear speedup w.r.t the number of workers and minibatch size.  
With the same bounded gradient assumption as in~\cite{localSgdConverge19}, Yu \textit{et al.}~\cite{paraRestartedSgd19} ensure convergence for non-convex objectives.
However, the convergence rate is not proven a linearly speedup w.r.t the number of workers.
Eliminating this bounded gradient assumption, both Haddadpour \textit{et al.}~\cite{localSgdAdaptive19} and Yu and Jin~\cite{dynamicBatchLocalSgd19} enhance the analysis of~\cite{paraRestartedSgd19}. 
They offer tighter bounds on both the convergence rate and synchronization round than previous analyses and prove a linear speedup for non-convex objectives under the Polyak-{\L}ojasiewicz (PL) condition~\cite{plCondition16}, which can be viewed as a generalization of strong convexity for non-convex optimization. 
Compared to \cite{localSgdAdaptive19}, whose asymptotic lower bound on the synchronization round is $\Omega(N^\frac{1}{3}T^\frac{1}{3})$, the analysis of \cite{dynamicBatchLocalSgd19} provides a tighter bound of $\Omega(log(T))$ for local SGD with dynamically increasing minibatch size, implying a promising direction to scale local SGD.
Woodworth \textit{et al.}~\cite{localVsMinibatch20} concentrate on comparing the convergence of minibatch distributed SGD and local SGD through a theoretical analysis.
In comparison to minibatch distributed SGD, local SGD exhibits superior convergence performance for quadratic objectives. 
Yet, for general convex objectives, local SGD demonstrates better convergence than minibatch distributed SGD with a large number of workers, but worse with a large minibatch size.

Some studies investigate the convergence guarantee of an extreme case of local SGD known as one-shot averaging.
This method uses very few, or even only a single, synchronization rounds at the very end of the distributed training process. 
In smooth and strongly convex problems under the PL condition, Spiridonoff \textit{et al.} (2021)~\cite{oneShotSgd21} demonstrate that one-shot averaging can converge in $\Omega(N)$ synchronization rounds, irrespective of the total number of gradient iterations $T$. 
This finding suggests the potential application of one-shot averaging to large models, typically requiring numerous training iterations.  

Some studies focus on building a unified framework for the convergence analysis of local SGD.
Wang and Joshi~\cite{cooperativeSgd21} present a unified framework named Cooperative SGD to design variants and analyze convergence performance of local SGD, expanding the design space significantly by incorporating various model averaging protocols and adjusting the number of local updates. 
Cooperative SGD reveals that vanilla local SGD, EASGD~\cite{elasticAvergeSGD15}, and decentralized local SGD are all special cases within this unified framework.
Leveraging the convergence analysis of this unified framework, Cooperative SGD offers the first convergence guarantee for EASGD with non-convex objective functions, applicable to both IID and non-IID data. 

\begin{table*}[!t]
\caption{A Comparison of convergence analyses of asynchronous SGD}\label{tab:asgdConvergence}
\centering
\begin{tabular}{ccl}
\Xhline{2\arrayrulewidth}
\cellcolor{blue!15}\textbf{Ref.} & \cellcolor{blue!15}\textbf{Synchronization Iteration} & \cellcolor{blue!15}\textbf{Setting}\\
\Xhline{2\arrayrulewidth}
Stich and Karimireddy (2020)~\cite{delayedSgdConverge20} & $\Omega(\frac{\sigma^2}{\epsilon^2}  + \frac{\tau_{max}}{\epsilon})$ &  General quasi-convex and smooth non-convex\\
Aviv \textit{et al.} (2021)~\cite{asgdDelayed21} & $\Omega(\frac{\sigma^2}{\epsilon^2}  + \frac{\tau_{avg}}{\epsilon})$ &  Strongly convex; delay-adaptive learning rate\\
Cohen \textit{et al.} (2021)~\cite{asyncDelaySgdCohen21} & $\Omega(\frac{\sigma^2}{\epsilon^4}  + \frac{\tau_{avg}}{\epsilon^2})$ &  Smooth non-convex\\
Koloskova \textit{et al.} (2022)~\cite{asgdConverge22} & $\Omega(\frac{\sigma^2}{\epsilon^2}  + \frac{\tau_{avg}}{\epsilon})$ &  Smooth non-convex; delay-adaptive learning rate\\
Mishchenko \textit{et al.} (2022)~\cite{asyncSgdDelay22} & $\Omega(\frac{\sigma^2}{\epsilon^2}  + \frac{\tau_{avg}}{\epsilon})$ &  Non-convex, convex, or strongly convex; Lipschitz-continuous gradients;\\
& & delay-adaptive learning rate\\
\Xhline{2\arrayrulewidth}
\end{tabular}
\begin{tablenotes}
\item $\tau_{max}$, the maximum synchronization delay; \\
\item $\tau_{avg}$, the average synchronization delay; \\
\item $\sigma^2$, the upper bound on the gradient variance within a worker;\\
\item $\epsilon$, the approximate stationary point, which bounds the squared gradient norm.
\end{tablenotes}
\end{table*} 

\subsubsection{Convergence of asynchronous SGD} \label{sec:asyncConvergence}
Several studies offer theoretical convergence guarantees for asynchronous SGD, emphasizing the the synchronization iteration bound concerning the gradient delay, upper bound on gradient noise, and approximate stationary point for training termination. 
Table~\ref{tab:asgdConvergence} presents a comparison of analysis results of various asynchronous SGD algorithms. 

Stich and Karimireddy~\cite{delayedSgdConverge20} conduct a convergence analysis for asynchronous SGD involving gradient compression and error feedback, whose mechanisms will be introduced in Section~\ref{sec:compression}, and incorporating bounded gradient noise in the model for both general quasi-convex and smooth non-convex objectives. 
The complexity of synchronization iteration is bounded linearly by the maximum synchronization delay, indicating the maximum model staleness.
However, these findings lack robustness in heterogeneous environments, where the maximum delay can surpass the average delay significantly. 
Aviv \textit{et al.}~\cite{asgdDelayed21} introduce a delay-adaptive learning rate scheme for asynchronous SGD. 
Smaller learning rates are assigned to updates with a larger delay, providing a tighter synchronization iteration bound proportional to the average synchronization delay rather than the maximum delay.  
However, the result is confined to strongly convex problems, and the proof relies heavily on the assumption of an upper bound on the variance of delays, which is can be closely relevant to the maximum delay. 
In contrast, Cohen \textit{et al.}~\cite{asyncDelaySgdCohen21} focus on smooth non-convex objectives and present a synchronization iteration bound proportional to the average synchronization delay.
However, this analysis requires twice as many communication rounds at every step and relies on the appropriate tuning of hyperparameters.
Eliminating the assumption of the upper bound on the delay variance in~\cite{asgdDelayed21} and the hyperparameter tuning in~\cite{asyncDelaySgdCohen21}, Koloskova \textit{et al.}~\cite{asgdConverge22} apply another delay-adaptive learning rate scheme, achieving the same synchronization iteration bound as in~\cite{asgdDelayed21}.
Mishchenko \textit{et al.}~\cite{asyncSgdDelay22} employ analyzing techniques and achieve convergence guarantees similar to those of~\cite{asgdConverge22}, but cover a broader range of optimization problems assuming Lipschitz-continuous gradients, including non-convex, convex, and strongly convex. 

\subsection{Model Synchronization in FL} \label{sec:flSgd}
Model synchronization in the domain of FL presents additional challenges. 
Firstly, and most importantly, there can be a huge amount of workers, represented by various types of devices, exhibit heterogeneity in geographical locations, computational and communication-resource capacities, model structures and objectives, and data distributions. 
Secondly, the membership of workers participating in the synchronization can be dynamic and unstable.
Thirdly, concerns arise regarding privacy and data security issues.
These challenges are addressed by numerous studies employing approaches in various directions.

\subsubsection{Randomly selected workers} \label{sec:flRandom}
To reduce communication overhead among numerous heterogeneous workers in FL, FedAvg~\cite{fedAvg17} randomly selects only a fraction of workers in each synchronization round of local SGD. 
In the context of decentralized SGD, NetMax~\cite{selectedWorkerFL21} enables each worker to select a peer stochastically based on a fine-tuned probability for model synchronization in each round. 
The optimal selection probabilities for all workers are derived by a network monitor, aiming to minimize the total convergence time w.r.t the network bandwidth capacity and utilization status.
While the method of selecting workers randomly has become the state-of-the-art practice for distributed SGD in FL, it does not address various other heterogeneity issues such as non-IID data and heterogeneous models.
This limitation has prompted a series of studies on this topic, focusing on different strategies tailored to cope with heterogeneity in FL, some of which are variants of FedAvg. 

\subsubsection{Breaking down the model}  \label{sec:flBreakdown}
A branch of variants~\cite{layerSgdFL20,paramFreezeFL21,paramFreezeFL23,layerAggrFL23} distinguishes model components and synchronizes specific components intermittently, enabling varied synchronization frequencies for different components and enhancing communication efficiency for large models. 
Leveraging the observation that shallow layers of DNN models capture general features, while deep layers acquire ad-hoc features specific to data sets, ASTW\_FedAvg~\cite{layerSgdFL20}
synchronizes shallow layers more frequently than deep layers, excluding deep layers in certain synchronization rounds. 
This approach enhances FedAvg for heterogeneous data sets; however, the coarse division of the so-called shallow and deep layers is empirical and not robust. 
Exploiting the observation that many parameters stabilize before the model reaches its ultimate convergence, Adaptive Parameter Freezing (APF)~\cite{paramFreezeFL21} introduces the parameter-wise freezing scheme.
In this scheme, specific stable local parameters are fixed in local feedforward and backpropagation iterations and excluded during synchronization for specified periods. 
The periods of parameter-wise freezing are not predetermined based on prior knowledge of model convergence behaviors.
Instead, they are adjusted dynamically in an Additive-Increase Multiplicative-Decrease (AIMD) manner, depending on the stability of parameters that were previously frozen in subsequent iterations. 
Nevertheless, empirical observations~\cite{overParam19} suggest that some parameters remain unstable even when the model reaches convergence, especially in the case of over-parameterized large models.
This instability weakens the ability of these parameters to freeze and to reduce communication overhead in APF.
APF\# and APF++~\cite{paramFreezeFL23} extend APF to tackle the jittering issue in over-parameterized large models by employing aggressive methods for parameter freezing.
In APF\#, unstable parameters have a probability of being frozen for one round, and in the more more aggressive APF++, both the probability and the number of freezing rounds increase over time.

However, these fine-grained distributed SGD algorithms can be computationally expensive.
APF and its extensions offer precise controls over the trade-off between model consistency and communication overhead.
However, maintaining parameter-wise freezing period introduces memory and computation overhead, with complexity is linear to the model size.
This suggests a careful consideration of these methods for large models. 
To achieve efficient and fine-grained SGD for FL, YOGA~\cite{layerAggrFL23} adopts a layer-wise gradient aggregation approach based on ranks in decentralized SGD.
YOGA assigns ranks to both layers and peers, employing a greedy algorithm to aggregate gradients from layers and peers selectively. 
Layers are ranked according to their learning speed and discrepancy, identifying the significance of each layers effectively.
Peers are ranked based on data distribution divergence and available bandwidth, addressing the issues of non-IID data and heterogeneous resource capacities in FL.

\subsubsection{FL-tailored aggregation strategies} \label{sec:flAggr}
Some studies focus~\cite{fedProxSgd20,fedNovaFLObj20,crossDecentralizedNonIID21,decentralFLNonIID23} on diverse aggregation strategies to determine local and global parameters, addressing challenges related to heterogeneity in both data and resources within FL environments.
FedProx~\cite{fedProxSgd20}, akin to FedAvg, presents a new local updating strategy among workers. 
Instead of updating local parameters through SGD, FedProx identifies local parameters that minimize a local proximal objective inexactly, which restricts the magnitude of local updates to alleviate model inconsistency arising from data heterogeneity. 
Addressing resource heterogeneity involves adjusting the inexact level of local parameters in each worker.
This adjustment correlates with the number of local iterations and reduces computation and communication overhead of each worker during a synchronization round. 
However, the computation of finding these inexact local parameters in FedProx remains sophisticated, hindering its convergence rate in large-scale FL. 
FedNova~\cite{fedNovaFLObj20} addresses this issue by dealing with inconsistent local objectives arising from the heterogeneity in the number of local updates and non-IID data in FL. 
Gradients for updating the global model are computed as a scaled weighted sum of the normalized local gradients from randomly selected workers. 
These weights serve as flexible hyperparameters unique to each worker, and the scale is associated with the number of local update iterations in each worker.
An illustrative example of the scale is the weighted sum of the numbers of local update iterations. 

In contrast to the studies above that focus on centralized SGD, some studies address FL heterogeneity through decentralized SGD approaches.
Cross-Gradient Aggregation (CGA)~\cite{crossDecentralizedNonIID21} tackles non-IID data by computing the so-called cross gradients during model aggregation. 
In addition to computing local gradients by training local models at each worker, each worker retrieves neighboring models and derives cross gradients, which are computed by the neighboring models using local data on this worker. 
These local and cross gradients are then projected into aggregated gradients using quadratic programming and updated into each model via SGD with momentum.
CGA exhibits superior performance over many other decentralized SGD algorithms, especially for non-IID data. 
On the other hand, AsyNG~\cite{decentralFLNonIID23} tackles non-IID data through the dynamical selection of suitable neighbors for P2P model synchronization. 
Selection of neighbors relies on a priority-based algorithm, wherein the priority of a neighbor is determined by the difference in loss and the gap in training iterations compared to the worker.
A neighbor is deemed more favorable for synchronization where there is a larger loss difference and a smaller training iteration gap.

\subsubsection{Hierarchical gradient aggregation} \label{sec:flHierarchical}
Some of studies~\cite{scaleFL19,hierarchiSgdFL23,semiDecentralFL21,decentralAllReduceFL21} focus on clustering workers into groups for gradient aggregation, similar to~\cite{groupGossipSgd22,adaptiveSgdJIS22}, but with considerations for the heterogeneity issues in FL.
To address membership and data security issues, Bonawitz \textit{et al.}~\cite{scaleFL19} propose a two-level hierarchical aggregating scheme with encrypted computation for model synchronization among numerous mobile devices in FL, based on FedAvg. 
This scheme employs a global aggregator that is capable of spawning several sub-aggregators, each being responsible for membership maintenance and model synchronization within a subset of devices.  
The scheme adopts the Secure Aggregation protocol~\cite{secureAggr17}, performing model aggregation in a logical incorruptible third party. 
This ensures encrypted model updating, preventing the exposure of the model to untrusted devices and cloud providers.
To optimize the topology for hierarchical SGD,
FedCH~\cite{hierarchiSgdFL23} proposes a two-level gradient aggregating architecture comprising a specific number of groups.
Each group aggregates gradients from different workers within it through local SGD.
Additionally, there is a global PS that aggregates gradients from different groups using asynchronous SGD. 
FedCH aims to determine the optimal allocation of workers to groups, known as the cluster topology, to minimize training loss while considering resource and time budget constraints within this architecture.

In contrast to the grouping methods above that focus on the centralized SGD, certain studies focus on decentralized SGD, wherein each group employs gossip-based gradient aggregation.
TT-HF~\cite{semiDecentralFL21} examines FL across wireless edge devices, where devices within an edge function naturally as workers belonging to the same group. 
Workers apply local SGD, synchronize their models aperiodically through gossip within an edge, and synchronizes models aperiodically across edges via a global PS. 
Drawing on the convergence analysis of this synchronization strategy, TT-HF adjusts the learning rate and synchronization frequencies within the edge and globally adaptively to achieve the optimal trade-off among delay, model accuracy, and energy consumption. 
To mitigate the impact of constantly joining, leaving, and failing participating workers with convergence guarantees, Moshoit SGD~\cite{decentralAllReduceFL21} clusters them into different groups for gradient aggregation iteratively in every synchronization round.
This ensures that any pair of workers will not be clustered into the same group in consecutive rounds. 
Moshoit SGD is designed for cloud-based and edge-based distributed DL environments with unreliable devices and unstable networks.
It is theoretically and experimental proven to be more effective than the completely gossip-based approach in these environments.

\subsubsection{Adaptive hyperparameters throughout training} \label{sec:flAdaptive}
Various studies~\cite{adatpiveLocalFLRes19,adaptiveSyncCompFL23,freqBatchFL23,adaptiveResFL23,adaptiveSamplingFL23,adjustBatchEpochFL22} address issues of heterogeneous data and resources in FL by focusing on the adapting hyperparameters of distributed SGD, including the synchronization frequency, minibatch size, learning rate, and data-sampling strategy. 
Wang \textit{et al.}~\cite{adatpiveLocalFLRes19} analyze the convergence bound of local SGD incorporating non-IID data initially. 
Based on this analysis, they propose a local SGD algorithm with an adaptive synchronization frequency, optimized to minimize errors in optimization with heterogeneous resource constraints in workers and the PS, common in edge-based distributed DL paradigms. 
FedLamp~\cite{adaptiveSyncCompFL23} and AdaSFL~\cite{freqBatchFL23} employ approaches similar to~\cite{adatpiveLocalFLRes19} for determining the optimal synchronization frequency based on convergence bounds within heterogeneous resource constraints adaptively. 
The difference is that FedLamp optimizes the synchronization frequency and data compression ratio jointly, as detailed in Section~\ref{sec:compression}, aimed at minimizing training time while considering convergence and resource constraints.
Conversely, AdaSFL optimizes the synchronization frequency and minibatch size jointly to achieve the same objective under identical constraints.
In Adaptive Asynchronous Federated Learning (AAFL)~\cite{adaptiveResFL23}, the number of local updates in each synchronization round is determined adaptively by an experience-driven deep reinforcement learning algorithm, minimizes the maximum local accumulated training time of workers while considering bandwidth budget and convergence performance constraints.

Recently, several studies have focused on reconstructing data based on sampled data on each worker to address the challenge of non-IID heterogeneous data in FL.
For instance, FAST~\cite{adaptiveSamplingFL23} has developed an online learning algorithm that optimizes the synchronization frequency and data sampling on each device jointly to minimize global training loss within resource and time budgets.
This online algorithm can offer the optimized data sampling results in each iteration dynamically without prior knowledge of the data distribution on each worker.

There are empirical approaches in addition to the aforementioned optimization-based methods.
In AMBLE~\cite{adjustBatchEpochFL22}, the learning rate, minibatch size, and synchronization frequency are adjusted concurrently for local updating in heterogeneous workers, employing empirical strategies.
AMBLE formulates the local updating time for workers, assigning greater values to the number of local updating iterations and minibatch size for faster workers. 
The learning rate is then adjusted linearly in relation to the adaptive synchronization frequency and minibatch size.
AMBLE demonstrates superior experimental prediction results compared to FedAvg in both IDD and non-IDD scenarios.

\subsubsection{Ability to forget over time} \label{sec:flUnlearning}
Some studies~\cite{forgetFL22} focus on FL heterogeneity in the temporal direction, distinct from the heterogeneity observed in the spatial direction. 
During incremental training, where the model retains previously learned knowledge while acquiring new information, the data distribution may be non-stationary over time.
The model must learn to forget old knowledge during the retraining to acquire new knowledge;  this ability to forget is known as unlearning.
However, unlearning the FL model is challenging because training data are not shared with a central PS server. 
Investigating the unlearning problem in FL, Liu \textit{et al.}~\cite{forgetFL22} introduce a time-saving and energy-efficient retraining method.
They utilize the first-order Taylor expansion to approximate the objective and the diagonal empirical Fisher Information Matrix~\cite{secondOrderOpt21} to approximate the inverse Hessian in the Quasi-Newton method.
This rapid retraining method allows different FL parties to collaborate efficiently in the unlearning process without sharing data, thus erasing data samples from a trained FL model completely.


\subsection{Lessons Learned toward Communication-efficient Large-scale Model Synchronization} \label{sec:sgdLessons}
In this section, we summarize lessons learned from developing communication-efficient algorithms for model synchronization in large-scale distributed DL.
These insights have ensued a number of interesting research topics.

$\bullet$ \textit{The bottleneck lies in communication overhead, and employing large-batch training can accelerate large-scale distributed DL.} When dealing with large data sets and models in distributed DL with numerous powerful workers, using a minibatch size that is significantly greater than those traditionally used can diminish communication overhead and expedite training.
Because distributed SGD has been demonstrated to exhibit linear convergence speedup w.r.t the number of workers in various optimization settings, large-batch training can fully exploit the computational power offered by high parallelism in a large-scale cluster. 
However, large-batch training poses challenges that need attention, including increased memory requirements, potential decay in model generalization, and the need for more complicated hyperparameter tuning. 
For instance, a staleness-aware learning rate is proposed to adjust the learning rate of different models adaptively during different training stages, aiming to coordinate the model consistency across workers in large-batch distributed training.

$\bullet$ \textit{Heterogeneity is inevitable, and distributed SGD must adapt to heterogeneous environments dynamically.} 
Various forms of heterogeneity are unavoidable in large-scale distributed DL. 
For optimal performance, distributed SGD should adjust its hyperparameters to match the characteristics of data, models, resources, and even the temporal dynamics of these factors.
The analysis of theoretical convergence guarantees for new adaptive distributed SGD algorithms can follow existing analyzing frameworks presented in other literature, controlling certain hyperparameters as fixed variables while keeping other dynamic.
Challenges of this approach include efficient profiling of these characteristics, the development of efficient algorithms to search for optimal solutions in dynamic training environment, and ensuring robustness across diverse workloads, including those with non-IID and non-stationary data sets. 
The isolation and opacity of environmental characteristics in FL scenarios further complicate finding a practical solution to address these challenges.

$\bullet$ \textit{Stragglers pose obstacles, and hierarchical SGD can alleviate the straggler problem caused by heterogeneous device resources effectively.} 
In the large-scale distributed DL scenario with a considerable number of heterogeneous devices having diverse computational and communication resources, devices can be clustered into different groups based on their resource capacities, localities, and data profiles.
Hierarchical SGD, which applies different model-synchronization strategies for different groups at different levels, can limit the impact of stragglers to a minimal extent. 
However, this strategy also poses challenges pending to be tackled. 
Examples include the complexity in devising the optimal clustering topology, balancing workload, and ensuring model consistency across groups and levels.
Additionally, the strategy must demonstrate adaptability to dynamic device membership throughout the entire training process.

\newcommand\cCate[1]{\parbox[t]{2.0cm}{\centering #1}}
\newcommand\cRef[1]{\parbox[t]{2.2cm}{#1}}
\newcommand\cYear[1]{\parbox[t]{0.6cm}{\centering #1}}
\newcommand\cContent[1]{\parbox[t]{10.5cm}{#1 \vspace{1.5pt}}}
\begin{table*}[!t]
     \caption{Studies on communication-efficient compression techniques for large-scale distributed DL}\label{tab:compression}
    \centering
    \begin{tabular}{|c|c||l|l|l|}
    \Xhline{2\arrayrulewidth}
        \multicolumn{2}{|c||}{\cellcolor{blue!15}\textbf{Category}} & \cellcolor{blue!15}\textbf{Strategy\&Ref.} & \cellcolor{blue!15}\textbf{Year} & \cellcolor{blue!15}\textbf{Highlight}\\
        \Xhline{2\arrayrulewidth}
        \parbox[t]{2mm}{\multirow{20}{*}{\rotatebox[origin=c]{90}{Quantization (\ref{sec:quan})}}} & \multirow{2}{*}{\cCate{Error-feedback (\ref{sec:quanErrorFeedback})}} & \cRef{1-bit SGD~\cite{1bitQuan14}} & \cYear{2014} & \cContent{The first quantization for distributed DL, reducing gradients to 1 bit with error feedback.} \\ \cline{3-5}
        ~ &  & \cRef{SignSGD~\cite{signsgdQuan18}} & \cYear{2018} & \cContent{Quantizing gradients to 1-bit signs based on majority votes among workers.} \\ \cline{3-5}
        ~ & & \cRef{EF-SignSGD~\cite{errorFeedbackSignSGD19}} & \cYear{2019} & \cContent{SignSGD with error feedback.} \\ \cline{3-5}
        ~ & ~ & \cRef{1-bit Adam~\cite{1bitAdamQuan21}} & \cYear{2021} & \cContent{Applying 1-bit quantization with error feedback to the Adam optimizer.} \\ 
        \cline{2-5}
        ~ & \multirow{2}{*}{\cCate{Stochastic (\ref{sec:quanStochastic})}} & \cRef{Gupta \textit{et al.}~\cite{fixPoint16bitQuan15}} & \cYear{2015} & \cContent{Stochastically rounding to 16-bit computation.} \\ \cline{3-5}
        ~ & & \cRef{DoReFa-Net~\cite{dorefaLowBitwidth16}} & \cYear{2016} & I\cContent{ntroducing noise in uniform distribution to compensate quantization errors.} \\ \cline{3-5}
        ~ & ~ & \cRef{QNN~\cite{qnn17}} & \cYear{2017} & \cContent{Stochastic binary quantization based on a hard sigmoid probability function.} \\ \cline{3-5}
        ~ & ~ & \cRef{ZipML~\cite{zipmlQuan17}} & \cYear{2017} & \cContent{Stochastic quantization based on empirical distributions derived via double sampling.} \\ \cline{3-5}
        ~ & ~ & \cRef{NaturalComp~\cite{naturalQuan22}} & \cYear{2022} & \cContent{Stochastically rounding gradients to the upper or lower nearest powers of two.} \\ \cline{3-5}
        ~ & ~ & \cRef{Suresh \textit{et al.}~\cite{distributedmeanEst17}} & \cYear{2017} & \cContent{Quantizing rotated gradients multiplied by a random rotation matrix, eliminating gradient distribution assumption.}\\
        \cline{2-5}
        ~ & \multirow{2}{*}{\cCate{Matrix decomposition (\ref{sec:quanMatrix})}} & \cRef{ATOMO~\cite{atomoSVDQuan18}} & \cYear{2018} & \cContent{Applying entry-wise or singular value decomposition (SVD) on gradients.} \\ \cline{3-5}
        ~ & & \cRef{PowerSGD~\cite{powersgdQuan19}} & \cYear{2019} & \cContent{Applying two low-rank matrices for decomposing and composing gradients efficiently.} \\ \cline{3-5}
        ~ &  & \cRef{Vogels \textit{et al.}~\cite{practicalLowRankQuan20}} & \cYear{2020} & \cContent{Increasing the number of rank 1 power iterations to enhance PowerSGD for higher rank quantization approximation.} \\ \cline{3-5}
        ~ &  & \cRef{PCA-AWFL~\cite{wirelessFLPCA23}} & \cYear{2023} & \cContent{Leveraging principle component analysis to reduce uplink gradients in wireless FL scenarios.} \\
        \cline{2-5}    
        ~ & \cCate{Guarantee (\ref{sec:quanBounds})}& \cRef{\cite{qsgdQuan17,randomizedMeanEst18,terngradQuan17,errorCompensatedSGD18,nonUniformQSGD21,adaptiveQuan20,scaledIntQuan22}} & \cYear{2017-2022} & \cContent{Analyses of theoretical convergence guarantees of quantization w.r.t. the quantization level.} \\ 
        \cline{2-5}
        ~ & \multirow{2}{*}{\cCate{Quantization for FL (\ref{sec:quanFL})}}& \cRef{AQG~\cite{adaptiveQuanFL22}} & \cYear{2022} & \cContent{Adaptive quantization level; skipping stable quantized gradients and amplifying certain quantized gradient regarding device dropouts.}\\
        \cline{3-5}
        ~ & & \cRef{AdaGQ~\cite{adaptiveQuanHeteFL23}} & \cYear{2023} & \cContent{Adaptive quantization level for each training iteration on each device.} \\ 
        \Xhline{2\arrayrulewidth}
        \parbox[t]{2mm}{\multirow{34}{*}{\rotatebox[origin=c]{90}{Sparsification (\ref{sec:spar})}}} & \cCate{Threshold (\ref{sec:sparThreshold})}& \cRef{Strom~\cite{fixedSparse15}} & \cYear{2015} & \cContent{The first sparsification method for DL, selecting gradients beyond a fixed threshold bound and computing residuals of these gradients and the threshold.}\\
        \cline{3-5}
        ~ &  & \cRef{top-$k$~\cite{ratioSparse17}} & \cYear{2017} & \cContent{top-$k$ sparsification dropping a significantly large portion of gradients.} \\ \cline{3-5}
        ~ & ~ & \cRef{Mem-SGD~\cite{sparseMem18}} & \cYear{2018} & \cContent{Sparsification with error feedback for distributed SGD.} \\ \cline{3-5}
        ~ & ~ & \cRef{DGC~\cite{momentumSparse18}} & \cYear{2018} & \cContent{top-$k$ sparsification with momentum correction, local gradient clipping, momentum factor masking, and warmup training for SGD with momentum.}\\
        \cline{3-5}
        ~ & ~ & \cRef{EGC~\cite{entropyThresholdSpar21}} & \cYear{2021} & \cContent{Sparsification based on layer-wise thresholds determined by the entropy of gradient bins of each layer.}\\
        \cline{3-5}
        ~ & ~ & \cRef{MIPD~\cite{mipdAdaptiveSparse22}} & \cYear{2022} & \cContent{Sparsification based on layer-wise thresholds determined by the L2-Norm of gradients of each layer.}\\
        \cline{2-5}
        ~ & \multirow{2}{*}{\cCate{Scalability considerations (\ref{sec:sparScale})}}& \cRef{Global top-$k$~\cite{globalTopDSparse19}} & \cYear{2019} & \cContent{Selecting a portion of globally largest gradients across all workers and aggregating sparsified gradients hierarchically.} \\
        \cline{3-5}
        ~ &  & \cRef{ScaleCom~\cite{scalecomScaleSparse20}} & \cYear{2020} & \cContent{Workers taking turns to be the leading worker and using its local top-$k$ selection indices for the selection in all other workers.}\\
        \cline{3-5}
        ~ &  & \cRef{SIDCo~\cite{SIDCoStatBasedSparse21}} & \cYear{2021} & \cContent{Sampling and matching gradients to some heuristic distribution, whose optimal sparsification threshold is determined empirically.}\\
        \cline{3-5}
        ~ & ~ & \cRef{MSTopK~\cite{sparseCommCloud21}} & \cYear{2021} & \cContent{Using binary search to determine an approximate sparsification threshold efficiently.} \\ \cline{3-5}
        ~ & ~ & \cRef{Ok-Topk~\cite{OkTopkSparse22}} & \cYear{2022} & \cContent{Global top-$k$ with adaptive threshold.} \\ \cline{3-5}
        ~ & ~ & \cRef{JointSpar~\cite{largeBatchSparse22}} & \cYear{2022} & \cContent{Dropping certain layers based on a probability distribution.} \\ 
        \cline{2-5}
        ~ & \cCate{Guarantee (\ref{sec:sparConverge})} & \cRef{\cite{gsparRandomOpt18,convergenceSparse18,fftSparse20,hardThresholdSparse21}} & \cYear{2018-2021} & \cContent{Analyses of theoretical convergence guarantees of sparsification w.r.t. the sparsification level.} \\ 
        \cline{2-5}
        ~ & \multirow{2}{*}{\cCate{Communication-computation trade-off (\ref{sec:sparCommCom})}}& \cRef{OMGS-SGD~\cite{optMergeSparsification20}} & \cYear{2020} & \cContent{Optimizing scheduling of layer-wise aggregation of top-$k$ sparsification to overlap communication with computation.}\\
        \cline{3-5}
        ~ &  & \cRef{DRAGONN~\cite{dragonnRandomSparse22}} & \cYear{2022} & \cContent{Applying sparsification when the communication benefit surpasses computation overhead.} \\
        \cline{2-5}
        ~ & \cCate{Sparsification for FL (\ref{sec:sparFL})}& \cRef{STC~\cite{robustSparseFL20}} & \cYear{2020} & \cContent{top-$k$ sparsification and ternary quantization for bidirectional communication data compression in FL.}\\
        \cline{3-5}
        ~ & & \cRef{GossipFL~\cite{gossipFLSparse22}} & \cYear{2022} & \cContent{Synchronizing sparsified gradients among workers in a gossip manner based on a bandwidth-aware matrix.}\\
        \cline{3-5}
        ~ & ~ & \cRef{QSFL~\cite{clientLevelSparse22}} & \cYear{2022} & \cContent{Worker-level sparsification and model-level sparsification.} \\ \cline{3-5}
        ~ & ~ & \cRef{FAB-top-$k$~\cite{adaptiveSparseFL20}} & \cYear{2020} & \cContent{Optimizing the sparsification level to minimize training time with constraints of the fairness of sparsity among workers.}\\
        \cline{3-5}
        ~ & ~ & \cRef{FedDD~\cite{fedDDDiffDropout23}} & \cYear{2023} & \cContent{Optimizing the sparsification level to minimize training time with constraints of heterogeneous resources, data, and models.}\\
        \Xhline{2\arrayrulewidth}
        \parbox[t]{2mm}{\multirow{11}{*}{\rotatebox[origin=c]{90}{Others (\ref{sec:compOther})}}} & \cCate{Combined methods (\ref{sec:compOtherCombined})}& \cRef{\cite{fixedSparse15,ratioSparse17,redSyncSpar19,fftSparse20,mipdAdaptiveSparse22}} & \cYear{2015-2022} & \cContent{Sequentially applying sparsification and quantization.}\\
        \cline{2-5}
        ~ & \cCate{Guarantee (\ref{sec:compOtherConverge})}& \cRef{\cite{periodQuan18,qsparseLocal19,acsgdAdapt22}} & \cYear{2018-2022} & \cContent{Analyses of theoretical convergence guarantees of combining quantization and sparsification.}\\
        \cline{2-5}
        ~ & \multirow{2}{*}{\cCate{Sparsification with encoding (\ref{sec:compSparEncoding})}}& \cRef{3LC~\cite{3lcQuan19}} & \cYear{2019} & \cContent{Combining sparsification and quantization with lossless encoding algorithms.} \\ \cline{3-5}
        ~ & & \cRef{DFS~\cite{sparseBlkFormat23}} & \cYear{2023} & \cContent{Encoding sparsified gradient blocks into a zero-compacted format for bidirectional communication.}\\
        \cline{2-5}
        ~ & \cCate{Residual (\ref{sec:compResidual})}& \cRef{ResFed~\cite{ResFedFL23}} & \cYear{2023} & \cContent{Compressing residuals based on the difference of a sequence of updated model for bidirectional communication.}\\
        \cline{2-5}
        ~ & \cCate{Autoencoder (\ref{sec:compAutoencoder})}& \cRef{LGC~\cite{autoencodeGradComp21}} & \cYear{2021} & \cContent{Auto-encoding the common component of sparsified gradients.} \\ 
        \Xhline{2\arrayrulewidth}
    \end{tabular}
\end{table*}

\section{Communication-Efficient Data Compression}\label{sec:compression}
\begin{figure}[!t]
\centering
\includegraphics[width=1\columnwidth]{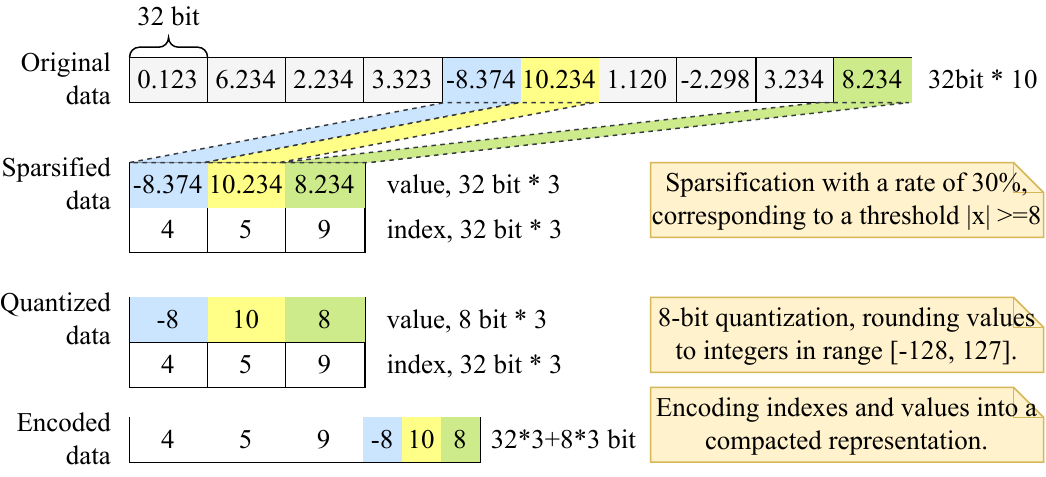}
\caption{An example of applying various communication data compression technologies to the data in distributed DL}
\label{fig:compression}
\end{figure}

Compressing data, including model parameters and gradients, can reduce communication overhead across workers and the PS during distributed training and inference effectively. 
This process entails two crucial trade-offs: the balance between computation and communication, and that between accuracy and workloads.
The former trade-off expects that allocating computational resources for compression can significantly enhance communication performance and consequently, overall efficiency. 
This requires compression algorithms to be computationally efficient. 
The latter anticipates reduced communication overhead or training and inference time, without significantly compromising model convergence performance and prediction accuracy. 
As listed in Table~\ref{tab:compression}, communication data compression technologies for large-scale distributed DL mainly fall into the following categories: gradient quantization, gradient sparsification, and others.
Fig.~\ref{fig:compression} illustrates an example of these compression technologies.

This section reviews existing works in these categories, with a special focus on applying communication-efficient data compression technologies for distributed DL in large-scale scenarios. 
We also summarize lessons learned from these works for future high-performance communication data compression technologies in large-scale distributed DL.

\subsection{Gradient Quantization}\label{sec:quan}
Gradient quantization~\cite{CommCompSurvey2023}, or quantization, is a lossy compression technique used in distributed training, particularly the data-parallel training with distributed SGD. 
Exchanging high-precision gradients across the cluster and updating the model iteratively consume computational and communication resources intensively. 
Quantization methods reduce the data, typically gradients, originally represented by 32- or 64-bit floating-point values into lower-precision values with fewer bits.
This helps mitigate computation and communication overhead for large-scale distributed DL. 

Various quantization methods primarily differ in the number of bits used and the data-rounding strategy.
Common bit numbers for quantized gradients, also referred to as quantization levels, include 16~\cite{fixPoint16bitQuan15}, 9~\cite{naturalQuan22}, 8~\cite{8bitQuan18}, 2~\cite{terngradQuan17}, 1~\cite{signsgdQuan18}, or of variable length~\cite{distributedmeanEst17}.
Typical data-rounding strategies encompass deterministic rounding with error feedback, stochastic rounding, and matrix decomposition. 
This subsection initially introduces quantization methods categorized by various data-rounding strategies, subsequently presents theoretical analyses on the convergence performance of quantization, and finally, discusses special considerations for applying quantization to FL in large-scale and heterogeneous settings.

\subsubsection{Deterministic quantization with error feedback} \label{sec:quanErrorFeedback}
Some quantization methods utilize deterministic functions to reduce model parameters or gradients into lower-precision values. 
The 1-bit SGD method~\cite{1bitQuan14} is recognized as the first quantization method with error feedback, or error compensation, for distributed DL. 
1-bit SGD condenses 32-bit gradients into one bit per value by using a constant quantization threshold of 0. 
To uphold model accuracy, its error-feedback mechanism ensures that previous quantization errors are reflected in the subsequent minibatch gradient quantization. 
Similarly, SignSGD~\cite{signsgdQuan18} transmits only 1-bit signs of gradients.
It employs a majority-vote strategy to aggregate these gradient signs in the PS before broadcasting them back to workers. 
SignSGD demonstrates efficient convergence in non-convex problems using optimizers of SGD and SGD with momentum, particularly when gradients exhibit density comparable to, or greater than than stochasticity and curvature.
However, Karimireddy \textit{et al.}~\cite{errorFeedbackSignSGD19} show that SignSGD may fail to converge in certain convex problems or exhibit poor generalization due to the biased nature of the sign quantization.
To address this bias, they introduce EF-SignSGD, incorporating an error-feedback mechanism akin to 1-bit SGD~\cite{1bitQuan14} into SignSGD.
This modification results in the model converging as rapidly as SGD and demonstrate superior generalization compared to SignSGD.

The quantization with error feedback needs adaptation when applied to large models.
Though error feedback can compensate for accumulated gradient errors for SGD and SGD with momentum, it cannot address the non-linear gradient errors in the Adam optimizer.
The Adam optimizer is utilized widely in Transformer-based models such as Bert~\cite{bert19} and LLMs~\cite{lamda22}. 
To overcome this limitation in quantization for Adam optimizers, 1-bit Adam~\cite{1bitAdamQuan21} capitalizes the observation that Adam's variance remains stable in the early stage of distributed training.
It divides the distributed training process into two phases: Adam warmup and momentum quantization. 
The warmup phase employs vanilla Adam for a few epochs.
Subsequently, the Adam's variance term becomes fixed, and the momentum quantization phase utilizes the SGD with momentum optimizer, applying 1-bit quantization with error feedback. 

Without randomization, deterministic quantization methods are easy to compute.
However, biased quantization errors are commonly acknowledged as a prevalent issue in most deterministic methods. 
To mitigate quantization bias, the error-feedback mechanism is one approach; 
nevertheless, the focus of numerous studies is on stochastic quantization methods to address this problem.

\subsubsection{Stochastic quantization}\label{sec:quanStochastic}
Stochastic quantization methods usually make distribution assumptions on the data and employ stochastic functions to ensure that quantized gradients serve as unbiased approximations to the original gradients. 
To the best of our knowledge, Gupta \textit{et al.}~\cite{fixPoint16bitQuan15} are the first to introduce a stochastic rounding scheme to gradient quantization, which condenses distributed training into 16-bit fixed point computation. 
The likelihood of rounding a floating-point value to a specific fixed-point value is proportional to the proximity of these two values. 
Deterministic and stochastic methods can also be combined.
DoReFa-Net~\cite{dorefaLowBitwidth16} quantizes CNN model parameters and activations deterministically, while quantizing gradients stochastically. 
When quantizing gradients, DoReFa-Net introduces gradient noise in a uniform distribution to compensate for the quantization error.
However, assuming a uniform distribution for gradient errors in the stochastic functions of these methods may not fully eliminate quantization bias. 

To address the limitation of the uniform distribution assumption on gradient errors, some studies opt for alternative distributions to ensure unbiased quantization. 
QNN~\cite{qnn17} also quantizes gradients into binary values, similar to~\cite{1bitQuan14}, but bases it on a hard sigmoid probability function. 
This stochastic binarization method is verified to reduce training time and memory consumption significantly without compromising the model prediction accuracy.
ZipML~\cite{zipmlQuan17} employs an empirical distribution of gradients, derived through a double sampling strategy, to minimize stochastic quantization variance in linear models. 
Nevertheless, its performance on non-linear models is not well investigated.  
Natural Compression~\cite{naturalQuan22} rounds gradients stochastically to the nearest powers of two, whether upper or lower. 
In contrast, Suresh \textit{et al.}~\cite{distributedmeanEst17} propose a stochastic rotated quantization method that makes no assumption on the data distribution. 
This method generates a random rotation matrix to rotate gradients and quantizes the rotated results, yielding a lower mean square error (MSE) between the original and quantized gradients. 
To further reduce the MSE, a variable-length coding approach is approach, encoding quantized data into variable-length formats.

\subsubsection{Matrix decomposition quantization}\label{sec:quanMatrix}
Recently, some studies have proposed to apply quantization to the decomposed matrices of gradients.
ATOMO~\cite{atomoSVDQuan18} applies entry-wise or singular value decomposition (SVD) on gradients, and the decomposed results are sparsified subsequently. 
ATOMO is an unbiased compression scheme and represents a generalized version of QSGD~\cite{qsgdQuan17} and TernGrad~\cite{terngradQuan17}. 
However, deriving SVD iteratively is computationally expensive. 
To address this issue, PowerSGD~\cite{powersgdQuan19} employs a low-rank gradient compressor based on the power iteration~\cite{powerIteration75}. 
This method introduces two low-rank matrices for quantization and the decomposition and composition involve only low-rank matrix multiplication and orthogonalization, ensuring computation efficiency. 
Claiming to be the pioneering gradient-quantization method, it achieves end-to-end wall-clock training speedup compared to vanilla SGD, enhancing scalability to large clusters, and exhibits superior generalization performance. 
On one hand, in contrast to prior works such as sign-based quantization with a majority vote~\cite{signsgdQuan18,errorFeedbackSignSGD19}, which require a gather operation, PowerSGD allows the hierarchical addition of the decomposed matrix. 
This approach enables efficient leverage of communication optimization for the \texttt{allreduce} primitive in a large-scale cluster. 
On the other hand, a low-rank gradient update can be interpreted as a spectral regularization~\cite{geometryBiasOpt18}, contributing to the model's enhanced generalization. 
A follow-up study by Vogels \textit{et al.}~\cite{practicalLowRankQuan20} enhances the practicality of PowerSGD by achieving a higher-rank quantization approximation through increasing the number of rank-1 power iterations. 
In the scenarios of wireless federated learning, where the uplink transmission from the worker to the PS is a communication bottleneck, PCA-AWFL~\cite{wirelessFLPCA23} leverages principle component analysis to reduce the dimension of uploaded gradients of Nesterov's momentum. 

\subsubsection{Convergence guarantees for quantization} \label{sec:quanBounds}
Lower-precision computation through quantization can degrade the convergence performance of DL models, thereby reducing prediction accuracy. 
Consequently, minimizing communication costs when ensuring theoretic guarantees on convergence performance becomes a primary focus of quantization methods.
Alistarh \textit{et al.}~\cite{qsgdQuan17} analyze tight bounds on the trade-off between precision and quantization variance trade-off, and propose QSGD, a family of stochastic gradient-quantization algorithms.
QSGD provides theoretical guarantees on convergence rate with a bound on the gradient precision. 
In line with QSGD, Kone{\v{c}}n{\`y} and Richt{\'a}rik~\cite{randomizedMeanEst18} investigate the trade-off between the communication cost and MSE through a family of stochastic quantization methods.
These methods vary in terms of variable-size or fixed-size encoding, as well as dense or sparse communication protocols. 
This study provides quantization MSE bounds associated with various communication costs.
Wen \textit{et al.}~\cite{terngradQuan17} further prove the convergence from the perspective of statistical bound on gradients.
Based on this convergence analysis, they propose a special case of QSGD, known as TernGrad, which map gradients stochastically into ternary values: positive, zero, and negative. 
Workers and the PS synchronize these ternary values along with a gradient scalar.
Subsequently, this scalar multiplies the ternary values in each worker, producing actual gradients for updating models.
ECQ-SGD~\cite{errorCompensatedSGD18} further combines the error-feedback mechanism with stochastic gradient quantization, ensuring a tighter bound on the worst-case error compared to stochastic quantization methods without error feedback, such as QSGD.
NUQSGD~\cite{nonUniformQSGD21} offers strictly tighter bounds on the trade-off between communication cost and quantization variance than QSGD.
This improvement is achieved by substituting the uniform quantization scheme in QSGD with an unbiased nonuniform logarithmic scheme. 

In the pursuit of adaptive quantization, Faghri \textit{et al.}~\cite{adaptiveQuan20} suggest quantization methods that adjust quantization levels among different iterations dynamically based on runtime statistics of data distribution. 
IntSGD~\cite{scaledIntQuan22} employs an adaptive scaling factor to multiply gradients before quantizing them stochastically into integer values. 
These values are then scaled back for model updating by the reciprocal of the scaling factor.
This scaling factor is calibrated using the moving average of model parameters, facilitating computation efficiency of quantization.
Theoretical and empirical demonstrations show that these adaptive methods offer tighter variance and precision bounds than non-adaptive ones. 

\subsubsection{Gradient quantization in FL}\label{sec:quanFL}
Recently, emerging studies have focuses on gradient quantization in FL, which presents unique challenges. 
First, the heterogeneity of computing devices and non-IID nature of data in FL amplify the challenges for unbiased quantization. 
Second, as the norm of gradients varies during training in FL, a fixed quantization level failed to minimize the communication cost while guaranteeing convergence throughout the training process.
Third, ensuring the reliability of devices in FL is challenging, and frequent device dropouts pose a significant concern. 
To tackle these challenges, Mao \textit{et al.}~\cite{adaptiveQuanFL22} introduce Augmented Adaptive Quantized Gradient (Augmented AQG).
Augmented AQG enables devices to skip transmitting slowly varying quantized gradients, addressing the heterogeneity issue.
It determines the quantization level during training adaptively to aim for the minimum communication cost and amplifies certain quantized gradients appropriately to mitigate potential device dropouts. 
On the other hand, AdaGQ~\cite{adaptiveQuanHeteFL23} addresses the fluctuating gradient norm problem on diverse edge devices in FL.
It employs an adaptive quantization method, allowing each device to decide its own suitable quantization level for every training iteration. 
Both approaches improve the distributed training performance when dealing with non-IID data. 

\subsection{Gradient Sparsification}\label{sec:spar}
Gradient sparsification, or sparsification, is another lossy mechanism orthogonal to gradient quantization for communication data compression in distributed training, particularly the data-parallel training with distributed SGD.
Sparsification takes advantage of the observation known as gradient sparsity during backpropagation when updating model parameters, where gradients exhibit positive skewness, with many being close to zero or insignificant.
It involves discarding less significant gradients and sparsifying more substantial gradients, aiming to reduce communication costs in distributed training. 
Typically, sparse data are encoded into index and values pairs, which can undergo additional quantization and compression. 

In this subsection, we introduce various sparsification methods, consider their scalability issues, provide theoretical analyses of convergence guarantees, and examine the trade-off between computation and communication. 
Additionally, we discuss optimizations for sparsification in high-performance FL within large-scale and heterogeneous settings.

\subsubsection{Sparsification with a threshold}\label{sec:sparThreshold}
To the best of our knowledge, Strom~\cite{fixedSparse15} introduces the first gradient-sparsification method for distributed training. 
This method only selects gradients with absolute values above a fixed threshold and sends sparse residual values calculated by the difference between these gradients and the threshold. 
This threshold represents the sparsification level, indicating the level of sparsity after sparsification. 
Since an appropriate fixed threshold is hard to determine in practice, Aji and Heafield~\cite{ratioSparse17} propose an approach known as top-$k$ sparsification, whereby a large portion (e.g., 99\%) of gradients with smaller values are dropped. 
Similar to the quantization error, sparsification can also accumulate gradient errors during iterative training.  
Mem-SGD~\cite{sparseMem18} incorporates the error-feedback mechanism into sparsification to compensate for sparsification errors, thus avoiding error explosion. 
However, these methods are designed for SGD but may result in significant gradient errors if applied to SGD with momentum, which is a pervasive replacement of vanilla SGD.
To ensure the convergence performance of sparsification when applied to SGD with momentum, DGC~\cite{momentumSparse18} utilizes four other techniques on top of the top-$k$ sparsification: momentum correction, local gradient clipping~\cite{gradientClipping13}, momentum factor masking~\cite{momnetumFactorMask16}, and warmup training. 
The momentum correction technique sparsifies momentum terms instead of gradients to maintain the accumulated discounting factor in SGD with momentum.
In addition, DGC applies local gradient clipping to prevent exploding gradients by limiting the upper bounds of gradients. 
It employs momentum factor masking to address staleness problems by preventing momentum for stale gradients.
Furthermore, warmup training accelerates training by utilizing a smaller learning rate and less aggressive gradient sparsity during the early stage. 

However, the top-$k$ sparsification methods above have limitations in identifying the significance of gradients.
Initially, the threshold is determined based on the gradient distribution of the entire DNN model, which is a coarse metric because gradients of different layers of the model can follow different distributions.
Moreover, the magnitude of gradients may not be the best measurement for gradient significance.
Addressing these limitations, EGC~\cite{entropyThresholdSpar21} employs the layer-wise threshold, determined by the entropy of gradient bins of each layer, indicating the significance of that layer.
In contrast, MIPD~\cite{mipdAdaptiveSparse22} determines the layer-wise sparsification threshold adaptively across different layers and employs the layer-wise L2-Norm as the indicator of gradient significance. 
These finer-grained approaches for determining the sparsification threshold are expected to contain aggregated sparsification errors within a small range and preserve prediction accuracy; however, calibrating the threshold for every layer in every training iteration adds to computation complexity, which may exacerbate scalability problems.

\subsubsection{Scalability considerations for sparsification} \label{sec:sparScale}
Threshold-based sparsification methods may encounter scalability problems when applied in large-scale clusters for three main reasons.
Firstly, the local top-$k$ methods mandate every worker to select a specific portion of gradients, resulting in an accumulated gradient volume that increases linearly with the number of workers and becomes non-negligible in large clusters. 
Secondly, frequent determination of the dynamic threshold in top-$k$ sparsification is computationally expensive, which hinders its application in large DNN models.
Thirdly, as training with large batch sizes~\cite{largeMinibatch17} becomes popular for large models in large clusters, issues such as the increasing computation cost per iteration and Layer-wise Adaptive learning Rate Scaling (LARS)~\cite{scaleBatchSize17,largeBatch19} pose new challenges to gradient sparsification; however, there is a lack of evidences that these sparsification methods work well in this scenario.

To tackle the first issue of large gradient volume, Shi \textit{et al.}~\cite{globalTopDSparse19} propose a global top-$k$ sparsification method to reduce the communication cost in large-cluster environments.
This method selects a portion of globally largest gradients among all workers and synchronizes through a hierarchical aggregation strategy. 
Notably, this method demonstrate significant improvements in scalability efficiency compared to earlier top-$k$ sparsification methods.
Another relevant development is ScaleCom~\cite{scalecomScaleSparse20}, which capitalizes on the observation that distributions of sparsification errors are similar among workers.
ScaleCom employs a Cyclic Local top-$k$ (CLT-k) method, where workers take turns as the leading worker, utilizing its local top-$k$ selection indices for the gradient selection in all other workers. 
The calculation of CLT-k is commutative, ensuring efficiency in \texttt{allreduce} implementations. 

To tackle the second issue of calculating overhead, SIDCo~\cite{SIDCoStatBasedSparse21} employs the heuristic that DNN gradients can be approximated by three representative sparsity-inducing distributions. 
The thresholds of these heuristic distributions can be determined efficiently at a designated compression ratio.
SIDCo concentrates on sampling gradients and matching them to one of the distributions, thus obtaining the threshold for sparsification naturally.
On the other hand, MSTopK~\cite{sparseCommCloud21} introduces an approximate top-$k$ sparsification approach, which uses binary search to determine an approximate sparsification threshold efficiently. 
Similar to~\cite{globalTopDSparse19}, Ok-Topk~\cite{OkTopkSparse22} also adopts the global top-$k$ method but only updates the threshold every few iterations instead of every iteration to further reduce the threshold computation overhead. 
The rationale of this periodic strategy derives from empirical observations that gradients change very slowly across multiple consecutive training iterations.
In addition, collective communication libraries tailored for sparse data enhance collective operations of sparsified gradients during distributed training, as detailed in Section~\ref{sec:infraCCSpar}.

To address the third issue arising with LARS in large-batch training, where learning rates scale differently in different model layers, ScaleCom additionally applies low-pass filtering to accumulated sparsification errors locally to accommodate rapid model changes and large gradient noise caused by the scaled learning rate.
The scalability of ScaleCom is ensured because its sparsification performance is independent of the number of workers and the batch size per worker.
To reduce computation costs in large-batch training, JointSpar~\cite{largeBatchSparse22} sparsifies gradient computation and communication jointly to avoid redundant gradient computation for those deemed to be dropped.
JointSpar first decomposes gradients into gradient blocks layer by layer, and then decides which gradient blocks to drop based on a probability distribution set over gradient blocks, where each element indicates the probability of dropping a specific gradient block.
JointSpar avoids both computation and communication of these blocks to drop. 
The convergence performance of JointSpar has been demonstrated theoretically and experimentally for large-batch training with LARS.

\subsubsection{Convergence guarantees for sparsification} \label{sec:sparConverge}
Similar to quantization, there is a trade-off between convergence and communication in sparsification. 
The communication performance is associated with the sparsification compression ratio, i.e., sparsification level.
The convergence performance of the sparsification methods above is supported by experimental results on commodity cloud infrastructures, but many methods lack theoretical guarantees. 
In an effort to develop sparsification methods with theoretical convergence guarantees, 
GSpar~\cite{gsparRandomOpt18} adopts a random approach, which drops gradients with probabilities and amplifies the remaining gradients to ensure unbiased sparsification. 
The probabilities are calibrated to minimize the communication cost after sparsification while adhering to a specific gradient error variance constraint.
Alistarh \textit{et al.}~\cite{convergenceSparse18} provide the first convergence theoretical analysis of the top-$k$ sparsification. 
By giving non-trivial upper bounds on the convergence rate, this analysis demonstrates that top-$k$ sparsification methods provide convergence guarantees for distributed SGD with either convex or non-convex training objectives.
Wang \textit{et al.}~\cite{fftSparse20} analyze the convergence performance of top-$k$ sparsification with respect to the compression ratio. 
They propose a Fast-Fourier-Transform-based (FFT-based) method for top-$k$ sparsification, employing FFT to transform gradients into the frequency domain and drops low-energy gradients according to a compression ratio that is bound to converge.  
Compared to previous gradient quantization and top-$k$ sparsification methods, the distribution of reconstructed gradients using this FFT-based method is closer to that of the original gradients prior to sparsification, indicating that the FFT-based method preserves more relevant information. 

When local or global top-$k$ sparsification methods apply constant thresholds across different training iterations, they result in sub-optimal convergence or communication performance. 
To guarantee the convergence performance in the presence of varying thresholds across different training iterations,
Sahu \textit{et al.}~\cite{hardThresholdSparse21} propose an adaptive top-$k$ sparsification method with a communication constraint. 
Given a communication budget for the entire training, this method selects the threshold adaptively for each iteration, with the goal of minimizing the accumulated sparsification error of the entire training. 
A theoretical analysis shows that it converges for both convex and non-convex objectives.

\subsubsection{Communication-computation trade-off optimization} \label{sec:sparCommCom}
Besides concerns about convergence, there are also concerns about the computation overhead introduced by sparsification, such as the additional encoding and synchronization overhead for sparsified gradients.
To tackle the trade-off between computation and communication in gradient sparsification,
OMGS-SGD~\cite{optMergeSparsification20} defines top-$k$ sparsification as an optimization problem of scheduling a set of layer-wise aggregation tasks of sparsified gradients.
These tasks of different gradient layers involve computation and communication stages and can be scheduled in a pipeline, allowing the communication stage of an earlier task can overlap with the computation stage of subsequent tasks.
OMGS-SGD find the optimal scheduling of layer-wise gradient aggregation, aiming to minimize the training time. 
From an analytical perspective, DRAGONN~\cite{dragonnRandomSparse22} implements sparsification selectively, activating it only when the estimated communication time saved exceeds the computing overhead associated with sparsification. 
DRAGONN also minimizes the sparsification computing overhead through the use of an efficient hashing algorithm to obtain gradients above the threshold, instead of relying on the mask-based (e.g., Hadamard-product-based) selection algorithm widely adopted by other sparsification methods. 

\subsubsection{Sparsification in FL}\label{sec:sparFL}
Similar to quantization, challenges regarding heterogeneity of computing devices, non-IID data, and communication patterns in FL also present in gradient sparsification. 
Numerous gradient-sparsification methods have been developed to address the FL heterogeneity from various perspectives.
STC~\cite{robustSparseFL20} represents the pioneering work in addressing bidirectional communication data compression in FL. 
When many studies focus gradient quantization in the worker-to-server direction,
STC employs top-$k$ sparsification and ternary quantization for compressing communication in both the worker-to-server uploading and server-to-worker downloading. 
Focusing on the heterogeneity in bandwidth and communication pattern in FL, GossipFL~\cite{gossipFLSparse22} utilizes a bandwidth-aware gossip matrix to guide how a peer worker synchronizes sparsified gradients with a single other peer in a gossip-based P2P communication pattern in FL.
This gossip matrix is designed based on the bandwidth information among peer workers for high-bandwidth communication, thus requiring little communication time in the synchronization. 
Concerning the heterogeneity of gradient contributions among workers in FL,
QSFL~\cite{clientLevelSparse22} employs a two-level gradient-sparsification method: worker-level sparsification for worker-to-server communication and model-level sparsification for server-to-worker communication. 
On the worker level, QSFL selects qualified workers for uploading gradients to the PS server considering both the worker's contribution to the loss and the relevance between the local worker model and the global server model.
On the model level, QSFL divides the global synchronized model into segments and sends one segment to each qualified worker. 
This worker-level sparsification is considered exclusive for FL scenarios where worker models can be heterogeneous, distinguishing it from the standard tensor-level~\cite{ratioSparse17} or layer-level~\cite{largeBatchSparse22} sparsification used in homogeneous model synchronization. 

To address the trade-off between computation and communication in FL,
Han \textit{et al.}~\cite{adaptiveSparseFL20} present a Fairness-Aware Bidirectional top-$k$ (FAB-top-$k$) sparsification method, which formulates sparsification as an online learning algorithm.
This approach determines the optimal sparsification level to minimize overall training time, considering factors including the estimated derivative sign, adjustable search interval, and fairness of sparsity among workers, thus addressing non-IID data issues.
Similarly, FedDD~\cite{fedDDDiffDropout23} formulates sparsification in FL as a convex optimization problem to find the optimal sparsification level for minimizing the overall training time, considering the heterogeneity of device computational resources, data distribution and quality, and model size and structure.

\subsection{Other Gradient Compression Technologies}\label{sec:compOther}
Other gradient compression technologies mainly include combining quantization and sparsification, integrating sparsification with compression encoding methods, residual compression, and autoencoder compression. 
This subsection introduces recent advancements in these technologies and provides an analysis of the convergence guarantees on some hybrid compression methods.

\subsubsection{Combining quantization and sparsification}\label{sec:compOtherCombined}
Gradient sparsification is usually applied in conjunction with other orthogonal compression technologies, such as gradient quantization.
In~\cite{fixedSparse15,ratioSparse17}, sparsified gradients are quantized using 1-bit~\cite{1bitQuan14} and 2-bit quantization, respectively, to further reduce communication costs. 
Similarly, RedSync~\cite{redSyncSpar19} applies gradient sparsification sequentially with a binary-searched threshold, similar to~\cite{sparseCommCloud21}, and 1-bit quantization, similar to~\cite{1bitQuan14}.
In~\cite{fftSparse20}, after FFT-based sparsification, the sparsified gradient frequency data are packed into a dense vector with another status vector, indicating the index locations. 
The dense vector is further quantized by a range-based quantization method, which converts 32-bit IEEE-754-format data to a lower-precision format with a range that matches the original format and has a number of bits constrained by training convergence requirements.
In~\cite{mipdAdaptiveSparse22}, following layer-wise sparsification, MIPD further applies quantization based on layer-wise gradient distributions, aiming to reduce the aggregated approximation error based on the gradient distribution of the entire model. 

\subsubsection{Convergence guarantees for quantization and sparsification combination}\label{sec:compOtherConverge}
Theoretical analyses are used demonstrate the convergence performance of the combination of sparsification and quantization.
PQASGD~\cite{periodQuan18} presents a special combination form of sparsification and quantization, where the synchronization of quantized gradients is triggered only periodically but not iteratively. 
This periodic synchronization can be perceived as the sparsification over training iterations but not over the model, as discussed earlier. 
Therefore, PAQSGD can be viewed as a combination of quantization and iteration-wise sparsification. 
Theoretically, this combination has been proven to converge at a rate inversely proportional to the square root of the product of the total minibatch size across all workers and the number of iterations within a period, indicating its ability to scale linearly as the number of worker increases. 
Similar to PAQSGD, Qsparse-local-SGD~\cite{qsparseLocal19} also synchronizes gradients periodically. 
The difference is that in the local computation between synchronization intervals, each worker additionally undergoes ordinary gradient sparsification and quantization sequentially, while maintaining a local error memory for the combined error compensation. 
Qsparse-local-SGD has been demonstrated theoretically to converge for smooth non-convex objectives with a fixed learning rate and smooth convex objectives with a decaying learning rate. 
To adjust the compression level of the combined methods adaptively to meet the convergence guarantee, AC-SGD~\cite{acsgdAdapt22} optimizes the quantization and sparsification level jointly to minimize the gradient variance, regarding factors including the communication budget, gradient norm, and remaining number of iterations.
This jointly optimized compression level guarantees convergence for non-convex and quadratic objectives. 

\subsubsection{Combining sparsification with compression encoding}\label{sec:compSparEncoding}
Gradient sparsification is also usually combined with efficient encoding algorithms. 
To maximize the compression ratio, both STC~\cite{robustSparseFL20} (which has been introduced) and 3LC~\cite{3lcQuan19} combine sparsification and quantization with lossless encoding algorithms. 
3LC first quantizes gradients into ternary values using a sparsity multiplier, which indicates the threshold for rounding gradients to zero, one of the ternary values. 
It then uses lossless based-3$^5$ encoding to compact every five ternary values into a one-byte representation, followed subsequently by another lossless zero-run-length encoding~\cite{runLengthEncode67} to shorten consecutive runs of the same bytes.
DFS~\cite{sparseBlkFormat23} segments sparsified gradients dynamically into data blocks and encodes them into a zero-compacted format for bidirectional communication of model synchronization. 
Given the increased compression complexity, 
designing algorithms that integrate various lossy and lossless compression technologies for distributed SGD requires a careful trade-off between computation and communication performance.
It is essential to ensure that the communication benefit provided by compression surpasses its computation overhead.

\subsubsection{Residual compression}\label{sec:compResidual}
Some compression technologies for distributed training adopt the residual-based approach.
This approach exploits the heuristic that residuals calculated from the difference between updated and predicted models within a specific period contain denser information than gradients. 
ResFed~\cite{ResFedFL23} predicts a model in local iterations in the worker through memorizing the model trajectory, which consists of a sequence of updated models across multiple training updates, and then derives the residuals.
It communicates residuals between workers and the PS server through compression methods such as sparsification and quantization. 
This residual-based compression is used for both uploading and downloading communications, making it appropriate for FL scenarios with constrained bandwidth. 
The residual-based sparsification and quantization can be perceived as a generalized version of the gradient sparsification and quantization, because when the predicted model is set to the last updated model, residuals essentially become gradients. 

\subsubsection{Autoencoder compression}\label{sec:compAutoencoder}
Some gradient-compression methods adopt a DNN autoencoder to encode gradients. 
By exploiting the observation that gradients share similarities across workers during distributed training~\cite{scalecomScaleSparse20},
LGC~\cite{autoencodeGradComp21} conceptually decomposes gradient values into two components: the common and innovation components, which represents the parts making gradients in a worker similar to or distinguished from those in other workers, respectively. 
The innovation component is captured by sparsified gradients with a very aggressive sparsification rate, e.g., 0.001\%, 
The common component is generated by inputting 0.1\% spasified gradients into the encoder side of a lightweight autoencoder, which has been trained for extracting the similarity among gradients of different workers during the early stage of distributed training. 
During gradient synchronization, workers and the PS server exchange the innovation and encoded common components, which are highly compacted for fast communication and can be decoded by the decoder side of the lightweight autoencoder efficiently.

\subsection{Lessons Learned toward Communication-efficient Large-scale DL Compression}
In this section, we explore lessons learned from developing communication-efficient compression algorithms for large-scale distributed DL, highlighting potential research problems for future studies in this area.

$\bullet$ \textit{The environment is complex and dynamic, and environment-aware adaptive compression can optimize the communication overhead in diverse settings.} 
The compression ratio can be adjusted based on diverse environment information dynamically, including but not limited to the data distribution, model architecture, and communication topology. However, the diversity and heterogeneity of these factors pose challenges to finding efficient methods for discovering the optimal solution and providing real-time adaption.

$\bullet$ \textit{Gradients vary in significance, and finer-grained assessment of gradient significance leads to improved overall performance in terms of convergence and communication in distributed DL with large models.} 
In large models with different learning characteristics in different components, a coarse criterion for gradient significance can fail to capture important information for model updating. 
Utilizing different strategies to calibrate gradient significance at the granularity of edges, workers, model components, or even model layers can optimize the convergence performance with the minimal communication overhead.
Yet, fine-grained compression entails a high computation overhead. 
The primary challenge is finding a balance between the computation and communication overhead.

$\bullet$ \textit{Various compression methods can be compatible with each other, and hybrid and hierarchical compression methods can collaborate to reduce communication overhead significantly.}
For large-scale distributed DL, a combination of various lossy and lossless compression algorithms can be utilized to optimize the communication at different hierarchical levels of model synchronization.
The convergence performance of the hybrid and hierarchical approach is guaranteed theoretically in certain optimization settings. 
However, there is a lack of detailed studies on the specific contributions of each component in the hybrid and hierarchical approach to the overall convergence and communication performance.
This approach approach usually relies on empirical combinations of different compression technologies, and achieving the optimal combination is challenging due to the large search space.

\begin{figure*}[!t]
\centering
\includegraphics[width=0.8\textwidth]{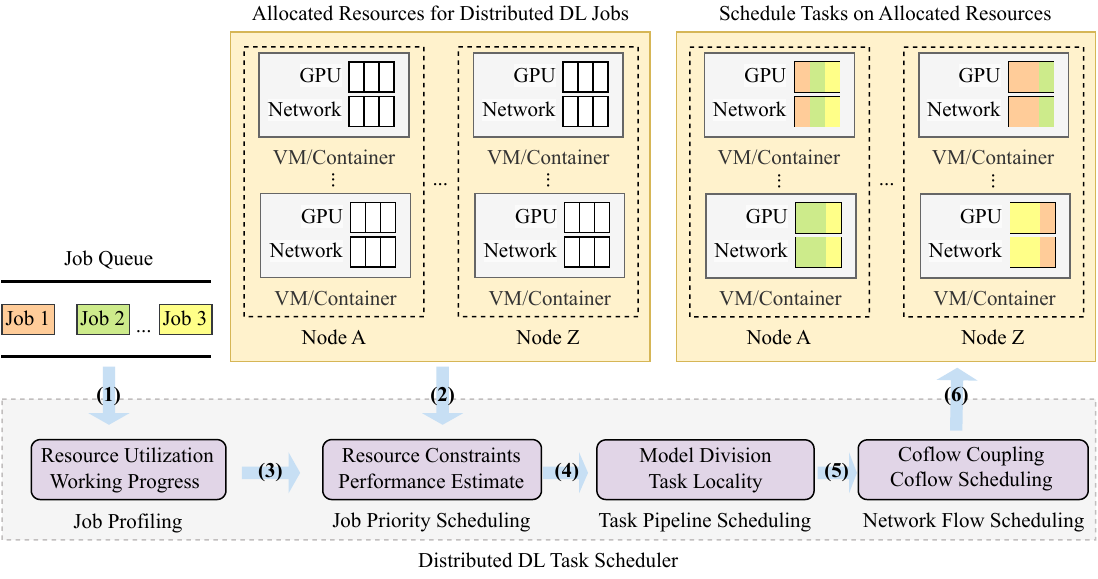}
\caption{An illustration of resource-management and task-scheduling mechanisms in large-scale distributed DL}
\label{fig:resource_task}
\end{figure*}

\section{Large-Scale Resource Allocation and Task Scheduling}\label{sec:resourceTask}
This section introduces resource-allocation and task-scheduling strategies for high-performance distributed DL at a large scale. 
These strategies are typically integrated in cluster-level and distributed-DL-level frameworks.
Fig.~\ref{fig:resource_task} illustrates a procedure of resource-allocation and task-scheduling mechanisms for large-scale distributed DL within a cluster. 
This procedure comprises six major steps.
First, for a queue of distributed DL jobs, the task scheduler conducts job profiling based on various workload characters, such as resource-utilization status and working progress. 
Second, the resource manager allocates GPU and network resources distributed within the cluster for jobs based on their characteristic profiling. The resources can be represented in a physical or virtual manner, and virtual resources can be encapsulated in virtual machines or containers.
Third, the job-level scheduling determines job-execution priorities based on resource constraints and job performance estimation.
Fourth, the task-pipeline-level scheduling divides the job into subtasks and locates them onto available resources for the pipeline execution of the subtasks, aiming to increase task parallelism and overlap computational and communication workloads.
Fifth, the network-flow-level scheduling optimizes the coflows of numerous subtasks by considering the relation and dependency of network flows.
Sixth, the scheduled jobs, task pipelines, and network flows run efficiently on the allocated GPU and network resources within the cluster.

In this section, we first introduce large-scale GPU management aimed at achieving efficient computational and communication-resource utilization in distributed DL. 
We then introduce task scheduling to overlap computational and communication tasks of distributed DL, thereby increasing parallelism. 
During the discussion of these strategies, we focus on their applications in the large-scale setting, with heterogeneous resource capacities and task workloads.
Finally, we summarize some lessons learned from these strategies to assist in uncovering promising research directions within this scope.

\newcommand\gCate[1]{\parbox[t]{2.0cm}{\centering #1}}
\newcommand\gRef[1]{\parbox[t]{2.2cm}{#1}}
\newcommand\gYear[1]{\parbox[t]{0.6cm}{\centering #1}}
\newcommand\gContent[1]{\parbox[t]{10.8cm}{ #1 \vspace{1.5pt}}}
\begin{table*}[!t]
        \caption{Studies on resource-allocation strategies for large-scale distributed DL}\label{tab:GPUmanage}
        \centering
        \begin{tabular}{|c|c||l|c|l|}
                \Xhline{2\arrayrulewidth}
                \multicolumn{2}{|c||}{\cellcolor{blue!15}\textbf{Category}} & \cellcolor{blue!15}\textbf{Technology\&Ref.} & \cellcolor{blue!15}\textbf{Year} & \cellcolor{blue!15}\textbf{Highlight} \\ 
                \Xhline{2\arrayrulewidth}
                \multirow{30}{*}{\rotatebox[origin=c]{90}{\parbox[t]{3cm}{\centering Training (\ref{sec:GPU_train})}}} & \multirow{2}{*}{\gCate{GPU Sharing. Focus: (1)~production cluster; (2)~context switching; (3)~performance estimate; (4)~elasticity; (5)~hyperparameter tuning}} & 
                
                \gRef{Gandiva \cite{xiao2018gandiva}} & \gYear{2018} & \gContent{(1) Using the profiles of the DL workload to improve efficiency of training DL models and latency in a GPU cluster.}\\
                \cline{3-5} 
                ~ & ~ & \gRef{AntMan \cite{xiao2020antman}} & \gYear{2020} & \gContent{(1) Introducing co-designing the cluster scheduler and dynamic scaling mechanisms.} \\ 
                \cline{3-5}
                ~ & ~ & \gRef{FGD \cite{weng2023beware}} & \gYear{2023} & \gContent{(1) Monitoring the individual evaluation functions of DL jobs at runtime to make placement decisions and resource allocations elastically.} \\  
                \cline{3-5}
                ~ &  & \gRef{TGS \cite{wu2023transparent}} & \gYear{2023} & \gContent{(1) Designing adaptive rate-control and transparent unified-memory mechanisms .} \\
                \cline{3-5} 
                ~ & ~ & \gRef{Salus \cite{yu2019salus}} & \gYear{2019} & \gContent{(2) Achieving fine-grained GPU sharing among multiple DL applications.} \\ 
                \cline{3-5}
                ~ & ~ & \gRef{PipeSwitch \cite{bai2020pipeswitch}} & \gYear{2020} & \gContent{(2)Exploiting the profiles of DL applications to achieve millisecond-scale context switching.} \\ 
                \cline{3-5}
                ~ & ~ & \gRef{Optimus \cite{peng2018optimus}} & \gYear{2018} & \gContent{(3) Estimating a DL task’s remaining execution time and designing a marginal gain-based allocation algorithm.} \\
                \cline{3-5}
                ~ & ~ & \gRef{Harmony \cite{Harmony2019}} & \gYear{2019} & \gContent{(3) Placing training jobs in a manner that minimizes interference  and maximizes performance.} \\  
                \cline{3-5}
                ~ & ~ & \gRef{Horus \cite{yeung2021horus}} & \gYear{2021} & \gContent{(3) Proposing a data-driven approach to predict the GPU utilization of heterogeneous DL tasks.} \\  
                \cline{3-5}
                ~ & ~ & \gRef{Pollux \cite{qiao2021pollux}} & \gYear{2021} & \gContent{(4) Combining system throughput with statistical efficiency and introducing a formulation of goodput.} \\  
                \cline{3-5}
                ~ & ~ & \gRef{Zico \cite{2021Zico}} & \gYear{2021} & \gContent{(4) Monitoring the memory-usage pattern of individual DL jobs by tracking computational progress of training jobs.} \\  
                \cline{3-5}
                ~ & ~ & \gRef{AFS \cite{hwang2021elastic}} & \gYear{2021} & \gContent{(4) Handling future jobs requires proactive preparation based on current share calculations.} \\  
                \cline{3-5}
                ~ & ~ & \gRef{FlowCon \cite{FlowCon2023}} & \gYear{2023} & \gContent{(4) Minimizing the growth of GPU fragmentation through packing tasks.} \\  
                \cline{3-5}
                ~ & ~ & \gRef{Fluid \cite{MLSYS2021c0987e6b}} & \gYear{2021} & \gContent{(5) Utilizing a water-filling approach to accelerate the hyperparameter optimization process.} \\  
                \cline{3-5}
                ~ & ~ & \gRef{Titan \cite{gao2022titan}} & \gYear{2022} & \gContent{(5) Merging several fine-tuning workloads into one to improve resource utilization.} \\  
                \cline{3-5}
                ~ & ~ & \gRef{DISC \cite{liu2022adaptive}} & \gYear{2022} & \gContent{(5) Leveraging adaptive scaling to adjust the size of GPU time slices and formalizing the dynamic allocation of GPU time slices into an optimization problem.}\\
                \cline{3-5}
                ~ & ~ & \gRef{Hydro \cite{hu2023hydro}} & \gYear{2023} & \gContent{(5) Extending resources of hyperparameter tuning workloads by interleaving them with pipeline-enabled large-model training tasks.} \\  
                \cline{2-5}
                ~ & \multirow{2}{*}{\gCate{Network Bandwidth Sharing. Granularity: (1) job; (2) gradient block; (3) coflow}} &  
                \gRef{Liquid \cite{gu2021liquid}} & \gYear{2021} & \gContent{(1) Proposing intelligent cluster network-efficient scheduling methods in both immediate and batch modes.} \\   
                \cline{3-5}
                ~ & ~ & \gRef{Prophet \cite{zhang2021prophet}} & \gYear{2021} & \gContent{(2) Employing the monitored network bandwidth and the profiled gradient time interval to predict the number of gradients into gradient blocks.} \\  
                \cline{3-5}
                ~ & ~ & \gRef{Parrot \cite{li2020efficient}} & \gYear{2020} & \gContent{(3) Using a linear program (LP) solution to derive a weighted bandwidth scaling
                        strategy to minimize the time cost in the communication stage.} \\ 
                
                \Xhline{2\arrayrulewidth}
                \multirow{11}{*}{\rotatebox[origin=c]{90}{\parbox[t]{2.5cm}{\centering Inference (\ref{sec:GPU_inference})}}} & \multirow{2}{*}{\gCate{Spatial Sharing}} & \gRef{GSLICE \cite{dhakal2020gslice}} & \gYear{2020} & \gContent{Developing self-learning and adaptive GPU-resource allocation and batching schemes.} \\ \cline{3-5}
                ~ &  & \gRef{iGniter \cite{xu2022igniter}} & \gYear{2022} & \gContent{Leveraging inference performance model to calculate the appropriate batch size and lower bound of allocated GPU resources.} \\ \cline{3-5}
                ~ & ~ & \gRef{SLO-aware \cite{cho2022sla}} & \gYear{2022} & \gContent{Distributing inference requests to the deployed functions based on the autoscaling decision.}\\
                \cline{2-5}
                ~ & \multirow{2}{*}{\gCate{Temporal Sharing}} & \gRef{Nexus \cite{shen2019nexus}} & \gYear{2019} & \gContent{Applying a heuristic approach to select the requests.} \\ \cline{3-5}
                ~ & ~ & \gRef{INFaaS \cite{romero2021infaas}} & \gYear{2021} & \gContent{Identifying the colocation interference caused by the shared hardware resources.}\\
                \cline{3-5}
                ~ &  & \gRef{Cocktail \cite{gunasekaran2022cocktail}} & \gYear{2022} & \gContent{Building a distributed-weighted auto-scaling policy that utilizes the importance sampling technique.}\\
                \cline{2-5}
                ~ & \multirow{2}{*}{\gCate{Hybrid Sharing}} & \gRef{Gpulet \cite{choi2022serving}} & \gYear{2022} & \gContent{Allowing heterogeneous ML models to be mapped to multiple gpulets in the most cost-effective way.} \\ \cline{3-5}
                ~ &  & \gRef{FaST-GShare \cite{gu2023fast} } & \gYear{2023} & \gContent{Introducing the FaST-Manager to limit and isolate spatio-temporal resources.}\\
                \Xhline{2\arrayrulewidth}
        \end{tabular}
\end{table*}

\subsection{Resource Management}\label{sec:GPU_manage}

According to the different stages of DL, we categorize distinct resource-management techniques applicable to training and inference stages of DL. Table \ref{tab:GPUmanage} provides a summary of relevant computational and communication-resource-management technologies in the domain of large-scale distributed DL.

\subsubsection{Resource management for distributed training}\label{sec:GPU_train}

The training process of distributed DL requires an intensive consumption of computing power and memory of GPUs and network communication bandwidth across GPUs. Therefore, GPU and network bandwidth sharing is the focus of our discussion.

$\bullet$ \textit{GPU Sharing:}\label{sec:GPUSharing} Although GPUs have found extensive applications in distributed DL, a prevalent issue of underutilization is observed in production clusters. The recorded low GPU utilization typically ranges from 25\% to below 50\%~\cite{narayanan2020heterogeneity, hu2021characterization,weng2022mlaas,li2023lyra, weng2023beware, cheng2023towards}. This concern is particularly noteworthy in large-scale distributed computing environments. To address this issue, various distributed techniques have been developed to enable DL tasks to run efficiently on numerous devices~\cite{gao2022deep}. 

GPU-sharing techniques leverage partial resource allocation through virtualization, presenting a feasible strategy to mitigate the challenge of low GPU utilization in large-scale distributed DL. NVIDIA, acknowledged as the leading GPU provider, introduces Multiple Process Sharing (MPS)~\cite{NvidiaMPS} that offers an operating-system-level virtualization solution. Nevertheless, its implementation requires application-specific expertise to define resource limits for ensuring performance isolation. Moreover, MPS lacks compatibility with various DL frameworks. 
To address the performance isolation issue with MPS, another NVIDIA technology, Multi-Instance GPU (MIG)~\cite{NvidiaMIG}, enables the partitioning of a GPU into multiple discrete instances, each with dedicated resources. However, MIG is exclusively available to only a certain specifications of GPUs (e.g., A100 and H100) and primarily supports resource sharing at a coarse-grained level.

Some comprehensive solutions have been proposed to tackle complex characteristics of production clusters for large-scale distributed DL. Gandiva~\cite{xiao2018gandiva} leverages the profiles of distributed DL tasks and addresses the issue of GPU underutilization in three key ways. Initially, Gandiva allows incoming jobs to time-share GPUs with existing jobs when overloaded. Then, it permits time-sliced jobs to migrate to other GPUs. Lastly, it supports elastic GPU capacity, increasing the number of GPUs during idle times and reducing the number of GPUs as the load grows dynamically, thereby utilizing idle GPUs effectively. 
The performance of Gandiva is demonstrated on production clusters at Microsoft. 
AntMan~\cite{xiao2020antman} is a production solution for distributed DL clusters at Alibaba.
It analyzes the cause of GPU underutilization in distributed DL clusters for production use in three aspects: hardware, cluster scheduling, and job behavior. 
Exploiting the profiles of fluctuating resource demands from distributed training jobs, AntMan co-designs the cluster scheduler and distributed DL framework with dynamic scaling mechanisms for GPU resources during job execution. This approach ensures jobs' service-level objectives (SLOs) in large-scale clusters while enhancing cluster utilization through opportunistic scheduling.
Leveraging the analysis of the production trace at Alibaba, Fragmentation Gradient Descent (FGD)~\cite{weng2023beware} addresses severe GPU fragmentation in large clusters.
FGD minimizes GPU fragmentation growth through task packing to achieve maximum GPU allocation rates. 
TGS~\cite{wu2023transparent} provides transparent GPU sharing at OS layer for distributed DL tasks in production clusters of containers.
TGS addresses challenges of the lack of application profiling knowledge and the potential oversubscription of GPU memory during the sharing of GPU resources.
TGS tackles the first challenge by monitoring and controlling the rate of sending GPU kernels to the GPU for each container adaptively, aiming to maximize the rate of opportunistic jobs while not affecting that of production jobs.
TGS tackles the second challenge by unifying GPU memory and host memory in a single address space via CUDA unified-memory allocation~\cite{cudaUnifiedMem} that enables both performance isolation and transparency of GPU memory allocation. Oversubscribed memory of opportunistic jobs is evicted to the host memory automatically, ensuring the performance of production jobs. 

Some works use fast-context switching to improve GPU utilization. Salus~\cite{yu2019salus} achieves fine-grained GPU sharing with flexible scheduling policies, exposing two GPU-sharing primitives: fast job switching and memory sharing. The former enables rapid preemption and efficient time sharing for the currently active DL job on a GPU, whereas the latter packs smaller distributed DL tasks on the same device to ensure high memory utilization and prevent memory fragmentation. In contrast, PipeSwitch~\cite{bai2020pipeswitch} supports fast-context switching for task pipelines of distributed DL jobs. PipeSwitch optimizes the context switching overhead through model-aware grouping for pipelines and proactive allocating of GPU memory.
The model-aware grouping of layers aims to minimize the overhead of transferring the model between CPUs and GPUs during context switching. 
The proactive allocation of GPU memory for standby workers before it should be active expedites the speed of context switching. To prevent job interference, PipeSwitch enforces process-level isolation, by initialing a new separate process for each active-worker task. 

Some works employ performance-estimate-guided approaches to enhance GPU-resource allocation. Both the performance-estimation methods and the performance goals can vary in these approaches. Optimus~\cite{peng2018optimus} introduces a dynamic allocation algorithm based on marginal gains, estimating the remaining execution time of a distributed DL task. In this greedy policy, a job's larger marginal gain results in a higher allocation of resources. Harmony~\cite{Harmony2019} uses a deep reinforcement learning algorithm to place distributed DL jobs on allocated GPU resources with minimum interference and minimum training time. The learning rewards for unseen placements are guided by historical allocation samples. Horus~\cite{yeung2021horus} builds a model to predict GPU utilization of heterogeneous distributed DL tasks from the computation graph features. It identifies GPU utilization as a general proxy metric for making optimal placement decisions.

Elastic training, which involves extending and shrinking resource capacity dynamically, is an important strategy to improve resource utilization and save costs for distributed DL in the cloud environment. Some studies focus on elastic resources at the granularity of GPU memory. 
Pollux~\cite{qiao2021pollux} adjusts the GPU resources available to distributed DL jobs dynamically, aiming to maximize the overall training goodput within the cluster. To improve the efficiency of GPU memory sharing, Zico~\cite{2021Zico} monitors the memory-usage patterns of individual distributed DL jobs by tracking computational progress during training. Based on this monitoring, Zico allocates and deallocates memory among concurrent jobs automatically, ensuring no exceeding of the memory budget. 
AFS~\cite{hwang2021elastic} points out that handling future jobs requires proactive preparation of resources based on current share calculations. When the resource scheduler estimates that the resource contention will be heavy in the future, it allocates more resources to long-lasting jobs; otherwise it allocates more resources to short jobs.
In contrast, some studies focus on elastic-container resources. For instance, FlowCon~\cite{FlowCon2023} introduces a container-placement strategy based on growth efficiency and dynamic resource configuration for elastic allocation and withdrawal of resources during runtime. 

Several studies explore strategies for improving resource utilization during hyperparameter tuning in distributed DL clusters. Fluid~\cite{MLSYS2021c0987e6b} is a distributed DL hyperparameter tuning execution engine that abstracts the hyperparameter tuning process as a sequence of trial groups. It employs a water-filling approach to expedite the hyperparameter tuning process, thereby enhancing resource utilization. 
Titan~\cite{gao2022titan} adopts a different approach by consolidating multiple fine-tuning workloads into one, aiming to improve resource utilization. This consolidation is particularly advantageous given that multiple fine-tuning workloads often share the same model parameters. 
DISC~\cite{liu2022adaptive} leverages adaptive scaling to adjust the size of GPU time slices occupied by hyperparameter-tuning jobs at runtime. 
This dynamic allocation of GPU time slices for each hyperparameter tuning job is based on its potential to create a steep increase in the inference accuracy.
Hydro~\cite{hu2023hydro} addresses cluster-wide resource utilization and tuning efficiency by incorporating a heterogeneity-aware allocation strategy. This method extends the resources of hyperparameter-tuning workloads by interleaving them with pipeline-enabled large-model training tasks. By effectively utilizing idle time intervals on each node, caused by the gaps between the forward and backward processing of microbatches, Hydro enhances overall resource utilization and tuning efficiency in large-scale distributed DL clusters.

$\bullet$ \textit{Network bandwidth sharing:}\label{sec:rm_Network_Bandwidth} In large-scale distributed environments, network bandwidth is another significant factor determining the efficiency of distributed training. 
The focus of network bandwidth sharing can be in various granularities, including jobs and tasks, gradient blocks, and network coflows.

Some work focuses on optimizing network bandwidth sharing in the granularity of jobs and tasks.
Liquid~\cite{gu2021liquid} proposes a computational and communication-resource-estimation algorithm and a network-efficient job-placement strategy for distributed training jobs. 
The resource-estimation algorithm models resource requirements of distributed training jobs, including GPU computing power, GPU memory, and network-bandwidth requirements. 
The job-placement strategy assigns distributed training jobs to a cluster of computing nodes and containers, finding a best-fit job placement solution that satisfies the estimated computational and communication-resource requirements and exhibits less GPU fragmentation and network communication cost across containers.

Some work focuses on network bandwidth sharing in the granularity of gradient blocks.
For instance, Prophet~\cite{zhang2021prophet} groups into certain gradient blocks based on the profiled time interval and models the distributed training time in terms of the network bandwidth and order of network transfers of gradient blocks. 
Based on this model, Prophet searches for an optimal order of the network transfers of gradient blocks, aiming to minimize the distributed training time. 
This optimal order of gradient block transfers optimizes both the network bandwidth sharing among gradient blocks and the overlapping between network transfers and GPU computation.

Some work focuses on network bandwidth sharing in the granularity of coflows.
For instance, Parrot~\cite{li2020efficient} perceives the communication pattern of a distributed training job as a series of dependent coflows and estimates the remaining processing time of distributed training jobs based on the amount of information carried per coflow.
Parrot allocates network bandwidth to active coflows of concurrent jobs within the cluster, so that the effective completion time of coflows of the job with a shorter remaining processing time has a higher priority to be minimized.


\subsubsection{Resource management for distributed inference}\label{sec:GPU_inference}
In contrast to distributed training that caters to long-term offline workloads, distributed inference necessitates real-time execution with more stringent requirements on latency and accuracy. This difference in demands necessitates the development of novel resource management solutions to address the distinct characteristics of inference workloads effectively. In the distributed inference process, GPU sharing is the focus of various resource-allocation approaches that can be divided into two major categories: spatial sharing and temporal sharing. 
In the context of multiple distributed DL jobs, 
the spatial sharing of GPUs involves the sharing of GPU space partitions while the temporal sharing involves the sharing of computation time slices of an entire GPU.
Hybrid approaches combine techniques from these two categories.

$\bullet$ \textit{Spatial Sharing:}\label{sec:SGPUSharing} Many existing works exploit spatial sharing of GPUs to optimize the performance of distributed inference tasks. GSLICE~\cite{dhakal2020gslice} introduces an inference system that achieves safe and efficient GPU sharing through spatial GPU multiplexing systematically. It utilizes MPS~\cite{NvidiaMPS}, a GPU spatial-multiplexing framework with virtualization, to handle various inference requests. iGniter \cite{xu2022igniter} employs an inference performance model to calculate an appropriate batch size and the lower bound of allocated GPU resources. Subsequently, it allocates GPU resources for each inference workload by employing a greedy approach to identify the placement GPU devices that can achieve minimal performance interference. 
The SLO-aware ML Inference Framework~\cite{cho2022sla} designs a resource auto-scaling strategy in the cloud by leveraging rich and precise workload-specific metrics, with a special consideration of the heterogeneity in the GPU computational capability. This effective and elastic management of resources ensures meeting the SLO for diverse inference workloads in the cloud.

$\bullet$ \textit{Temporal Sharing:}\label{sec:TGPUSharing} 
Recent temporal-sharing approaches designed for specific distributed inference systems have shown improvements in GPU utilization, especially in cloud environments shared by numerous tenants. Nexus~\cite{shen2019nexus} employs a heuristic approach to select requests for co-location on the same GPU. Initially, it determines the most suitable batch size to meet throughput and SLO requirements for the existing inference workloads. Subsequently, Nexus identifies all possible combinations within a GPU's duty cycle on a single GPU in a best-fit manner, maximizing utilization without violating latency requirements. Focusing on inference services in the cloud, INFaaS~\cite{romero2021infaas} addresses the problem of co-location interference arising from shared hardware resources. It allocates available resources to interfered instances through workload migration or virtual-machine-level scaling, aiming to reduce monetary costs through GPU sharing while meeting latency requirements via virtual-machine-level scaling. Cocktail~\cite{gunasekaran2022cocktail} scales the virtual machine resources for various inference models in the cloud automatically and proactively based on the predicted workload and popularity of these models. This approach enhances the efficiency of resource allocation in distributed DL inference systems with a specific set of supported inference models.

$\bullet$ \textit{Hybrid Sharing:}\label{sec:HGPUSharing} Several works study the hybrid GPU-sharing approaches, considering both spatial and temporal sharing. Gpulet~\cite{choi2022serving} supports spatial sharing of GPUs via the abstraction of virtual GPUs that are split partitions derived from physical GPUs. 
Given allocated virtual GPU resources, Gpulet supports temporal sharing by scheduling the batch sizes of inference jobs of multiple tenants, with a goal to guarantee the SLO. This hybrid design enables cost-effective cloud-resource allocation for the inference of numerous heterogeneous DL models. FaST-GShare~\cite{gu2023fast} utilizes spatial and temporal sharing of GPUs to maximize inference function throughput in the Function-as-a-Service serverless architecture for distributed DL.
It supports auto-scaling of inference resources in the cloud based on the profiling of function throughput and resource allocation, maximizing GPU utilization while ensuring the SLO.

\newcommand\tCate[1]{\parbox[t]{1.3cm}{\centering #1}}
\newcommand\tRef[1]{\parbox[t]{2.0cm}{#1}}
\newcommand\tYear[1]{\parbox[t]{0.4cm}{\centering #1}}
\newcommand\tContent[1]{\parbox[t]{11.6cm}{ #1 \vspace{1.5pt}}}
\begin{table*}[!t]
    \caption{Studies on Task-Scheduling Strategies for Large-scale Distributed DL}\label{tab:scheduling}
    \centering
    \begin{tabular}{|c|c||l|c|l|}
    \Xhline{2\arrayrulewidth}
        \multicolumn{2}{|c||}{\cellcolor{blue!15}\textbf{Category}} & \cellcolor{blue!15}\textbf{Technology\&Ref.} & \cellcolor{blue!15}\textbf{Year} & \cellcolor{blue!15}\textbf{Highlight} \\ 
        \Xhline{2\arrayrulewidth}
        \multirow{50}{*}{\rotatebox[origin=c]{90}{\parbox[t]{5.0cm}{\centering Training Scheduling (\ref{sec:TraningSch})}}} & \multirow{2}{*}{\tCate{Efficiency: Job-level Scheduling (\ref{sec:Job})}} & \tRef{Amaral \textit{et al.}~\cite{Amaral_Polo_Carrera_Seelam_Steinder_2017}} & \tYear{2017} & \tContent{Proposing a new topology-aware workload-placement strategy to schedule DL jobs on multi-GPU systems.}\\ 
        \cline{3-5}
        ~ & & \tRef{Tiresias~\cite{Gu_Chowdhury_Shin_Zhu_Jeon_Qian_Liu_Guo_2019}} & \tYear{2019} & \tContent{Using LAS algorithm to prioritize jobs based on their service, a metric defined as the multiplication of requested GPU resources and execution time.} \\ 
        \cline{3-5}
        ~ & & \tRef{FfDL~\cite{Jayaram_Muthusamy_Dube_Ishakian_Wang_Herta_Boag_Arroyo_Tantawi_Verma_etal._2019}} & \tYear{2019} & \tContent{Using the operating lessons from the industry practice to guide the balance dependability with scalability, elasticity, flexibility, and efficiency.} \\ 
        \cline{3-5}
        ~ & & \tRef{Philly~\cite{jeon2019analysis}} & \tYear{2019} & \tContent{Correlating scheduler logs with logs from individual jobs and conducting a thorough analysis about the impact of gang scheduling and locality constraints on the queuing delay and job runtime.} \\ 
        \cline{3-5}
        ~ & & \tRef{E-LAS~\cite{Sultana_Chen_Xu_Yuan_2020}} & \tYear{2020} & \tContent{Using real-time epoch progress rates specific to distributed training jobs, as well as services obtained from the temporal and spatial domains, to guide scheduling decisions} \\ 
        \cline{3-5}
        ~ & & \tRef{SMD~\cite{Yu_Wu_Ji_Liu_2021}} & \tYear{2021} & \tContent{Allowing multiple jobs to compete for the communication bandwidth} \\ 
        \cline{3-5}
        ~ & & \tRef{OSDL~\cite{WANG2022109191}} & \tYear{2022} & \tContent{Designing job-placement and scheduling algorithms in networks that involve both OCS and EPS.} \\ 
        \cline{3-5}
        ~ & & \tRef{Sched$^2$~\cite{luan2019sched2}} & \tYear{2019} & \tContent{Using DRL to perform smart locality-aware scheduling of DLT jobs.} \\ 
        \cline{3-5}
        ~ & & \tRef{MLFS~\cite{Wang_Liu_Shen_2020}} & \tYear{2020} & \tContent{Leveraging the data from the heuristic scheduling method for training a DRL model and making decisions on job scheduling using this trained DRL model automatically.} \\ 
        \cline{3-5}
        \cline{2-5}
        ~ & \multirow{2}{*}{\tCate{Efficiency: Task-pipeline-level Scheduling (\ref{sec:Pipeline})}} & \tRef{PipeDream~\cite{narayanan2019pipedream}} & \tYear{2019} & \tContent{Partitioning DNN layers among workers automatically to balance workload and reduce communication.} \\
        \cline{3-5}
        ~ &  & \tRef{GPipe~\cite{huang2019gpipe}} & \tYear{2019} & \tContent{Employs an innovative batch-splitting pipelining algorithm, achieving nearly linear convergence speedups when a model is distributed across multiple accelerators.} \\ 
        \cline{3-5}
        ~ &  & \tRef{Piper~\cite{tarnawski2021piper}} & \tYear{2021} & \tContent{Leveraging tensor parallelization techniques within a single layer to reduce the search space significantly.} \\ 
        \cline{3-5}
        ~ &  & \tRef{Chimera~\cite{li2021chimera}} & \tYear{2021} & \tContent{Integrating bidirectional pipelines for the efficient training of large-scale models, striking an optimal balance among pipeline efficiency, memory cost, and convergence friendliness.} \\ 
        \cline{3-5}
        ~ &  & \tRef{AutoPipe~\cite{liu2022autopipe}} & \tYear{2022} & \tContent{Introducing a method for achieving balanced partitioning and reducing startup overhead automatically.} \\ 
        \cline{3-5}
        ~ &  & \tRef{OOO BackProp~\cite{oh2022out}} & \tYear{2022} & \tContent{Leveraging gradient computation dependencies to reorder pipeline executions, which maximize GPU resource utilization.} \\ 
        \cline{3-5}
        ~ &  & \tRef{DeAR~\cite{zhang2023dear}} & \tYear{2023} & \tContent{Addressing the issues of startup latency and sub-optimal training performance.} \\ 
        \cline{3-5}
        ~ &  & \tRef{HetPipe~\cite{park2020hetpipe}} & \tYear{2020} & \tContent{Presenting a DNN training system that integrates PMP with DP.} \\ 
        \cline{2-5}
        ~ & \multirow{2}{*}{\tCate{Efficiency: Network-flow-level Scheduling (\ref{sec:Network-Flow})}} & \tRef{JPAS~\cite{Zhou_He_Luo_Yu_Sun_2020}} & \tYear{2020} & \tContent{Using a simple greedy mechanism to order all DDL jobs periodically.} \\ 
        \cline{3-5}
        ~ & & \tRef{Geryon~\cite{wang2020geryon}} & \tYear{2020} & \tContent{Employing multiple flows with varying priorities to transfer parameters of different urgency levels.} \\
        \cline{3-5}
        ~ &  & \tRef{TensorExpress~\cite{kang2020tensorexpress}} & \tYear{2020} & \tContent{Enables each switch to transmit tensor packets according to their priority using multiple queues.} \\ 
        \cline{3-5}
        ~ &  & \tRef{Beamer~\cite{he2021beamer}} & \tYear{2021} & \tContent{Reducing the SCT by considering stage information in its scheduling approach.} \\ 
        \cline{3-5}
        ~ &  & \tRef{Tereis~\cite{chen2023tereis}} & \tYear{2023} & \tContent{Exploring the utilization of idle GPU computational resources during data transmission periods.} \\ 
        \cline{3-5}
        ~ &  & \tRef{Mercury~\cite{duan2023accelerating}} & \tYear{2023} & \tContent{Working on packet granularity to shift priority scheduling to the transport layer.} \\ 
        \cline{2-5}

        ~ & \multirow{2}{*}{\tCate{Cost-Effective (\ref{sec:TrainSchCost})}}& \tRef{Cynthia~\cite{Zheng_Xu_Chen_Zhou_Liu_2019}} & \tYear{2019} & \tContent{Providing predictable distributed training performance and reducing the training budget.} \\ 
        \cline{3-5}
        ~ & & \tRef{FC$^2$~\cite{Ta_2019}} & \tYear{2019} & \tContent{A scheduler that recommends cost-effective cloud-resource allocations for distributed training tasks with a PS.} \\ 
        \cline{3-5}
        ~ & & \tRef{Jahani~\cite{Jahani_Lattuada_Ciavotta_Ardagna_Amaldi_Zhang_2019}} & \tYear{2019} & \tContent{Modeling the scheduling process as a MILP problem to reduce the leasing cost in a global manner while maintaining the job latency.} \\ 
        \cline{3-5}
        ~ & & \tRef{GPOEO~\cite{Wang_Zhang_Lai_Hao_Wang_2022}} & \tYear{2022} & \tContent{Saving power in GPU data centers and using a customized scheduler to orchestrate jobs.} \\ 
        \cline{3-5}
        ~ & & \tRef{STS~\cite{10328678}} & \tYear{2023} & \tContent{Exploiting the probability distribution of early termination and adapting the resource assignment during the execution of the jobs to minimize the expected energy cost} \\ 
        \cline{2-5}
        ~ & \multirow{2}{*}{\tCate{Deadline Guarantee (\ref{sec:TrainSchDea})}} & \tRef{GENIE~\cite{8778770}} & \tYear{2020} & \tContent{Proposing a prediction model derived from lightweight profiling to estimate the processing rate and response latency for diverse DL workloads.} \\
        \cline{3-5}
        ~ &  & \tRef{Chronus~\cite{Gao_Ye_Sun_Wen_Zhang_2021}} & \tYear{2021} & \tContent{Providing deadline guarantee for SLO jobs and maximizing the performance of best-effort jobs.} \\ 
        \cline{3-5}
        ~ &  & \tRef{Hydra~\cite{10036352}} & \tYear{2023} & \tContent{Adopting a sampling approach that exploits the inherent iterative periodicity of DL jobs to estimate job completion times accurately on heterogeneous GPUs.} \\ 
        \cline{3-5}
        \Xhline{2\arrayrulewidth}
        \multirow{12}{*}{\rotatebox[origin=c]{90}{\parbox[t]{4cm}{\centering Inference Scheduling (\ref{sec:InferenceSch})}}} & \multirow{2}{*}{\tCate{Efficiency (\ref{sec:InferenceSchEff})}} & \tRef{Sniper~\cite{10.1145/3489517.3530474}} & \tYear{2022} & \tContent{Using non-invasive performance characterization network based on neural network similarity (NNS) to predict the inference time of DNNs accurately.} \\ \cline{3-5}
        ~ & ~ & \tRef{AutoDeep~\cite{Li_Han_Zhang_Li_Tan_2020}} & \tYear{2020} & \tContent{Leveraging Bayesian Optimization and Deep Reinforcement Learning to unearth the optimal cloud configuration and device placement with limited search time adaptively.}\\
        \cline{3-5}
        ~ & ~ & \tRef{Clipper~\cite{crankshaw2017clipper}} & \tYear{2017} & \tContent{Observing the corresponding accuracy and latency feedback to make a choice to use a best-effort search approach.}\\
        \cline{2-5}
        ~ & \multirow{2}{*}{\tCate{Throughout capability (\ref{sec:InferenceSchThr})}} & \tRef{Rafiki~\cite{Wang_Gao_Zhang_Wang_Chen_Ng_Ooi_Shao_Reyad_2018}} & \tYear{2018} & \tContent{Using a practical AIMD algorithm to adjust inference batch size.} \\ \cline{3-5}
        ~ & ~ & \tRef{Nanily~\cite{Tang_Wang_Liu_Wang_Han_2019}} & \tYear{2019} & \tContent{Deriving the corresponding batch size so that the inference execution time is equal to or close to the maximum remaining time.}\\
        \cline{3-5}
        ~ & ~ & \tRef{RRL~\cite{Qin_Zawad_Zhou_Yang_Zhao_Yan_2019}} & \tYear{2019} & \tContent{Focusing on optimizing parallel configurations at different levels.}\\
        \cline{3-5}
        ~ & ~ & \tRef{Morphling~\cite{Wang_Yang_Yu_Wang_Li_Sun_He_Zhang_2021}} & \tYear{2021} & \tContent{Adapting the meta-model to a new inference service by sampling a small number of configurations and using it to find the optimal one.}\\
        \cline{3-5}
        \Xhline{2\arrayrulewidth}
    \end{tabular}
\end{table*}

\subsection{Training-Task Scheduling}\label{sec:TraningSch}
In large-scale GPU clusters with complex network connections, scheduling distributed DL workloads effectively is critical for ensuring the high performance of task execution, optimal hardware utilization, and achievement of various scheduling objectives. Training and inference stages of distributed DL are widely recognized as particularly resource-intensive~\cite{gao2022deep}. 
The following subsections study task-scheduling strategies on the training and inference workloads, respectively, and focus on providing efficient communication or overlapping computation and communication for overall efficiency in large-scale distributed DL.

As the distributed training of a model can consume a lot of computational and communication resources, efficient task-scheduling strategies for training workloads are crucial. 
The performance goals of scheduling can include efficiency, cost-effective, and deadline guarantees
while the granularity levels can be on jobs, task pipelines, and network flows.
In this section,
we survey the task scheduling of large-scale distributed training with various performance goals and granularity levels. 
Table~\ref{tab:scheduling} summarizes the related literature.

\subsubsection{Efficiency}\label{sec:TrainSchEff}
Efficiency is one of the most critical performance goals of the scheduling of distributed training~\cite{gao2022deep}. This subsection categorizes task-scheduling strategies based on various scheduling granularity levels, including jobs, task pipelines, and network flows.

$\bullet$ \textit{Job-level scheduling}\label{sec:Job} is one of the most common and effective scheduling methods that focus on reordering priorities of jobs~\cite{Gu_Chowdhury_Shin_Zhu_Jeon_Qian_Liu_Guo_2019,Jayaram_Muthusamy_Dube_Ishakian_Wang_Herta_Boag_Arroyo_Tantawi_Verma_etal._2019}.
Amaral \textit{et al.}~\cite{Amaral_Polo_Carrera_Seelam_Steinder_2017} introduce a novel topology-aware workload-placement strategy, designed for scheduling distributed DL jobs on multi-GPU systems. This strategy satisfies workload requirements efficiently while minimizing interference simultaneously. Building upon this foundation, Tiresias~\cite{Gu_Chowdhury_Shin_Zhu_Jeon_Qian_Liu_Guo_2019}, drawing inspiration from the classic Multi-Level Feedback Queue (MLFQ) algorithm~\cite{chowdhury2015efficient}, develops a priority discretization approach to mitigate issues related to frequent preemption. 
A Least Attained Service (LAS) algorithm is conceptualized to prioritize jobs based on their service level that is quantified by the product of requested GPU resources and execution time. FfDL~\cite{Jayaram_Muthusamy_Dube_Ishakian_Wang_Herta_Boag_Arroyo_Tantawi_Verma_etal._2019}, an open-source scheduling platform developed by IBM, incorporates operational insights from industry practices to strike a balance between dependability and scalability, while maintaining elasticity, flexibility, and efficiency. 
In a related study, Philly~\cite{jeon2019analysis} performs a comprehensive analysis by correlating scheduler logs with logs from individual jobs, examining the impact of gang scheduling and locality constraints on queuing delay and job runtime. Drawing on insights from this analysis, Philly advocates relaxing locality constraints to enhance job-time efficiency.
E-LAS~\cite{Sultana_Chen_Xu_Yuan_2020}, with the objective of reducing the average completion time for distributed training jobs, shifts the focus away from reliance on job runtime estimates or prior knowledge. Instead, it utilizes real-time epoch progress rates specific to distributed  training jobs, combined with service metrics derived from temporal and spatial domains, to inform scheduling decisions. This innovative approach enables E-LAS to surpass the efficiency of Tiresias.
Additionally, SMD~\cite{Yu_Wu_Ji_Liu_2021} presents a resource-scheduling analytical model that accommodates multiple jobs vying for communication bandwidth. This model treats the scheduling problem as a non-convex integer non-linear program with bin-packing constraints and introduces an $\epsilon$-approximation algorithm, termed the sum-of-retios multi-dimensional-knapsack decomposition, for its resolution.
OSDL~\cite{WANG2022109191} designs and proposes algorithms for job placement and scheduling in networks that involve both optical circuit switching (OCS) and electrical packet switching (EPS). Simulation results demonstrate that OSDL surpasses multiple well-established scheduling methods in terms of performance.

Several scheduling methods have integrated machine-learning techniques, particularly deep reinforcement learning (DRL), to enhance task-scheduling efficiency.
Sched$^2$~\cite{luan2019sched2} utilizes DRL to schedule distributed training jobs with a locality-aware approach intelligently. This method is capable of understanding both the locality sensitivity of jobs and the fragmentation condition of clusters comprehensively within the entire learning stack. Through this heightened awareness, the DRL model adjusts its scheduling decisions dynamically and adaptively, responding effectively to the varying locality sensitivities of individual jobs and the evolving state of cluster fragmentation.
MLFS~\cite{Wang_Liu_Shen_2020} employs data from heuristic scheduling methods to train a DRL model, subsequently using this model to make informed decisions about job scheduling autonomously. 

$\bullet$ \textit{Task-pipeline-level scheduling}\label{sec:Pipeline} is specialized in managing the sequential processing of tasks within a pipeline architecture, especially for large distributed training jobs with large models. It orchestrates the flow of data and tasks across various stages or processes expertly in a pre-defined order. By ensuring each task is executed efficiently and in adherence to the specified sequence, the scheduler optimizes overall pipeline performance.
For example, PipeDream~\cite{narayanan2019pipedream} introduces a system that incorporates inter-batch pipelining into intra-batch parallelism. This approach enhances parallel training throughput by overlapping computation with communication and minimizing communication when feasible. Moreover, PipeDream partitions DNN layers among workers automatically to balance workload and reduce communication.
Addressing the need for efficient and task-independent model parallelism, GPipe~\cite{huang2019gpipe} emerges as a pipeline parallelism library. It facilitates scaling of any network describable as a sequence of layers. GPipe employs an innovative batch-splitting and pipelining algorithm, achieving nearly linear convergence speedups when a model is distributed across multiple accelerators. This method offers the flexibility to scale various DNN models to immense sizes efficiently.
Piper~\cite{tarnawski2021piper} aims to partition the DNN computation graph across numerous accelerators optimally while combining various parallelism modes and optimizations. As an efficient optimization algorithm based on two-level dynamic programming, Piper leverages tensor parallelization within a single layer to reduce the search space significantly.
Chimera~\cite{li2021chimera} integrates bidirectional pipelines for efficient training of large-scale models, striking an optimal balance among pipeline efficiency, memory cost, and convergence friendliness. AutoPipe~\cite{liu2022autopipe} introduces a method for achieving balanced partitioning and reducing startup overhead automatically. It includes a planner for rapid and automated generation of balanced pipeline partition schemes and a micro-batch slicer that adjusts micro-batches in line with planner outcomes to minimize pipeline startup overhead.
Out-Of-Order (OOO) BackProp~\cite{oh2022out} leverages gradient computation dependencies to reorder pipeline executions, maximizing GPU-resource utilization. In data-parallel training, OOO reorders the sequence of gradient computations to optimize the overlap between computation and parameter communication. In pipeline-parallel training, it prioritizes critical gradient computations to minimize pipeline stalls.
DeAR~\cite{zhang2023dear} proposes decoupling the all-reduce primitive into two continuous operations, enabling both backpropagation and feed-forward computations without extra communication. This method addresses the issues of startup latency and sub-optimal training performance.
For heterogeneous pipelines, HetPipe~\cite{park2020hetpipe} presents a DNN training system that integrates pipelined model parallelism (PMP) with data parallelism (DP). In this system, a wave-synchronous-parallel approach is proposed to support both PMP and DP for virtual workers.
    
$\bullet$ \textit{Network-flow-level scheduling}\label{sec:Network-Flow} manages network bandwidth effectively, reduces latency, and ensures efficient utilization of network resources for network flows related to distributed DL jobs. By adeptly balancing these critical elements, the scheduler enhances both performance and reliability of network operations significantly.
JPAS~\cite{Zhou_He_Luo_Yu_Sun_2020} implements a straightforward greedy mechanism to organize all distributed training jobs periodically. This approach enables each host machine to prioritize its network flows according to the established job order, delegating the task of flow scheduling and rate allocation to the underlying priority-enabled networks.
Geryon~\cite{wang2020geryon} employs multiple flows with varying priorities to transfer parameters of different urgency levels. This approach coordinates multiple parameter servers effectively and gives precedence to urgent parameter transfers across the entire network fabric.
To address in-network delays, such as queuing delays, TensorExpress~\cite{kang2020tensorexpress} enables each switch to transmit tensor packets according to their priorities using multiple queues. This method ensures that high-priority data packets are handled efficiently to minimize delays.
Beamer~\cite{he2021beamer} focuses on reducing the stage-completion time (SCT) by considering stage information in its scheduling approach. It proposes a stage-aware coflow-scheduling method to minimize the average SCT.
Tereis~\cite{chen2023tereis} explores the utilization of idle GPU computational resources during data transmission periods. It predicts the execution time for a distributed DL job and its corresponding data transmission time, allowing for the simultaneous packaging of two jobs on the same GPU. This ensures that one job is completed before the other concludes its data transfer.
Mercury~\cite{duan2023accelerating} shifts priority scheduling to the transport layer innovatively, focusing on packet granularity. In this system, packets in the Mercury buffer with the highest priority are transmitted first. Additionally, Mercury incorporates immediate aggregation at the transport layer, enabling full overlapping of gradient push-and-pull operations. This approach not only streamlines data flow but also maximizes the efficiency of network resource utilization.

\subsubsection{Cost-Effective}\label{sec:TrainSchCost}
The primary objective of cost-effective scheduling is to minimize operational costs while ensuring optimal performance for distributed training. It achieves a balance between resource utilization, energy consumption, and monetary expenditures in the scheduling decisions. During the training phase, cost efficiency emerges as a significant goal for the scheduling of training workloads.
Cynthia~\cite{Zheng_Xu_Chen_Zhou_Liu_2019} offers predictable distributed training performance while reducing the training budget. This scheduler identifies the optimal resource type and maintains training throughput effectively, thereby minimizing monetary costs.
FC$^2$~\cite{Ta_2019}, similar to Cynthia, is a scheduler that recommends cost-effective cloud resource allocations for parameter servers in distributed training tasks. It prioritizes instances with the largest network bandwidth within the budget to circumvent communication bottlenecks. Furthermore, it introduces a heuristic named Scale-Opt for determining worker instances, ensuring job throughput, and maximizing cost savings.
Jahani~\cite{Jahani_Lattuada_Ciavotta_Ardagna_Amaldi_Zhang_2019} considers computing nodes with varying numbers of GPUs as distinct virtual machines. The scheduling process is modeled as a mixed-integer linear programming (MILP) problem, aiming to reduce leasing costs globally while maintaining job latency.
GPOEO~\cite{Wang_Zhang_Lai_Hao_Wang_2022} achieves significant power savings for training workloads. It can be integrated into GPU data centers easily, utilizing a customized scheduler to manage job orchestration.
STS~\cite{10328678} optimizes the scheduling of distributed training jobs from the perspective of cloud service providers operating data centers. 
It leverages the probability distribution of early job termination to adapt resource assignments during job execution, with the aim of minimizing the expected energy cost.

\subsubsection{Deadline Guarantee}\label{sec:TrainSchDea}
Diverging from the aforementioned scheduling methods, deadline-guaranteed scheduling focuses on ensuring that jobs are completed before a specified deadline. This approach is critical for tasks for which timing is a key factor.
GENIE~\cite{8778770}, a trailblazing deadline-aware scheduler for distributed training workloads, explores the key factors that impact the performance of distributed DL tasks. It introduces a predictive model based on lightweight profiling, enabling the accurate estimation of processing rates and response latencies for a variety of distributed DL workloads. A significant limitation of GENIE, however, is its inability to handle mixed workloads that include both deadline-sensitive tasks and best-effort tasks simultaneously~\cite{gao2022deep}.
Chronus~\cite{Gao_Ye_Sun_Wen_Zhang_2021}, an end-to-end scheduling system, meets SLOs by guaranteeing deadlines for SLO-aware jobs while also enhancing the performance of best-effort jobs. This dual-focused strategy enables Chronus to manage a wide range of workload requirements.
Furthering these developments, Hydra~\cite{10036352} emerges as a dynamic and multifaceted scheduler, equipped to tackle various scheduling challenges, including adhering to deadlines and reducing job completion times. Hydra introduces an innovative sampling approach that capitalizes on the iterative periodicity inherent in distributed DL jobs. This technique allows for the precise estimation of job completion times in heterogeneous GPU environments, thereby elevating the efficiency and effectiveness of scheduling for various distributed DL workloads.

\subsection{Inference Scheduling}\label{sec:InferenceSch}
As distributed-DL-based applications permeate our daily lives increasingly in the form of online services, the management and scheduling of large-scale inference workloads on GPUs have become critical. These inference jobs, distinct from the resource-intensive and long-duration training workloads mentioned earlier, possess unique characteristics and requirements. Consequently, they necessitate novel scheduling solutions tailored to their specific needs~\cite{gao2022deep, tang2023survey}.
In alignment with the approach of the preceding section, this section categorizes various scheduling strategies with a particular focus on two primary goals: efficiency and throughput. This categorization facilitates a comprehensive understanding of the different methodologies employed in inference scheduling, highlighting their significance in enhancing the overall performance of distributed inference workloads.

\subsubsection{Efficiency}\label{sec:InferenceSchEff}
Unlike the training phase, where the focus is often on maximizing accuracy and model robustness, the inference phase primarily emphasizes the reduction of latency and cost.

To maintain satisfactory latency, inference schedulers are designed to scale resources proactively in response to request density and to reorder execution sequences strategically at the job level. 
Sniper~\cite{10.1145/3489517.3530474} stands out as a self-updating cloud-edge collaborative inference scheduling system with a focus on time awareness. It employs a non-invasive performance characterization network based on neural network similarity to predict the inference time of DNNs accurately. This system achieves a stable increase in throughput successfully, demonstrating its effectiveness in optimizing the scheduling process in dynamic cloud-edge environments.

In practice, cost efficiency is another critical factor in the inference phase, prompting some schedulers to incorporate various mechanisms aimed at achieving cost-effective inference.
AutoDeep~\cite{Li_Han_Zhang_Li_Tan_2020} endeavors to automate cloud deployment for real-time online DNN inference, focusing on minimizing costs while maintaining acceptably low latency. To achieve this, AutoDeep utilizes Bayesian optimization combined with DRL. This innovative approach enables the adaptive discovery of the optimal cloud configuration and device placement, reducing the required search time significantly. Through this method, AutoDeep balances the trade-off between operational costs and latency in DNN inference workloads efficiently.

Meanwhile, latency and cost are recognized as interdependent factors in system design. Improving one aspect may inadvertently compromise the other if the solution is not designed meticulously. This interplay motivates researchers to develop scheduler systems that strike a balance between these objectives.
Clipper~\cite{crankshaw2017clipper} introduces an innovative model-selection abstraction, accommodating both single-model selection and model-integration selection. This system conducts inferences across all models and integrates their results. Clipper monitors accuracy and latency feedback continuously, employing a best-effort search approach to optimize model selection based on these parameters.
 
\subsubsection{Throughout capability}\label{sec:InferenceSchThr}
Throughput capability represents another crucial objective for scheduling inference workloads. Generally, the scheduling system is fine-tuned to enhance throughput through strategic batch execution and configuration adjustments.

Batching inference has been identified as an efficient method to enhance utilization and reduce system overhead~\cite{gao2022deep}. In recent times, various schedulers have incorporated heuristic methods to fine-tune the batch size for optimal performance.
For instance, Rafiki~\cite{Wang_Gao_Zhang_Wang_Chen_Ng_Ooi_Shao_Reyad_2018} employs a practical Additive-Increase Multiplicative-Decrease (AIMD) algorithm to adjust the inference batch size dynamically. This approach allows for responsive adaptation to varying workload conditions.
Nanily~\cite{Tang_Wang_Liu_Wang_Han_2019} establishes an upper limit on the batch size by calculating the maximum remaining time for a request. This is determined by subtracting the minimum queuing time of available resources from the remaining time. It then computes an appropriate batch size such that the inference execution time is equal to or approximates this maximum remaining time.

In addition to the aforementioned approaches, certain schedulers employ end-to-end configuration tuning to enhance system throughput.
RRL~\cite{Qin_Zawad_Zhou_Yang_Zhao_Yan_2019} emphasizes the optimization of parallel configurations at various levels, including both request-level parallelism and intra-request level parallelism. These optimizations play a significant role in reducing the overall system latency and improving throughput.
Morphling~\cite{Wang_Yang_Yu_Wang_Li_Sun_He_Zhang_2021}, on the other hand, presents a rapid and near-optimal auto-configuration framework designed specifically for cloud-native model serving. This framework adapts to new inference services by sampling a limited set of configurations and then employs a meta-model to identify the most optimal configuration. This strategy allows Morphling to adjust quickly and efficiently to varying service requirements while maintaining high system performance.

\subsection{Lessons Learned toward Large-scale Resource Allocation and Task Scheduling}
In this section, we discuss some lessons learned from designing resource-allocation and task-scheduling strategies for large-scale distributed DL. These lessons learned help to reveal promising directions for future studies.

$\bullet$ \textit{Fine-grained and elastic GPU memory and bandwidth sharing strategies are critical to improve GPU utilization.} 
Conventional GPU-allocation methods often assign individual distributed DL jobs exclusive access to a GPU and can lead to extremely low utilization. Exploring the resource-allocation optimization for diverse distributed training and inference workloads running on heterogeneous GPU networks in large-scale clusters comprehensively is necessary.
GPU-sharing strategies should consider various important factors, including performance isolation, elastic allocation, orchestration of computational and communication-resource allocation, and resource fragmentation.

$\bullet$ \textit{High-performance large-scale distributed DL requires the orchestration of efficient allocation of GPU and network resources.} 
The allocation of network resources can frequently be overlooked as a bottleneck for efficient resource utilization in distributed DL.
Many resource-allocation strategies of distributed DL focus on addressing computation issues, such as low utilization, load imbalance, and long queuing delays. However, with the increase of the cluster scale, the complexity of GPU network connections increases exponentially, and lack of consideration to efficient network-resource allocation can result in significant low job-execution performance of large-scale distributed DL. 
Efficient network-bandwidth-allocation strategies can alleviate communication contention.
Fully utilizing resources of both GPU and network bandwidth leads to enhanced overall performance of distributed training and inference at a large scale.


$\bullet$ \textit{These framework-level strategies can orchestrate with optimizations for distributed DL algorithms to meet common and distinct requirements of distributed training and inference.} 
Allocation and scheduling efficiencies are important for both distributed training and inference.
Examples of distinct performance objectives of distributed training include promoting fairness and increasing training throughput. In contrast, distributed inference requires achieving low latency and ensuring the SLO.
Efficient distributed DL is the result of the orchestration of efficient resource-allocation and task-scheduling strategies.
To tackle complex challenges posed by large data sets, large cluster scale, and large models,
distributed DL also requires the orchestration of these strategies and contemporary distributed DL algorithms, for developing more efficient, effective, and comprehensive distributed DL solutions for various large-scale scenarios. 



\newcommand\iCate[1]{\parbox[t]{1.5cm}{\centering #1}}
\newcommand\iRef[1]{\parbox[t]{2.4cm}{#1}}
\newcommand\iYear[1]{\parbox[t]{0.6cm}{\centering #1}}
\newcommand\iContent[1]{\parbox[t]{10.5cm}{ #1 \vspace{1.5pt}}}
\begin{table*}[!t]
    \caption{Studies on communication-efficient infrastructures for large-scale distributed DL}\label{tab:infrastructure}
    \centering
    \begin{tabular}{|c|c||l|c|l|}
    \Xhline{2\arrayrulewidth}
        \multicolumn{2}{|c||}{\cellcolor{blue!15}\textbf{Category}} & \cellcolor{blue!15}\textbf{Technology\&Ref.} & \cellcolor{blue!15}\textbf{Year} & \cellcolor{blue!15}\textbf{Highlight} \\ 
        \Xhline{2\arrayrulewidth}
        \multirow{4}{*}{\rotatebox[origin=c]{90}{\parbox[t]{1.5cm}{\centering GPU interconnects (\ref{sec:infraInterconnect})}}} & \multirow{2}{*}{\iCate{Intra-node (\ref{sec:infraInterGPU})}} & \iRef{PCIe~\cite{allreducePcie21}} & \iYear{-} & \iContent{Wired point-to-point link connection for high-speed serial communication between GPUs.}\\ 
        \cline{3-5}
        ~ & & \iRef{NVLink~\cite{nvlink}} & \iYear{-} & \iContent{Direct GPU-to-GPU interconnects based on wired mesh networking.} \\ 
        \cline{2-5}
        ~ & \multirow{2}{*}{\iCate{Inter-node (\ref{sec:infraInterNode})}} & \iRef{GPUDirect-RDMA~\cite{gpuDirectRDMA}} & \iYear{-} & \iContent{Direct GPU memory access between nodes within a cluster via PCIe.} \\
        \cline{3-5}
        ~ &  & \iRef{NVSwitch~\cite{nvlink}} & \iYear{-} & \iContent{All-to-all GPU communication across multiple nodes.} \\ 
        \Xhline{2\arrayrulewidth}
        \multirow{17}{*}{\rotatebox[origin=c]{90}{\parbox[t]{2.5cm}{\centering Programmable network devices (\ref{sec:infraDevice})}}} & \multirow{2}{*}{\iCate{Switch (\ref{sec:infraSwitch})}} & \iRef{SwitchML~\cite{switchMLInaNsdi21}} & \iYear{2021} & \iContent{Resource: reusing a fixed-size memory pool for INA within a rack.} \\ \cline{3-5}
        ~ &  & \iRef{ATP~\cite{atpInaNsdi21}} & \iYear{2021} & \iContent{Resource: allocating memory dynamically upon their arrival for multiple training jobs.} \\ \cline{3-5}
        ~ & ~ & \iRef{INAlloc~\cite{switchMemoryINA23}} & \iYear{2023} & \iContent{Resource: an additional switch memory manager to manage switch memory allocation for multiple jobs.}\\
        \cline{3-5}
        ~ & ~ & \iRef{GRID~\cite{gridInaRouting23}} & \iYear{2023} & \iContent{Routing: randomized-rounding-based gradient routing with INA to maximize the gradient sending rate with resource constraints.}\\
        \cline{3-5}
        ~ & ~ & \iRef{GOAT~\cite{goatInaRouting23}} & \iYear{2023} & \iContent{Routing: dividing the model into submodels and finding the optimal submodel gradient routing to minimize communication overhead.}\\
        \cline{3-5}
        ~ & ~ & \iRef{PANAMA~\cite{paranaIna21}} & \iYear{2021} & \iContent{Congestion: multiple aggregation tree to disperse gradient traffic and ensure fairness among multiple jobs.}\\
        \cline{3-5}
        ~ & ~ & \iRef{A$^2$TP~\cite{a2tpIna23}} & \iYear{2023} & \iContent{Congestion: two congestion windows for independent congestion control of switch resources and link bandwidth.}\\
        \cline{3-5}
        ~ & ~ & \iRef{NetReduce~\cite{inaWithTransport23}} & \iYear{2023} & \iContent{Congestion: transport-transparent gradient INA, reusing transport layer's congestion control.}\\
        \cline{2-5}
        ~ & \multirow{2}{*}{\iCate{SmartNIC (\ref{sec:infraNIC})}} & \iRef{Guo \textit{et al.}~\cite{DlrmSmartNIC23}} & \iYear{2023} & \iContent{Multi-level in-SmartNIC caching and computing system for recommendation models.} \\ \cline{3-5}
        ~ &  & \iRef{FCsN~\cite{FPGAInference22}} & \iYear{2022} & \iContent{Offloading control logic, scheduling, and computation of distributed inference jobs to SmartNICs.}\\
        \Xhline{2\arrayrulewidth}
        \multirow{22}{*}{\rotatebox[origin=c]{90}{\parbox[t]{2cm}{\centering Collective communication (\ref{sec:infraCC})}}} & \multirow{2}{*}{\iCate{Interface (\ref{sec:infraCCL})}} & \iRef{\cite{openmpi,fbGloo,horovod}} & \iYear{-} & \iContent{Point-to-point and collective communication interfaces for gradient computation and communication via GPU interconnects.}\\
        \cline{3-5}
        ~ & & \iRef{NCCL~\cite{nccl}} & \iYear{-} & \iContent{A high-bandwidth and low-latency collective communication library tailored for NVLink.} \\ \cline{3-5}
        ~ & ~ & \iRef{MSCCLang~\cite{mscclang23}} & \iYear{2023} & \iContent{A unified framework for writing customized collective communication algorithms.} \\ 
        \cline{2-5}
        ~ & \multirow{2}{*}{\iCate{Heterogeneous networks (\ref{sec:infraCCHete})}}& \iRef{BlueConnect~\cite{blueconnectCCL19}} & \iYear{2019} & \iContent{Decomposing \texttt{allreduce} into multiple reduce-scatter and all-gather primitives in symmetric topologies.}\\
        \cline{3-5}
        ~ &  & \iRef{Blink~\cite{blinkCCL20}} & \iYear{2020} & \iContent{Leveraging packing spanning trees to decomposing \texttt{allreduce} for optimal bandwidth utilization in asymmetric topologies.}\\
        \cline{3-5}
        ~ &  & \iRef{Plink~\cite{plinkLocalityCCL20}} & \iYear{2020} & \iContent{Probing the network to capture the physical network topology and bandwidth information to optimize collective communication.}\\
        \cline{2-5}
        ~ & \multirow{2}{*}{\iCate{Sparse data (\ref{sec:infraCCSpar})}}& \iRef{SparcML~\cite{sparcML19}} & \iYear{2019} & \iContent{Supporting sparse data streaming and sparse-to-dense switching for the minimal collective communication cost.} \\
        \cline{3-5}
        ~ &  & \iRef{OmniReduce~\cite{sparseCollectiveComm21}} & \iYear{2021} & \iContent{Splitting sparse data into blocks to increase INA parallelism for non-zero blocks.} \\ \cline{3-5}
        ~ & ~ & \iRef{Ok-Topk~\cite{OkTopkSparse22}} & \iYear{2022} & \iContent{Balancing sparse collective communication overhead across workers depending on consecutive buffer size.}\\
        \cline{3-5}
        ~ & ~ & \iRef{CommLib~\cite{sparseCommCloud21}} & 2021 & \iContent{A hierarchical aggregation and caching scheme for sparse collective communication to fully utilize GPU bandwidth and reduce I/O.}\\
        \cline{3-5}
        ~ & ~ & \iRef{DeepReduce~\cite{deepReduceSparseComm21}} & \iYear{2021} & \iContent{Decoupling indices and values of sparse data to apply different compression algorithms.} \\ 
        \cline{2-5}
        ~ & \multirow{2}{*}{\iCate{Synthesis (\ref{sec:infraSynthesisCC})}} & \iRef{SCCL~\cite{synthesizeCCL21}} & \iYear{2021} & \iContent{Synthesizing latency- and bandwidth-optimal collective implementations given a network topology with bandwidth constraints.}\\
        \cline{3-5}
        ~ &  & \iRef{TACCL~\cite{synthesisTACCL23}} & \iYear{2023} & \iContent{Synthesizing collective algorithms for heterogeneous multi-rack networks efficiently by dividing the problem into two independent sub-problems: routing and scheduling.}\\
        \Xhline{2\arrayrulewidth}
        \multirow{6}{*}{\rotatebox[origin=c]{90}{\parbox[t]{1.3cm}{\centering Topology (\ref{sec:infraTopo})}}} & \multirow{2}{*}{\iCate{Fixed (\ref{sec:infraFixedTopo})}} & \iRef{BML~\cite{bmlTopo20}} & \iYear{2020} & \iContent{ A BCube topology for fully distributed data-parallel training.} \\ \cline{3-5}
        ~ &  & \iRef{HammingMesh~\cite{hammingmeshTopo22}} & \iYear{2022} & \iContent{Local connectivity by affordable PCB-mesh interconnects and global connectivity by sparsely connected switches.}\\
        \cline{2-5}
        ~ & \multirow{2}{*}{\iCate{Configurable (\ref{sec:infraConfigurableTopo})}} & \iRef{SiP-ML~\cite{sipmlOpticalInterconnect21}} & \iYear{2021} & \iContent{Commercially affordable fully-connected and ring-based topologies powered by SiP interfaces.}\\
        \cline{3-5}
        ~ &  & \iRef{TopoOpt\cite{topoParallelismOpt23}} & \iYear{2023} & \iContent{Offline permuting \texttt{ring-allreduce} topologies for the optimal topology of multi-NIC nodes.}\\
        \Xhline{2\arrayrulewidth}
    \end{tabular}
\end{table*}

\begin{figure}[!t]
\centering
\includegraphics[width=0.8\columnwidth]{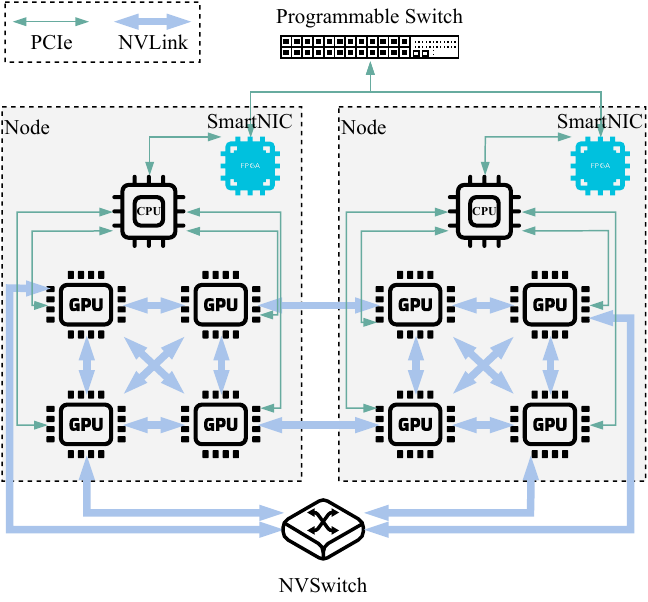}
\caption{A Torus topology for large-scale distributed DL with GPU interconnects, as well as programmable network interface cards and switches}
\label{fig:infrastructure}
\end{figure}

\section{Large-Scale Communication Infrastructures}\label{sec:infrastructure}
In this section, we introduce high-performance communication infrastructures for large-scale distributed DL, including GPU interconnects, programmable network devices, collective communication protocols, and communication topologies. 
Table~\ref{tab:infrastructure} summarizes the related technologies and studies on these topics.
We conclude this section with some lessons learned in this domain.
Fig.~\ref{fig:infrastructure} depicts an example Torus~\cite{torusTopo05} topology with high-bandwidth intra- and inter-node GPU interconnects, as well as programmable network devices. 
Implementing efficient collective communication built upon the high-performance communication connection and topology enhances distributed DL at a large scale.

\subsection{GPU Interconnects}\label{sec:infraInterconnect}
The communication between GPUs within a computing node and across nodes is significantly accelerated by high-speed GPU interconnects~\cite{pcieNvlinkEval19,interconnectionNetwork20}. 

\subsubsection{Intra-node} \label{sec:infraInterGPU}
In a scale-up solution, Peripheral Component Interconnect Express (PCIe)~\cite{allreducePcie21} and NVLink~\cite{nvlink} are prominent data link protocols for inter-GPU communication within a node. 
PCIe serves as a high-speed serial communication interface between GPUs via a wired point-to-point link connection. 
It supports up to 16 lanes per GPU, with the maximum bandwidth of 242 GBps for each GPU in PCIe version 7.0. 
NVLink is a direct GPU-to-GPU interconnect based on wired mesh networking.
It achieves significantly higher throughput compared to PCIe, supporting up to 18 links per GPU and providing a maximum bandwidth of 900 GBps for each GPU in NVLink version 4.0.
This elevated bandwidth enhances the speed and efficiency of inter-GPU communication, making NVLink particularly advantageous in scenarios demanding extensive inter-GPU data exchange.

\subsubsection{Inter-node} \label{sec:infraInterNode}
In a scale-out solution, the scalability of high-speed inter-node GPU interconnects can be achieved through the technologies of GPUDirect-RDMA~\cite{gpuDirectRDMA} and NVSwitch~\cite{nvlink}. 
GPUDirect-RDMA facilitates direct GPU memory access between nodes within a cluster via PCIe, thereby eliminating unnecessary memory copies through the CPU. 
NVSwitch extends the capability of NVLink by supporting all-to-all GPU communication across multiple nodes. 
NVSwitch can empower an ultra-large GPU server cluster with a capacity of up to 256 GPUs, delivering a staggering 57.6 TBps all-to-all bandwidth and supporting very large DL models with trillions of parameters. 

\subsection{Programmable Network Devices} \label{sec:infraDevice}
Programmable network devices, including programmable switches and network interface cards (NICs), facilitate software-hardware co-design to improve the communication efficiency of distributed DL algorithms. 

\subsubsection{Programmable switches} \label{sec:infraSwitch}
Recently, interest has been growing in leveraging programmable switches to execute in-network aggregation (INA)~\cite{inaSurvey23} for \texttt{allreduce} operations on gradients, aiming to mitigate network traffic during distributed training. 
Key considerations within this domain include on-switch resource utilization~\cite{switchMLInaNsdi21,atpInaNsdi21,switchMemoryINA23}, gradient routing~\cite{gridInaRouting23,goatInaRouting23}, and congestion control~\cite{paranaIna21,a2tpIna23,inaWithTransport23}.

$\bullet$ \textit{On-switch resource utilization:} 
The implementation of INA for gradients in the switch is significantly constrained by limited on-switch memory.
SwitchML~\cite{switchMLInaNsdi21} employs a fixed-size memory pool to aggregate gradients during \texttt{allreduce} communication within a rack switch. 
The reuse mechanism of the memory pool enables SwitchML to accommodate streaming gradients in a switch with limited computational and storage capabilities. 
However, SwitchML falls short in addressing challenges presented by practical scenarios featuring more complex topologies and workloads, such as a multi-rack cluster shared by multiple DL tenants.
This hinders its scalability across multiple racks accommodating multiple training jobs.
ATP~\cite{atpInaNsdi21} extends upon the work of SwitchML to enhance INA for multiple training jobs in a multi-rack setting by sharing limited switch resources across concurrently running training jobs efficiently. 
ATP employs a decentralized memory-allocation mechanism to allocate memory space on each switch to gradient packets from multiple training jobs dynamically upon their arrival at the switch. 
In contrast to SwitchML that conducts entire gradient aggregation in the rack switch, potentially encountering network bandwidth underutilization during heavy resource contention, 
ATP employs a best-effort approach to send gradients for aggregation. 
To further enhance the efficient utilization of switch resources for INA in a cluster, INAlloc~\cite{switchMemoryINA23} introduces a additional switch memory management layer.
This layer enables the distributed training job scheduler to optimize the management of physical memory on switches, thus improving job completion time. 

$\bullet$ \textit{Gradient routing:} Optimized gradient routing algorithms yield benefits for  INA in multi-rack environments, including improved load balancing, mitigation of bandwidth bottlenecks, and the optimal utilization of switch resources. 
However, gradient routing paths in SwitchML and ATP are fixed and not optimized. 
To tackle this issue, GRID~\cite{gridInaRouting23} presents a randomized-rounding-based algorithm for gradient routing with INA, aiming to maximize the gradient sending rate of workers with resource constraints. 
GOAT~\cite{goatInaRouting23} introduces a comparable approach with a focus on minimizing the network communication overhead among workers, switches, and the PS. 
This approach first partitions the DL model into submodels and then utilizes a knapsack-based randomized rounding algorithm to determine the switch responsible for aggregating specific submodel gradients. 

$\bullet$ \textit{Congestion control:} Because of the absence of an end-to-end byte stream abstraction in the transport layer for INA of distributed training, traditional congestion-control strategies at the transport layer, such as DCTCP~\cite{dctcp20} and pFabric~\cite{pfabric13}, are not readily applicable.
Many existing INA methods opt to implement their own straightforward congestion-control strategies.
For example, SwitchML implements basic retransmission timeout and self-clock mechanisms, 
while ATP utilizes the Explicit Congestion Notification (ECN) flag and Additive-Increase Multiplicative-Decrease (AIMD) mechanism, akin to TCP. 
However, these simplified congestion control implementations often fall short in optimizing performance for multiple jobs within large-scale shared cluster scenarios, or tend to couple the congestion control of switch resources and link bandwidth resources.  
To tackle the former issue, PANAMA~\cite{paranaIna21} employs multiple aggregation trees to disperse traffic, thus mitigating congestion hotspots and ensuring fair sharing of network resources across multiple jobs. 
To tackle the latter issue, A$^2$TP~\cite{a2tpIna23} employs two congestion windows for independent congestion control of switch resources and link bandwidth resources. 
Decoupling INA with the transport layer provides another direction in tackling the congestion control issue in this context.  
NetReduce~\cite{inaWithTransport23} preserves end-to-end abstracted connections by keeping packet data volume unchanged within the switch. 
As a result, the INA for gradients becomes transport-transparent, allowing the reuse of advanced congestion-control strategies of the transport layer, such as RoCEv2~\cite{RoCEv2_20}.

\subsubsection{Programmable NICs} \label{sec:infraNIC}
Commercial Field-Programmable-Gate-Array-based (FPGA-based) SmartNICs~\cite{nvidiaSmartNIC,intelSmartNIC} can server as computational resources for offloading domain-specific tasks from GPUs, as well as mitigating network communication overhead in distributed training and inference scenarios. 
Guo \textit{et al.}~\cite{DlrmSmartNIC23} have developed a heterogeneous SmartNIC system to address communication, memory, and computation bottlenecks inherent in the distributed training of Deep Learning Recommendation Models (DLRMs). 
To tackle the challenge of all-to-all communication in exchanging massive embedding tables within DLRMs, this system introduces a remote caching mechanism that buffers frequently used remote embedding lookup results on the SmartNIC, thereby reducing communication overhead during the feedforward process. 
This system further integrates local cache and in-SmartNIC computation mechanisms to access local embedding tables and execute irregular computations within the SmartNIC, obviating GPU intervention. 
These mechanisms alleviate memory bandwidth burdens and kernel launching overheads of GPUs effectively, optimizing both feedforward and backpropagation processing during the distributed training of DLRMs. 
On the other hand, FCsN~\cite{FPGAInference22} enhances distributed inference by offloading control logic, system scheduling, network communication, and neural network kernel computation to SmartNICs, eliminating CPU intervention.
The strategic offloading maximizes the overlap of computation and network communication during distributed inference, facilitating non-conflict streaming neural network kernel executions at line rate.

\subsection{Inter-GPU Collective Communication}\label{sec:infraCC}
The efficiency of collective communication among GPUs plays a pivotal role in determining the overall performance of distributed DL.
We introduce this topic in four dimensions: the collective communication library, optimization for heterogeneous network, optimization for sparse gradients, and the synthesis approach. 

\subsubsection{Collective communication libraries} \label{sec:infraCCL}
Collective communication libraries, such as Message Passing Interface (MPI)~\cite{openmpi}, Gloo~\cite{fbGloo}, and Horovod~\cite{horovod}, provide convenient programming interfaces for point-to-point and collective communication to exchange gradients on top of GPU interconnects. 
Specifically tailored to NVIDIA GPUs,
NVIDIA Collective Communications Library (NCCL)~\cite{nccl} is designed for compatibility with MPI
and optimized for delivering high bandwidth and low latency for multi-GPU and multi-node interconnects over PCIe and NVLink. 
In addition to point-to-point communication, NCCL supports various collective communication primitives, including but not limited to \texttt{allgather}, \texttt{allreduce}, and \texttt{broadcast}. 
Microsoft MSCCLang~\cite{mscclang23} provides a unified library for writing customized collective communication algorithms that can be compiled into NCCL built-in primitives. 
It enables high-performance customized collective implementations for various network environments. 
However, these collective communication libraries lack native-optimization for heterogeneous network environments and sparse gradient scenarios.

\subsubsection{Collective communication in heterogeneous networks} \label{sec:infraCCHete} 
The performance of collective communication can be improved by fully utilizing network bandwidth in heterogeneous networks.
BlueConnect~\cite{blueconnectCCL19} decomposes an \texttt{allreduce} operation into a series of reduce-scatter and all-gather primitives by leveraging the hierarchical topology information of communication bandwidth in a cluster. 
As a result, it minimizes the communication overhead for distributed training in heterogeneous networks and outperforms NCCL. 
However, BlueConnect is limited to symmetric topologies. 
Blink~\cite{blinkCCL20} further considers asymmetric topologies and heterogeneous links through the employment of packing spanning trees,
optimizing the utilization of high-link bandwidth in the presence of topology heterogeneity. 
In the cloud-based network environment with an opaque network topology, Plink~\cite{plinkLocalityCCL20} probes the network as to capture physical network topology and bandwidth/latency constraints that are then utilized to optimize collective communication in data center networks.

\subsubsection{Collective communication for sparse gradients}\label{sec:infraCCSpar} 
Gradient sparsity is a prevalent and substantial characteristic in numerous distributed DL models, and treating sparse gradients as dense data in collective communication can result in a waste of communication bandwidth. 
Unfortunately, most existing collective communication libraries are primarily designed for dense data communication and lack native supports for sparse data, including considerations for sparse data representation and aggregation in communication streams.
Addressing this limitation, SparcML~\cite{sparcML19} supports streaming sparse data in index-value pair representations natively and designs an efficient method for sparse-sparse and sparse-dense summation.
Sparse representations can switch to dense ones to ensure minimal communication cost during collective communication.
To increase the parallelism for sparse data aggregation, OmniReduce~\cite{sparseCollectiveComm21} introduces a streaming-aggregation mechanism that splits sparse data into blocks and achieves high INA parallelism for blocks containing non-zero values, thereby reducing communication overhead and maximizing communication bandwidth utilization effectively. 
Imbalanced workloads for key-value pair collective operations within the cluster lead to straggler performance for sparse gradients. 
To handle imbalance \texttt{allreduce} operations for top-$k$ sparsified gradients, Ok-Topk~\cite{OkTopkSparse22} generates a workload balancing scheme across workers based on the information about consecutive buffer sizes on all workers. 
This workload-balancing approach makes collective communication for sparse gradients more scalable, whereas the additional communication overhead is narrowly bounded.

The heterogeneity of cloud and FL environments poses new challenges to collective communication of sparse data. 
For cloud environments in which interconnects within a node are fast and those across nodes are slow in the GPU cluster, CommLib~\cite{sparseCommCloud21} adopts a hierarchical communication architecture for aggregating top-$k$ sparsified gradients to better utilize the GPU bandwidth within and across nodes.
It also introduces a multi-level data caching scheme to reduce I/O on public clouds.
For FL environments in which devices have constrained computational and communication resources and are separated geographically, the data volume of collective communication operations is a critical performance factor. 
To compress sparse gradients in FL environments, DeepReduce~\cite{deepReduceSparseComm21} decouples indices and values of sparse gradients into two sets, so each set can apply compressors that are optimized for its type independently. 
Specifically, DeepReduce introduces a Bloom-filter based compressor for indices and a curve-fitting based compressor for values, and significantly reduces the gradient volume compared to the traditional key-value pair representation.

\subsubsection{Synthesizing optimal collective communication} \label{sec:infraSynthesisCC}
Recently, a rising trend of synthesis approaches aims at crafting collective communication algorithms for distributed DL. 
Given a network topology with bandwidth constraints on GPUs and edges, SCCL~\cite{synthesizeCCL21} synthesizes Pareto-efficient latency- and bandwidth-optimal collective implementations from scratch.
However, SCCL lacks consideration for multi-rack network environments, impeding its scalability within a large-scale GPU cluster. 
To address this scalability issue, TACCL~\cite{synthesisTACCL23} enhances the synthesizer for multi-rack networks with heterogeneous links. 
On one hand, it uses a communication sketch containing information about the logical network topology and switch-hyperedge policy as the synthesizer input. 
On the other hand, TACCL adopts a heuristic approach, breaking down the problem into two independent steps: routing and scheduling. 
This division significantly reduces the search space, offering a pragmatic synthesis solution for collective communication in large-scale GPU clusters. 

\subsection{Communication Topologies}\label{sec:infraTopo}
Enhancing communication topologies is crucial for leveraging full computational and communication capabilities of a large-scale cluster in distributed DL. 
Conventional high-performance computing topologies for data centers can be used for reasonable-scale distributed DL~\cite{topologiesSurvey23} that include fix topologies such as Fat-tree~\cite{fatTreeTopo08}, BCube~\cite{bcubeTopo09}, Jellyfish~\cite{jellyfishTopo12}, and DCell~\cite{dcellTopo14}, and reconfigurable topologies such as Helios~\cite{heliosTopo10}, c-Through~\cite{cThroughTopo10}, and OSA~\cite{osaTopo13}.
However, these topologies are primary designed for electrical and optical networks and encounter challenges related to bandwidth and cost when dealing with all-to-all communication patterns inherent in distributed DL scenarios at ultra-scale, involving hundreds or even thousands of GPUs~\cite{topoImpactDML19,weng2022mlaas}.  
Recently, a number of communication topologies that address these challenges considering communication patterns of distributed DL and high bandwidth among GPU interconnects have been proposed. 

\subsubsection{Fixed topologies} \label{sec:infraFixedTopo}
To leverage the advantage of multiple NICs per computing node and provide a scalable and fault-tolerant network over Ethernet and commodity devices, BML~\cite{bmlTopo20} introduces a fully distributed gradient-synchronization algorithm on top of the BCube topology.
However, it is optimized exclusively for data-parallel training. 

To address the constraints of BML and pursue a communication topology that is both high-bandwidth while cost-effective for large-scale distributed training involving data and model parallelisms, HammingMesh~\cite{hammingmeshTopo22} integrates concepts from Fat-tree and Torus~\cite{torusTopo05}. 
This results in a two-dimensional topology with local and global connectivity. 
The local connectivity ensures high local bandwidth at low cost through the utilization of affordable Printed-Circuit-Board-mesh (PCB-mesh) interconnects, 
while the global connectivity can establish a global network by using a small number of sparsely connected switches to connect the meshes in rows and columns. 

\subsubsection{Reconfigurable topologies} \label{sec:infraConfigurableTopo}
The iterative nature of distributed training requires high bandwidth and low reconfiguration latency in reconfigurable topologies. 
To address these challenges, SiP-ML~\cite{sipmlOpticalInterconnect21} proposes two topologies driven by Silicon Photonic (SiP) interfaces: SiP-OCS and SiP-Ring. 
SiP-OCS adopts a fully connected topology to deliver high bandwidth using commercially available optical circuit switches, connecting GPUs to all switches through Tbps SiP interfaces. 
SiP-Ring employs a switch-less ring topology, minimizing reconfiguration latency through the use of Micro-ring resonators embedded in SiP interfaces. 
On the other hand, TopoOpt\cite{topoParallelismOpt23} adopts an offline method to find the best communication topology by exploring from \texttt{ring-allreduce} permutations across a set of computing nodes, each equipped with multiple NICs, and reconfigures switches to realize this target topology.
TopoOpt aims to co-optimize the network-topology and parallelism strategy for efficient large \texttt{allreduce} operations in both data and model parallelism modes.
By employing alternating optimization, TopoOpt narrows down the search space for co-optimization targets strategically, ensuring a network topology that provides ample bandwidth and a parallelism strategy that requires only a small hop count during model-parallelism transfers.

\subsection{Lessons Learned toward High-Performance Large-scale Communication Infrastructures}
We discuss some lessons learned from existing technologies introduced in this section, helping researchers working toward high-performance communication infrastructures for large-scale distributed DL. 

$\bullet$ \textit{High performance communication is monetarily costly, and economically affordable devices and topologies with best cost benefit are more practical.} 
The ultimate pursuit of the highest performance, regardless of cost, is only applicable in some specialized cases.
In most cases, we aim to achieve the best training throughput within a monetary budget for computational and communication resources. 
The optimal solution requires a co-design that involves adopting cost-effective communication devices and topologies for the high communication throughput and balancing computational and communication capacities for the overall training throughput. 

$\bullet$ \textit{INA is efficient, and co-designing communication infrastructures and algorithms can reduce communication overhead and increase scheduling efficiency.} 
Many modern communication devices also possess non-negligible computational capacity.
With a thorough understanding of the characteristics of model synchronization in large-scale distributed DL, designing a solution incorporating in-network computation to reduce communication traffic and implementing efficient transport scheduling at the package, flow, or coflow level to enhance communication throughput is feasible.
Challenges of such a solution include congestion-control strategies for INA, workload balancing among devices, fairness among multiple distributed DL jobs and tenants, and interference isolation of distributed DL traffic and other network traffic.

$\bullet$ \textit{Heterogeneous networks and sparse data are prevalent, and collective communication protocols should be fine-tuned for these scenarios.} 
Collective communication implementations should take into account the heterogeneity in the bandwidth of network links, the computational capacity of network devices, and the topology to  reduce communication overhead and enhance workload balancing, particularly in large-scale clusters.
Given the widespread occurrence of data sparsity in large DL models, implementing efficient collective communication for sparse data atop heterogeneous networks is imperative.
Implementing such efficient collective communication algorithms poses challenges, including discovering the true status of the underlying networks, routing and scheduling gradient aggregations in complex networks efficiently, and devising compatible libraries tailored for specialized high-performance interconnects.

\section{Large-Scale Distributed Training of Large Models: A Case Study on LLMs}\label{sec:caseStudy}
Recently, LLMs~\cite{gpt20,lamda22,llama23,llama2_2023,paLM23} have been used successfully in various applications of NLP. 
They have also initiated the trend of developing ultra-large foundation models for domain-specific applications~\cite{codeGen23,Bloomberggpt23} in diverse fields, including communications, computer science, and artificial intelligence.
As ultra-large foundation models can contain tens of billions of parameters and take hundreds of GPUs and tens of days to train, the distributed training of these models at scale encounters more and tougher challenges than conventional distributed DL. 
By inspecting the cases of training LLMs to examine how to apply those communication-efficient technologies in practical scenarios, researchers can identify more intrinsic and literal research challenges for achieving high performance in large-scale distributed DL.
 
This section discusses practical aspects of the distributed training of LLMs by examining the real cases in~\cite{llmGpuCluster21,decentralizedLlm22,lowCostdLlm22,llmMegatron22,cocktailSgdLlm23,swarmParallel23,OobleckResilientLlm23,llmEdgeAI24}.
The goal is to revisit key themes introduced earlier in this article and provide insights derived from related practices. 
Additionally, future trends in the use of large foundation models within the domain of communications are discussed.
The presentation adopts a question-and-answer style.

\subsection{Model Synchronization}
$\bullet$ \textit{Can data-parallel training be communication-efficient for LLMs?}
Data-parallel training alone is not a good choice for LLMs with slow GPU interconnects, but is acceptable with high-performance interconnects. 
According to~\cite{swarmParallel23}, concerning the volume of data to be transmitted across GPUs in data-parallel model synchronization, the time spent on communication can dominate the training time and leaves little chance for overlapping with the computation time.
For example, training a 100-billion-parameter model on 100 GPUs in the data-parallel mode, the volume of transmitted data can reach 10 trillion parameters in one model synchronization round.
Using 16-bit precision with slow interconnects of 1000 Mbps bandwidth, the time for one synchronization round is 1,600 seconds, which is completely unacceptable. 
But for very fast interconnects such as NVLink with 900 GBps point-to-point bandwidth, the synchronization time reduces to only 0.22 seconds, making it negligible if the synchronization frequency is reduced by local SGD.

\subsection{Communication Data Compression}
$\bullet$ \textit{What is the effect of communication data compression on LLMs?}
Though applying simple implementations of quantization and sparsification algorithms can reduce the communication overhead and training time of LLM, the effect of adaptive and fine-grained compression methods remains insufficiently explored. 
According to SWARM~\cite{swarmParallel23},
in the previous example with 100 billion parameters, the 16-bit precision can be reduced to 8-bit, resulting in the communication volume and time of model synchronization being only half of the origin values.
On the contrary, adaptive and fine-grained compression algorithms can introduce significant additional decision and computation overhead to LLMs. 
Considering this complexity, caution should be exercised when we applying these algorithms to the distributed training of LLMs.
The trade-off between computation and communication becomes more critical and requires further exploration in the large model scenario.

$\bullet$ \textit{Can hybrid communication-data-compression algorithms perform effectively on LLMs?}
We can utilize hybrid communication-data-compression algorithms to reduce communication overhead on slow networks aggressively.
For instance, CocktailSGD~\cite{cocktailSgdLlm23} employs random sparsification, top-$k$ sparsification, and quantization on the gradients sequentially when training an LLM with 20 billion parameters on a 500 Mbps network.
This hybrid method, achieving 117$\times$ compression, is only slightly slower than training on high-performance data center networks.

\subsection{Resource Allocation and Task Scheduling}
$\bullet$ \textit{How crucial is pipeline parallelism for LLM training?}
Pipeline parallelism is essential for LLM training.
As indicated in~\cite{llmGpuCluster21}, by diminishing the impact of the communication volume and worker idle time during pipeline flushes, heuristic pipeline-parallelism proves effective in practice with trillion-scale LLMs on more than 3,000 GPU.
In contrast to layer-slicing parallelism, multiple-layer-slicing pipeline parallelism only communicates end-of-layer activations and gradients, which can be 300$times$ smaller in the communication volume in a 2.2-billion-parameter example~\cite{lowCostdLlm22}.
It is a common practice to use what is known as 3D parallelism~\cite{llmMegatron22} for LLM training, which combines data, pipeline, and layer-slicing parallelisms, to maximize the training throughput.

$\bullet$ \textit{How to schedule tasks of LLM training on heterogeneous and low-bandwidth networks efficiently?}
LLMs can be trained on heterogeneous and low-bandwidth networks efficiently even with unstable worker devices, by employing fault-tolerant pipelines and redistributing tasks among workers.
Tackling the issue of unstable links between workers, SWARM parallelism~\cite{swarmParallel23} employs decentralized model-parallelism to randomize fault-tolerant pipelines and rebalance workers between pipeline stages dynamically.
The use of randomized fault-tolerant pipelines permits a worker in one pipeline stage to establish a connection to a stochastic worker in the next pipeline stage between different iterations.
This approach enables the rerouting of tasks from a disconnected worker to other workers within the same swarm in the next iteration.
This allows for the tolerance of worker failures in the pipeline, ensuring continuity in task execution.
The dynamic-rebalancing strategy enables workers to switch between pipeline stages for the purpose of balancing the throughput of different stages in instances of dynamic membership with unstable workers.

$\bullet$ \textit{How crucial is fault-tolerant scheduling for LLM training?}
Given the involvement of a large number of workers in prolonged training sessions for LLMs, ensuring fault-tolerance is of utmost importance for efficient scheduling. 
Frequent failure in devices or networks can potentially block the training process, degrade the convergence performance, and necessitate redundant restarting of failed tasks and pipelines. 
SWARM parallelism~\cite{swarmParallel23} incorporates the dynamic membership of unstable workers into account for efficient pipeline scheduling.
According to Oobleck~\cite{OobleckResilientLlm23}, a failed pipeline can be recovered by using pipeline replicas and templates swiftly.
We can instantiate some logically equivalent pipeline replicas, which possess replicated model states. 
Additionally, we define pipeline templates, which include information about the number of workers and stages in the pipeline, as well as the mapping of stages to GPUs.
Once a pipeline failure occurs, a new pipeline can be restored based on the pipeline template and replicas instantly.

$\bullet$ \textit{Can we utilize GPU resources in decentralized networks for LLM training?}
LLMs can undergo training on GPUs in decentralized, heterogeneous, and low-bandwidth networks using optimized data and model partitioning scheme, along with pipeline scheduling. 
According to~\cite{decentralizedLlm22}, there are numerous idle GPU clusters dispersed globally that can be utilized collectively for LLM training. 
We can divide the entire model into submodels, utilize a cost model to formulate the communication cost of both data and pipeline parallelisms based on the partitioning of these submodels, and partition submodels with a goal of minimizing this cost. 
For instance, when training GPT-3~\cite{gpt20} with 1.3 billion parameters, this scheduling strategy can achieve about 4$\times$ speedup in a global cloud across 8 worldwide regions, and is only 1.7-3.5$\times$ slower compared to the centralized data center solution with 100$\times$ faster networks.

\subsection{Communication Infrastructure}
$\bullet$ \textit{How can we construct cost-effective communication infrastructures for LLM training?}
Commodity communication infrastructures can be cost-effective for LLM training when we employ optimized technologies at various levels for distributed DL.
High-performance specialized communication infrastructures, such as NVIDIA A100 with NVLink interconnects, are frequently used for high-performance LLM training. 
However, considering the monetary cost, many researchers also use commodity networks, including data center networks, decentralized cloud infrastructures, and other low-bandwidth networks.
When data, model, and pipeline parallelism modes are combined, and optimized algorithms for communication data compression, resource allocation, and task scheduling are applied, commodity communication infrastructures can also achieve comparable performance to some specialized ones regarding the cost.

\subsection{Large foundation Models for Communications}
$\bullet$ \textit{What is the future trend of large foundation models within the domain of communications?}
Large foundation models can be trained with abundant domain-specific knowledge of communications to generate optimized strategies for diverse topics, including IoT, wireless networks, and network cybersecurity.
We refer to such models as large communications models here.
In the context of industrial IoT, large communications models can be exploited effectively in mobile edge computing (MEC) clusters to generate optimized codes for communication and scheduling strategies autonomously~\cite{llmEdgeAI24}. 
In the context of wireless networks, possessing capability to handle diverse multi-modal data, large communications models can conduct intricate data analyses for wireless sensor networks (WSN) based on unstructured mobile network logs, radio signals, and multimedia data, which are challenging for alternative models~\cite{dlWirelessSurvey19}.
Leveraging rich domain knowledge, large communications models can also be utilized to improve solutions for various optimization tasks of communications, such as routing, network-resource allocation, and radio control. 
In the context of network cybersecurity, particularly within the topics of the Internet, WSN, and IoT, large communications models can be used to detect anomalies, intrusions, and Distributed Denial of Service (DDoS) attacks~\cite{dlSecuritySurvey21}.  
They can also detect botnet attacks and generate verification strategies for anti-crawler protection. 
Overall, large communications models have the promising potential to extend their applications in various communications topics, enhancing the performance of existing technologies and innovating new solutions to address contemporary challenges.

\section{Conclusion}\label{sec:conclusion}
This paper surveys communication-efficient technologies at the algorithm, framework, and infrastructure levels for high-performance distributed DL in large-scale scenarios involving a large number of devices, large data, and large models, where heterogeneity and scalability are major research concerns.
We cover various topics, including distributed model synchronization, communication data compression, resource allocation, task scheduling, as well as communication interconnects, devices, protocols, and topologies.
During the discussion, We highlight the pros and cons of applying these technologies in a large-scale setting.
We present promising future research trends for each topic.
Moreover, a case study on the distributed training of large language models is showcased to illustrate the practical use of these communication-efficient technologies in real-world scenarios.

 



\bibliographystyle{IEEEtran}
\bibliography{IEEEabrv,surveyBib,surveySyncBib,surveyOptBib,surveyGPUBib,surveySchedulingBib,surveyLlmBib}




 

\begin{IEEEbiography}[{\includegraphics[width=1in,height=1.25in,clip,keepaspectratio]{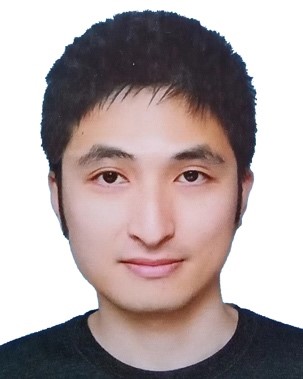}}]{Feng Liang}
(Member, IEEE) received his Ph.D. degree in computer science from the University of Hong Kong. He was a senior research engineer at Huawei and CTO of Jiufeng Tech Co., Ltd and is currently an associate professor with Shenzhen MSU-BIT University. His research interests are in broad areas of distributed computing, including distributed databases and computing systems, distributed machine learning, and interdisciplinary studies in bioinformatics. He has published papers in prestigious venues including IEEE TPDS, IEEE TNSM, Information Sciences, HPDC, IPDPS, and ACSAC.
\end{IEEEbiography}

\begin{IEEEbiography}[{\includegraphics[width=1in,height=1.25in,clip,keepaspectratio]{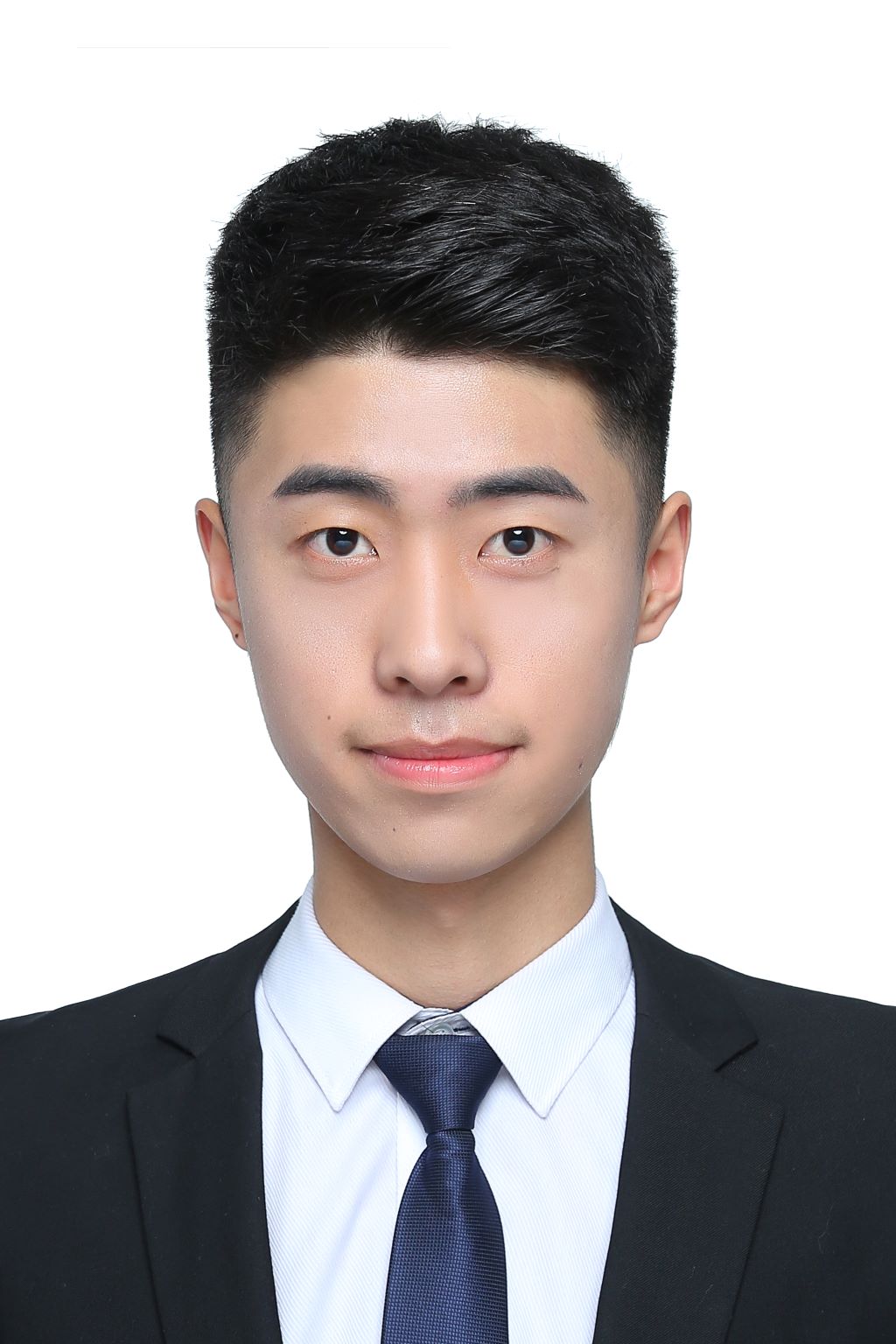}}]{Zhen Zhang}
is currently working toward the Ph.D. degree with the Gansu Provincial Key Laboratory of Wearable Computing, School of Information Science and Engineering, Lanzhou University, Lanzhou, China. He is also a guest student with the Shenzhen MSU-BIT University, Shenzhen, China. His research interests includes affective computing, person re-identification, and deep learning.
\end{IEEEbiography}

\begin{IEEEbiography}[{\includegraphics[width=1in,height=1.25in,clip,keepaspectratio]{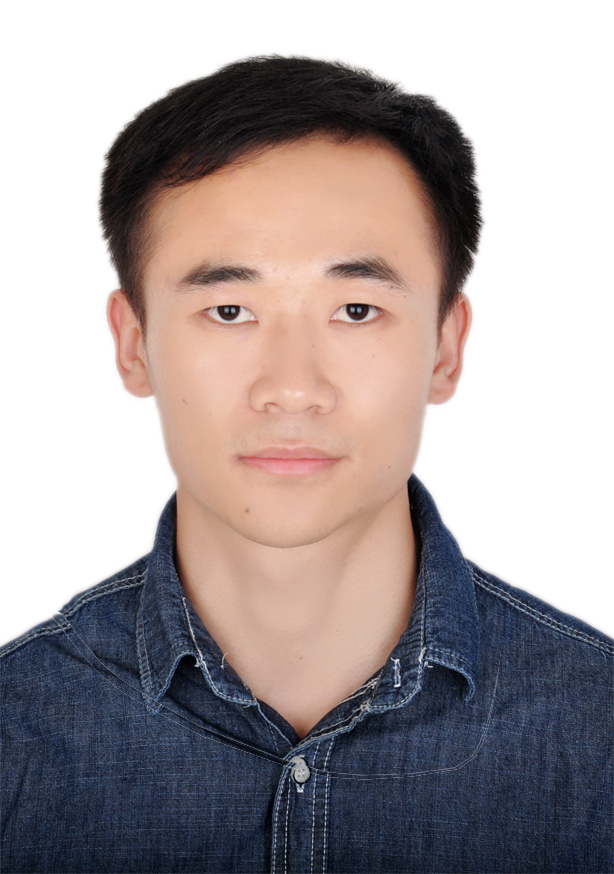}}]{Haifeng Lu}
received the the M.Sc. degree in School of Information Science and Engineering in 2021 from Lanzhou University, China. He is currently pursuing the Ph.D. degree with with the Gansu Provincial Key Laboratory of Wearable Computing, School of information Science and Engineering, Lanzhou University, Lanzhou, China. He is also a guest student with the Shenzhen MSU-BIT University, Shenzhen, China. His research interests include affective computing and machine learning.
\end{IEEEbiography}

\begin{IEEEbiography}[{\includegraphics[width=1in,height=1.25in,clip,keepaspectratio]{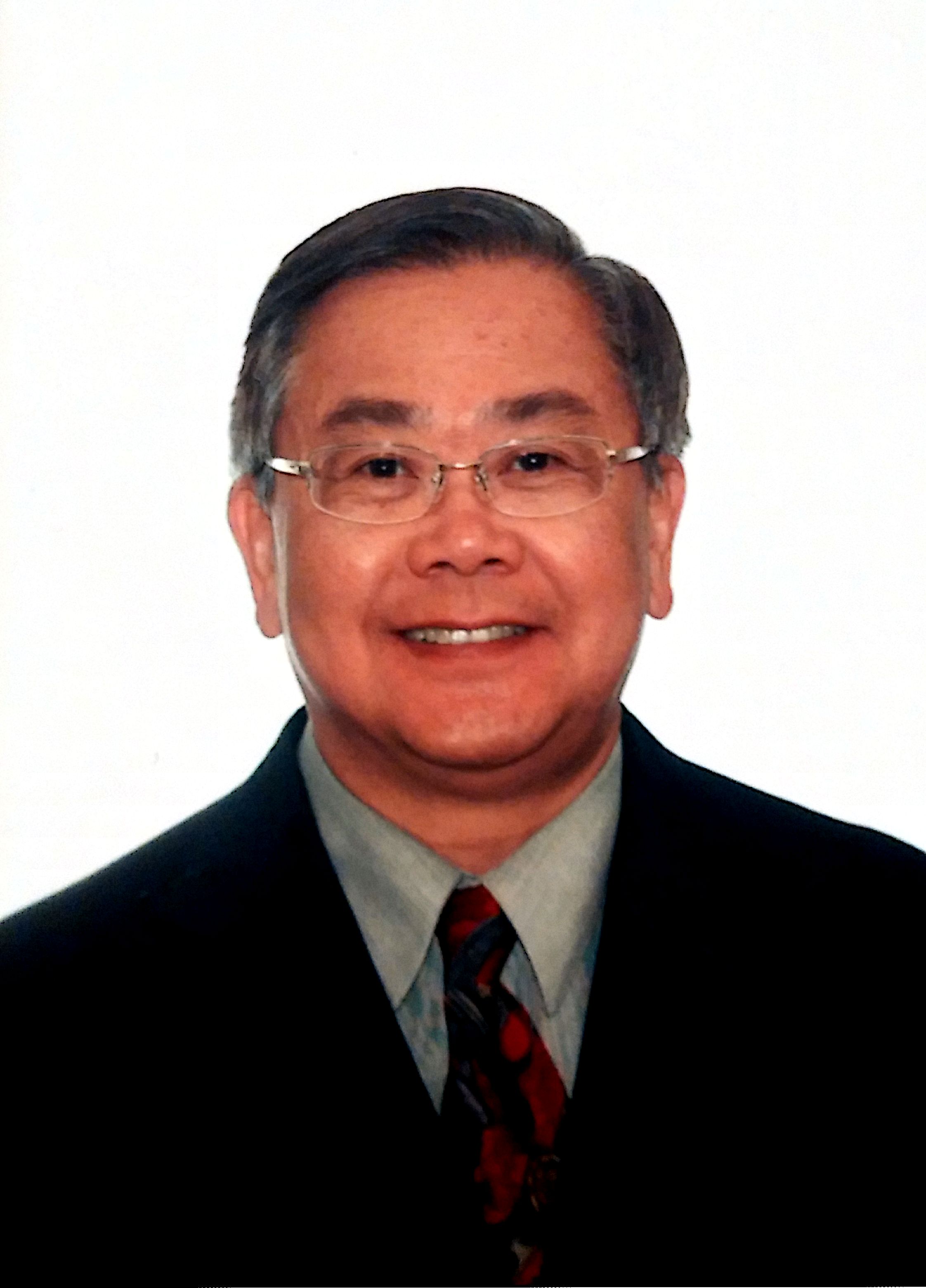}}]{Victor C. M. Leung}
(Life Fellow, IEEE) is currently a Distinguished Professor of computer science and software engineering with Shenzhen University, China. He is also an Emeritus Professor of electrical and computer engineering and the Director of the Laboratory for Wireless Networks and Mobile Systems, The University of British Columbia (UBC), Canada. His research interests include wireless networks and mobile systems. He has published widely in these areas. He is a Fellow of the Royal Society of Canada, the Canadian Academy of Engineering, and the Engineering Institute of Canada. He received the 1977 APEBC Gold Medal, the 1977–1981 NSERC Postgraduate Scholarships, the IEEE Vancouver Section Centennial Award, the 2011 UBC Killam Research Prize, the 2017 Canadian Award for Telecommunications Research, the 2018 IEEE TCGCC Distinguished Technical Achievement Recognition Award, and the 2018 ACM MSWiM Reginald Fessenden Award. He has coauthored papers that won the 2017 IEEE ComSoc Fred W. Ellersick Prize, the 2017 IEEE Systems Journal Best Paper Award, the 2018 IEEE CSIM Best Journal Paper Award, and the 2019 IEEE TCGCC Best Journal Paper Award. He has been serving on the editorial boards of the IEEE Transactions on Green Communications and Networking, IEEE Transactions on Cloud Computing, IEEE Access, IEEE Network, and several other journals. He is named in the current Clarivate Analytics list of “Highly Cited Researchers.”
\end{IEEEbiography}

\balance

\begin{IEEEbiography}[{\includegraphics[width=1in,height=1.25in,clip,keepaspectratio]{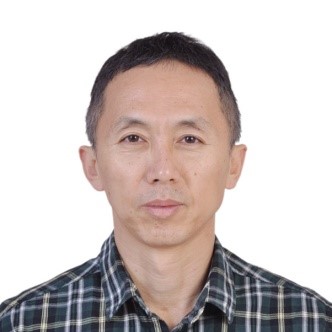}}]{Yanyi Guo}
received his Ph.D. degree in safety system engineering from Beiing Institute of Technology. He was an associate professor with BIT and is currently a senior researcher at Shenzhen MSU-BIT University. His research interests are mainly in the areas of safety modelling and evaluation. 
\end{IEEEbiography}

\begin{IEEEbiography}[{\includegraphics[width=1in,height=1.25in,clip,keepaspectratio]{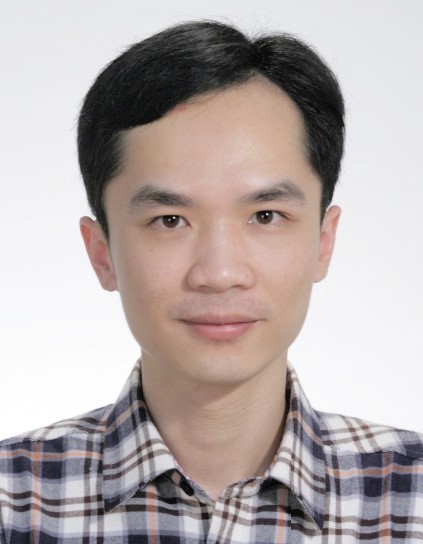}}]{Xiping Hu}
(Member, IEEE) received the Ph.D. degree from the University of British Columbia, Vancouver, BC, Canada. He is currently a professor with Beijing Institute of Technology, and with Shenzhen MSU-BIT University, China. He has more than 150 papers published and presented in prestigious conferences and journals, such as IEEE TPAMI/TMC/TPDS/TIP/JSAC, IEEE COMST, ACM MobiCom/MM/SIGIR/WWW, AAAI, and IJCAI. He has been serving as associate editor of IEEE TCSS, and the lead guest editors of IEEE IoT Journal and IEEE TASE etc. He has been granted several key national research projects as principal investigator. He was the Co-Founder and CTO of Bravolol Ltd., Hong Kong, a leading language learning mobile application company with over 100 million users, and listed as the top 2 language education platform globally. His research areas consist of mobile cyber-physical systems, crowd sensing and affective computing.
\end{IEEEbiography}



\vfill

\end{document}